\documentclass[%
 reprint,
superscriptaddress,
 amsmath,amssymb,
 aps,
]{revtex4-2}

\usepackage[english]{babel}
\usepackage{tabularx}
\usepackage{graphicx}
\usepackage{float}
\usepackage{dcolumn}
\usepackage{bm}
\usepackage[T1]{fontenc}
\usepackage[utf8]{inputenc}
\usepackage{bbm}
\usepackage[hidelinks]{hyperref}
\usepackage{xcolor}

\newcommand\numberthis{\addtocounter{equation}{1}\tag{\theequation}}
\DeclareUnicodeCharacter{202F}{\,}
\begin{document}
\setlength\extrarowheight{4pt}

\newcommand{\ion}[1]{\textcolor{teal}{(\textbf{Ion:} #1)}}

\newcommand{\bR}{\mathbf{R}}
\newcommand{\bRc}{\boldsymbol{\mathcal{R}}}
\newcommand{\bPhi}{\boldsymbol{\Phi}}
\newcommand{\rhos}{\rho_{\bRc,\bPhi}}

\preprint{APS/123-QED}

\title{Superconductivity in RbH$_{12}$ at low pressures: an \emph{ab initio} study}

	\author{{\DJ}or{\dj}e Dangi{\'c}}
	\email{dorde.dangic@ehu.es}
	\affiliation{Fisika Aplikatua Saila, Gipuzkoako Ingeniaritza Eskola, University of the Basque Country (UPV/EHU), 
		Europa Plaza 1, 20018 Donostia/San Sebasti{\'a}n, Spain}
	\affiliation{Centro de F{\'i}sica de Materiales (CFM-MPC), CSIC-UPV/EHU,
		Manuel de Lardizabal Pasealekua 5, 20018 Donostia/San Sebasti{\'a}n, Spain}	
        \author{Manex Alkorta}
		\affiliation{Fisika Aplikatua Saila, Gipuzkoako Ingeniaritza Eskola, University of the Basque Country (UPV/EHU), 
	Europa Plaza 1, 20018 Donostia/San Sebasti{\'a}n, Spain}
		\affiliation{Centro de F{\'i}sica de Materiales (CFM-MPC), CSIC-UPV/EHU,
	Manuel de Lardizabal Pasealekua 5, 20018 Donostia/San Sebasti{\'a}n, Spain}	
	\author{Yue-Wen Fang}
	\affiliation{Centro de F{\'i}sica de Materiales (CFM-MPC), CSIC-UPV/EHU,
		Manuel de Lardizabal Pasealekua 5, 20018 Donostia/San Sebasti{\'a}n, Spain}
	\author{Ion Errea}%
	\affiliation{Fisika Aplikatua Saila, Gipuzkoako Ingeniaritza Eskola, University of the Basque Country (UPV/EHU), 
		Europa Plaza 1, 20018 Donostia/San Sebasti{\'a}n, Spain}
	\affiliation{Centro de F{\'i}sica de Materiales (CFM-MPC), CSIC-UPV/EHU,
		Manuel de Lardizabal Pasealekua 5, 20018 Donostia/San Sebasti{\'a}n, Spain}
	\affiliation{Donostia International Physics Center (DIPC),
		Manuel de Lardizabal Pasealekua 4, 20018 Donostia/San Sebasti{\'a}n, Spain}

\date{\today}

\begin{abstract}
High-pressure polyhydrides are leading contenders for room temperature superconductivity. The next frontier lies in stabilizing them at ambient pressure, which would allow their practical applications. In this first-principles computational study, we investigate the potential for record-low pressure stabilization of binary superhydrides within the RbH$_{12}$ system including lattice quantum anharmonic effects in the calculations. 
We identify five competing phases for the pressure range between 0 and 100 GPa. Incorporating anharmonic and quantum effects on ion dynamics, we find the $Immm$ and $P6_3/mmc$ phases to be the most probable, potentially metastable even at pressures as low as 10 GPa. 
Notably, all phases exhibit metallic properties, with critical temperatures between 46 and 111 K within the pressure range they are dynamically stable. These findings have the potential to inspire future experimental exploration of high-temperature superconductivity at low pressures in Rb-H binary compounds.
\end{abstract}

\maketitle


\section{Introduction}

Superconductivity is one of the most intriguing phenomena in condensed matter physics. It is primarily manifested as a complete absence of electrical resistance below some critical temperature due to the pairing of two electrons into bosonic quasiparticles, Cooper pairs. This property of superconductors allows them to have almost unlimited technological utility. However, depending on the microscopic pairing mechanism, most of the critical temperatures are at cryogenic temperatures, severely limiting their technological application~\cite{superconductivity_challenge}. For example, prior to recent developments, the highest critical temperature for a conventional superconductor, where the attractive interaction between electrons is mediated by phonons, is measured for MgB$_2$ at 39 K~\cite{mgb2}, well below the temperature of liquid nitrogen (77 K). 

In recent years, however, we have seen a rise of a new material class of conventional superconductors, high-pressure hydrides~\cite{hydrides1, hydrides2, hydrides3, hydrides4, FLORESLIVAS20201}. Originally, high-pressure metallic hydrogen was predicted to be a high-temperature superconductor~\cite{Ashcroft1968}. Later on, this idea was revised in the sense that compounds containing large amounts of hydrogen could host high-temperature superconductivity at pressures lower than the one needed to synthesize metallic hydrogen~\cite{Ashcroft2004}. Finally, several years ago, an ab initio prediction of high-temperature superconductivity in H$_3$S was reported~\cite{H3Spred}. Experimental confirmation soon followed in the work of Eremets and co-workers~\cite{H3S}, who had independently been pursuing high-pressure studies of hydrogen sulfide. This led to the explosion of the field, with a large number of synthesized high-temperature superconductors in the last ten years, such as LaH$_{10}$~\cite{LaH10, LaH10_2}, YH$_9$~\cite{YH96}, YH$_6$~\cite{YH6}, CaH$_6$~\cite{CaH6, CaH6_2}, CeH$_{10}$~\cite{CeH10}, and LaBeH8~\cite{LaBeH8} to name a few. 

These high-temperature superconductors require large pressures to be synthesized, in the order of megabars, which significantly limits their technological applications. Hence, the focus of the field in the last few years has shifted from searching for the highest possible critical temperature to the possibility of finding dynamically and thermodynamically stable hydride materials at ambient pressures~\cite{hydrides_at_ambient1}. While thermodynamic stability is unlikely~\cite{hydrides_at_ambient1, gao2025maximumtcconventionalsuperconductors}, the hope is to find a metastable material at ambient pressure. These metastable materials would be formed at an elevated pressure at which they are on the enthalpy convex hull, but remain dynamically stable even after releasing pressure, as it happens with the paradigmatic case of diamond. Remarkably, dynamically stable ternary hydride perovskites at 0 GPa have been predicted recently, with a critical superconducting temperature above the temperature of liquid nitrogen that sparked new hopes for the possibility of discovering new high-temperature superconductors at normal conditions~\cite{MgIrH6_chris, MgIrH6_ion, RbPH3}. 

One of the common properties of high-pressure hydrides is that quantum effects due to the zero-point motion of constituent ions have a large influence on the stability of the structure~\cite{hydrides4}. These effects are exceptionally pronounced in hydrides due to the low mass and consequent large mean square displacements of hydrogen atoms, meaning they can explore large areas of the Born-Oppenheimer (BO) energy surface. As a consequence, this leads to pronounced anharmonic effects in the vibrational spectra of these materials, by changing phonon frequencies, and the total free energy landscape, renormalizing crystal structures. Both of these effects, quantum zero-point motion and anharmonicity, can be captured by the stochastic self-consistent harmonic approximation (SSCHA)~\cite{SSCHA1, SSCHA2, SSCHA3, SSCHA4, SSCHA5}, which already proved its utility in a large number of studies of high-pressure hydrides~\cite{H,LuNH1,Fang2024,H3S_ion2,LaH10_ion,PdH,Lucrezi2023,Lucrezi2024,belli2025,LaScH}. 

While the number of predicted potentially metastable ternary hydrides at low pressures keeps increasing~\cite{MgIrH6_chris, MgIrH6_ion, RbPH3, ThCeBeH, ThCeBeH2, LaBH8, RbBH8, BaSiH}, there has not been any prediction of binary hydrides that could be dynamically and thermodynamically stable at low pressures. Rubidium hydrides are possible candidates, with recent experimental synthesis of RbH$_9$ and RbH$_5$ at low pressures $\sim$ 10 GPa~\cite{RbH_exp, SemenokRbH}. Previous computational studies predicted metalic rubidium superhydrides to be dynamically stable at a relatively low pressure of 50 GPa~\cite{RbH,Hooper2012Rubidium} and thermodynamically stable at 100 GPa. They have also been predicted to be metallic and thus possibly superconducting at high temperatures. These calculations~\cite{RbH,Hooper2012Rubidium} were done purely at a density functional theory (DFT) level and did not consider the influence of the zero point motion and anharmonicity on their properties. 

The goal of the present study is exactly to explore the influence of quantum anharmonic effects on the structural and superconducting properties of RbH$_{12}$ by utilizing the SSCHA. We find that there are several competing phases at a DFT level in the 0-100 GPa pressure range. Quantum anharmonic effects do not significantly alter the energy landscape at 50 GPa, and the $Immm$ and $Cmcm$ phases therefore remain the most probable structures. On the other hand, these effects promote the dynamical stability of these phases, extending the stability of the $Immm$ phase down to 25 GPa, and $P6_3/mmc$ phase down to 10 GPa. All of the low-energy phases are metallic with the possibility of hosting high-temperature superconductivity (up to 100 K). We conclude the study with a presentation of possible structural and spectroscopic signatures of the competing phases in order to facilitate their identification in experiments. 

\section{Methods}

DFT and density functional perturbation theory (DFPT) calculations with
the Perdew-Burke-Ernzerhof parametrization~\cite{PBE} for the generalized gradient approximation were performed using the Quantum Espresso software package~\cite{QE1, QE2, QE3}. Ions were represented using ultrasoft pseudopotentials generated by the ``atomic'' code. Electronic-wave functions were defined in a plane-wave basis with an energy cutoff of 70 Ry, while the energy cutoff for the charge density was 280 Ry. The \textbf{k}-point grid used to sample the electronic states was $24\times 24\times 24$ for $Immm$ and $R\bar{3}m$ phases, $21\times 21\times 14$ for $C2/m$, $16\times 16\times 8$ for $Cmcm$, and $16\times 16\times 11$ for the $P6_3/mmc$ phase. Due to the metallic nature of these compounds, we used a Marzari-Vanderbilt-DeVita-Payne cold smearing~\cite{mv} for electronic states of 0.02 Ry in the self-consistent calculations. The crystal structure prediction was conducted using the CrySpy code~\cite{cryspy}. 

SSCHA calculations were done on $2\times 2\times 1$ supercells for the $Cmcm$ phase and $2\times 2\times 2$ supercell for all other phases. We used 600, 800, 1200, and 1600 configurations of random atomic positions per population for $R\bar{3}m, Immm, C2/m$, and $Cmcm$ and $P6_3/mmc$ phases, respectively. The third-order force constants and Hessian of the free energy were calculated using 8000, 6000, 8000, 10000, and 6000 configurations for $R\bar{3}m, Immm, C2/m$, $Cmcm$, and $P6_3/mmc$ phases, respectively.



The superconducting critical temperature was calculated solving isotropic Migdal-Eliashberg equations using SSCHA auxiliary phonons. We computed the isotropic Coulomb interaction for the $Immm$ phase of RbH$_{12}$ at 25 GPa. The details of the calculation are provided in the Supplementary Material~\cite{yang2024mattersim,Pellegrini_2022,IsoME,MLWF,MLWF2,wan90,disentangle,adaptive_smearing}. For the remaining phases, we employed $\mu^* = 0.118$, estimated from the results obtained for the $Immm$ phase. 
The electron-phonon coupling constants were calculated using DFPT as implemented in Quantum Espresso. They were calculated on $5\times 5\times 5$, $4\times 4\times 4$, $10\times 10\times 10$, $8\times 8\times 8$ and $6\times 6\times 3$ $\mathbf{q}$-point grids for $Immm$, $P6_3/mmc$, $R\bar{3}m$, $C2/m$ and $Cmcm$ phases, respectively. The double delta averaging on the Fermi surface was done using a Gaussian smearing of 0.008 Ry for most phases (0.012 Ry for $R\bar{3}m$) with electronic states calculated on $42\times 42\times 42$, $40\times 40\times 27$, $48\times 48\times 48$, $45\times 45\times 30$ and $40\times 40\times 20$ $\mathbf{k}$-point grids for each mentioned phase.

\section{Results}

\begin{figure}
    \centering
    \includegraphics[width=0.95\linewidth]{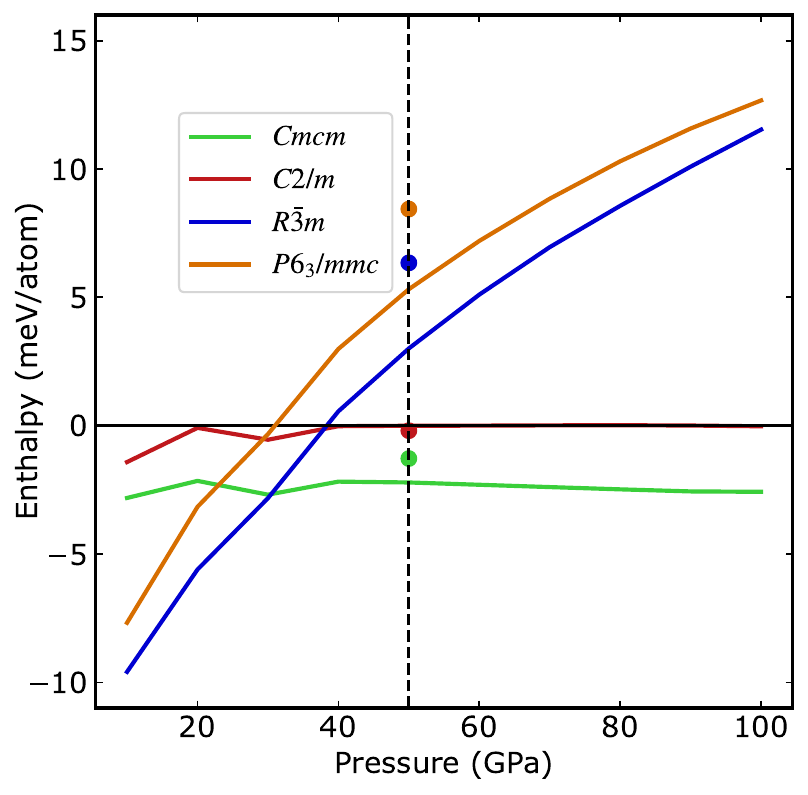}
    \caption{DFT enthalpies of competing phases of RbH$_{12}$ with respect to the $Immm$ phase as a function of pressure. The points represent the differences obtained with the SSCHA at 50 GPa and 0 K.} 
    \label{fig:enthalpies} 
\end{figure}

Ref.~\cite{RbH} identified three main competing phases of RbH$_{12}$ at 50 GPa: $Immm$, $C2/m$ and $Cmcm$. Out of these, the $Cmcm$ phase was found to be on the convex hull at 100 GPa, offering a viable route for its synthesis. Our independent crystal structure prediction calculations confirmed these results with the further identification of two more phases that are below 10 meV/atom from the lowest enthalpy structure at 50 GPa ($Cmcm$ in our crystal structure prediction): $R\bar{3}m$ and $P6_3/mmc$. Both $R\bar{3}m$ and $P6_3/mmc$ are found in Ref.~\cite{RbH} as being competitive at lower pressures, approximately 25 GPa. 
The enthalpy differences of these competitive structures calculated at a DFT level in the 0-100 GPa pressure range are given
in Fig.~\ref{fig:enthalpies} (a). Our results show that $R\bar{3}m$ has the lowest enthalpy between 10 and 30 GPa, while $Cmcm$ phase is the lowest lying structure above this presure. The enthalpy differences between all these competing phases are very small, on the order of few meV per atom. This would imply that the zero-point motion energy, neglected so far, can easily change the energy landscape of this material. 

To check this, we performed a SSCHA structural minimization for all these structures at 50 GPa. The SSCHA method minimizes the total Gibbs free energy in contrast to the BO potential energy $V(\bR)$, where $\bR$ represents the position of all ions, as it is the case in standard DFT calculations (with bold symbols we represent vectors and tensors). Besides the lattice vectors, the minimization is performed with respect to two additional sets of parameters, the SSCHA centroids $\bRc$, which represent the average positions of the ions, and the SSCHA auxiliary force constants $\bPhi$, which are related to the amplitude of the displacements of the ions around the centroid positions.
The SSCHA Gibbs free energy is given as 
\begin{equation}
    G = \langle K + V\rangle _{\rhos} - TS\left[\rhos\right] + P\Omega, \numberthis \label{eq:G}
\end{equation}
where $K$ is the ionic kinetic energy, $V$ the BO potential, $T$ the absolute temperature, $\rhos$ the SSCHA density matrix that is parametrized by $\bRc$ and $\bPhi$, $S\left[\rhos\right]$ the ionic entropy calculated with $\rhos$, $P$ the target pressure for the minimization, and $\Omega$ the volume. At 0 K, the Gibbs free energy reduces to the enthalpy, and the two terms are used interchangeably throughout the text.
Since the probability distribution defined by $\rhos$ is a Gaussian, we can further partition the Gibbs free energy as 
\begin{equation}
    G = F_h + E_{anh} + E_{BO} + P\Omega.
    \label{eq:G2}
\end{equation}
We will name $F_h$ as the phonon free energy. It has an analytical expression as a function of the SSCHA auxiliary frequencies $\omega _{\mu}$, which are obtained by diagonalizing the $\Phi_{ab}/\sqrt{M_aM_b}$ SSCHA auxiliary dynamical matrix ($a$ and $b$ represent both an ion and a Cartesian index): 
\begin{equation}
    F_h = \sum _{\mu}\left\{\frac{\hbar\omega _{\mu}}{2} - k_BT\ln\left[1 + n_{\mu}(T)\right]\right\},
    \label{eq:Fh}
\end{equation}
where $n_{\mu}(T)$ is the Bose-Einstein occupation of mode $\mu$. 
$E_{BO}$ is the value of BO energy surface for the equilibrium centroid positions: $E_{BO} = V(\bRc_{eq})$. Finally, $E_{anh}$ is
\begin{equation}
E_{anh} = \langle V - \mathcal{V}\rangle _{\rhos}
    \label{eq:Eanh}
\end{equation}
where $\mathcal{V}=1/2 (\bR-\bRc)\bPhi (\bR-\bRc)(\bR-\bRc)$ is the SSCHA auxiliary potential.
At the end of the SSCHA minimization, the obtained lattice parameters and centroid positions determine the crystal structure renormalized by anharmonicity and ionic quantum effects, $\bRc_{eq}$. 
The obtained $\bPhi(\bRc_{eq})$, however, does not tell us anything about the curvature of the total free energy at $\bRc_{eq}$, which need not be a local minimum of the total free energy. To check whether it is an actual minimum  and determine if the structure is dynamically stable, one needs to calculate the dynamical matrix defined by the Hessian of the SSCHA free energy and check if the eigenvalues of this matrix are positive~\cite{SSCHA4}.

\setlength{\tabcolsep}{0.6em}
\renewcommand{\arraystretch}{1.1}
\begin{table}[]
\caption{Total Gibbs free energy at 50 GPa and 0 K of different RbH$_{12}$ phases with respect to $Immm$ phase. Definition of each contribution is given in the text. Energy differences are given in meV/atom.} 
\begin{tabular}{ c  c  c c c c }
\hline
\hline
Phase     & $P\Delta\Omega$ & $\Delta F_h$ & $\Delta E_{anh}$ & $\Delta E_{BO}$ & $ \boldsymbol{\Delta G}$ \\ \hline
\multicolumn{6}{c}{SSCHA}                                                                                                                                                                                 \\ \hline
$Cmcm$ & 2.12  & 1.46 & -0.15  & -4.71 & \textbf{-1.28}     \\ 
$C2/m$ & 1.46  & 0.13 & 0.77  & -2.29 & \textbf{-0.20}       \\ 
$R\bar{3}m$ & 13.27 & 3.26 & 0.50 & -10.68 & \textbf{6.34}        \\ 
$P6_3/mmc$ & 11.49  & 3.27 & 0.37 & -6.69 & \textbf{8.44}        \\ \hline
\multicolumn{6}{c}{DFT}                                                                                                                                                                                         \\ \hline
$Cmcm$ & -0.4  & $-$           & $-$               & -1.8         & \textbf{-2.2}       \\ 
$C2/m$    & 0.0            & $-$           & $-$              & 0.0            & \textbf{0.0}        \\ 
$R\bar{3}m$ & 11.2            & $-$           & $-$               & -8.2           & \textbf{3.0}        \\ 
$P6_3/mmc$ & 10.4            & $-$           & $-$               & -5.1            & \textbf{5.3}        \\ 
\hline
\hline
\end{tabular}
\label{tab:G}
\end{table}

Fig.~\ref{fig:enthalpies} shows the Gibbs free energy (enthalpy) in the SSCHA of four phases with respect to the $Immm$ phase at 50 GPa ($\Delta G = G - G^{Immm}$, points) and in DFT in a pressure range from 0 to 100 GPa (lines). Since the plot shows only the relative enthalpy differences between the various $Immm$ phases, these values do not represent distances from the convex hull. The enthalpy differences, split into the different contributions in Eq. \eqref{eq:G2}, are summarized in Table \ref{tab:G} both at the SSCHA and DFT levels.

The biggest contribution to the total difference in every case is given by $P\Delta\Omega$ and $\Delta E_{BO}$ terms. These are also the only two terms that are considered in the DFT minimization of the structure. However, these terms are then renormalized during the SSCHA minimization by an amount which is comparable to the total differences in the DFT case. These renormalizations in this case are of different sign and in the end do not change significantly the enthalpy hierarchy. On the other hand, the phonon free energy $F_h$ has a much smaller effect on the total difference, and is only relevant when the free energy differences are on the meV/atom scale. Finally, the smallest contribution comes from $E_{anh}$, which is usually an order of magnitude smaller than the second smallest contribution, $F_h$. Including all of these contribution does not change the ordering of the structures in terms of stability, although it shifts the relative enthalpy somewhat. For example, $C2/m$ structure is degenerate with $Immm$ in the DFT case, while in the SSCHA case there is a slight difference in the final enthalpies (0.2 meV/atom). This difference comes from the stochastic sampling during SSCHA procedure. This can be verified by checking the symmetry of $C2/m$ structure with a lower tolerance on determining the symmetry which yields the $Immm$ structure. 
 
\begin{figure}
    \centering
    \includegraphics[width=0.95\linewidth]{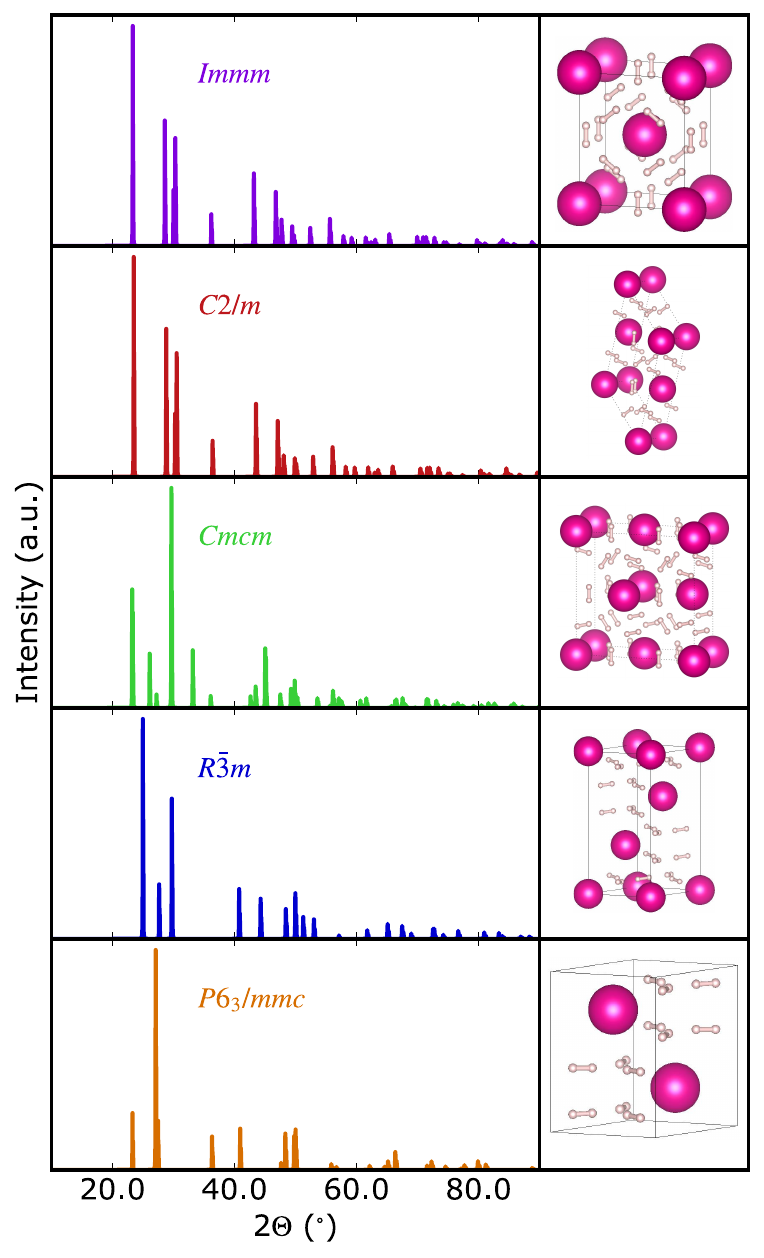}
    \caption{The XRD diffraction pattern for the representative phases in SSCHA at 50 GPa and 0 K. The crystal structure cartoons are generated by VESTA~\cite{VESTA}.}
    \label{fig:xrd} 
\end{figure}

Fig.~\ref{fig:xrd} shows the simulated x-ray diffraction (XRD) patterns of $P6_3/mmc$, $C2/m$, $Immm$, $Cmcm$ and $R\bar{3}m$ phases calculated with VESTA~\cite{VESTA}. XRD patterns are calculated for SSCHA structures at 50 GPa and 0 K. $C2/m$ phase has an almost indistinguishable XRD diffraction pattern to $Immm$, 
which further confirms the previous claim that these two are the same structures. The other four phases should be easily recognized and discriminated from the XRD pattern. In the side plots, we are also showing the structure of these phases in the conventional cell. Hydrogen forms molecules in all phases, without obvious cage structures that are usually associated with high-temperature superconductivity. The formation of hydrogen molecules instead is mostly associated with lower superconducting critical temperatures~\cite{hydrides3, H}. 

\begin{figure}
    \centering
    \includegraphics[width=0.95\linewidth]{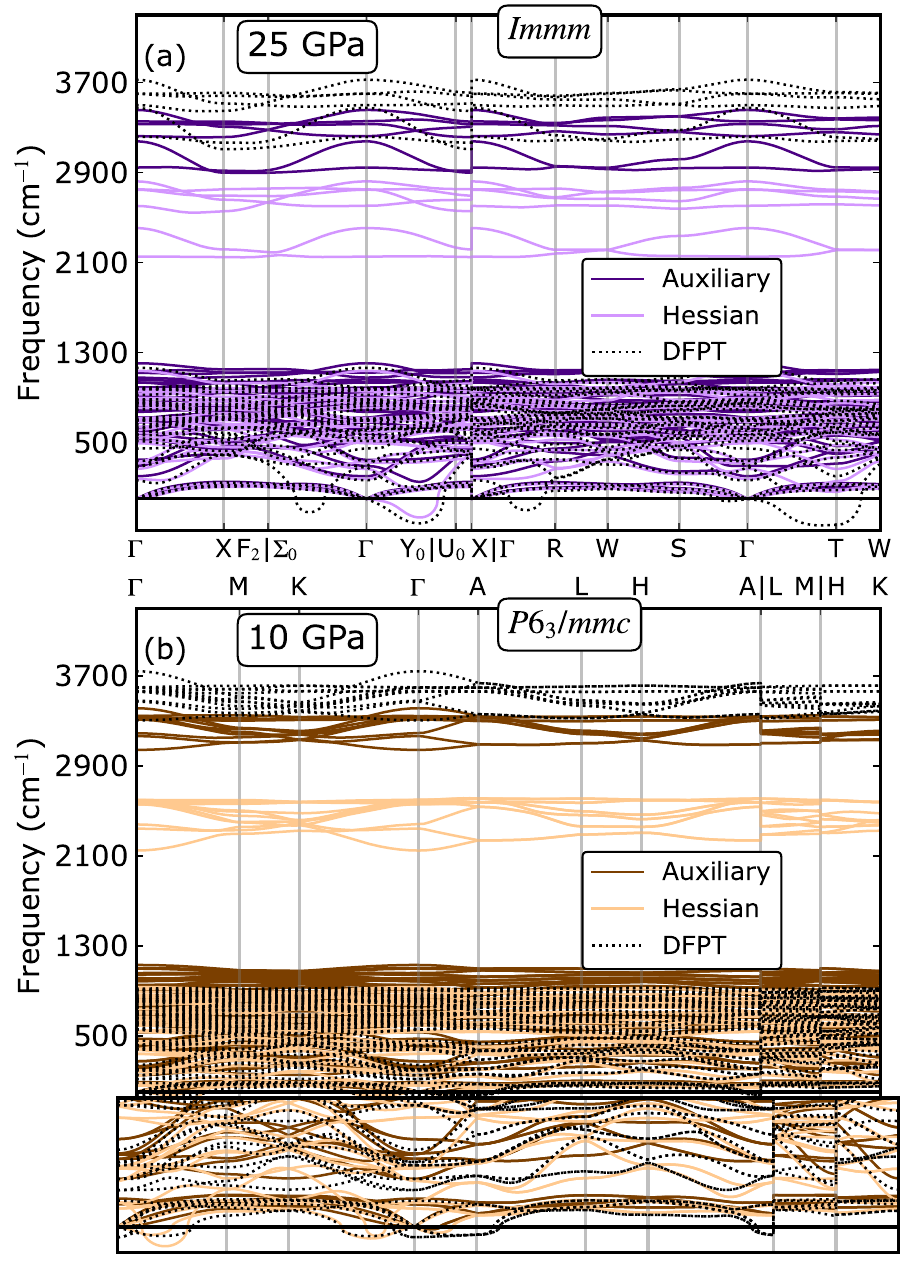}
    \caption{Phonon band structures for (a) $Immm$ phase calculated at 25 and (b) $P6_3/mmc$ phase calculated at 10 GPa and 100 K using SSCHA auxiliary and Hessian, and DFPT harmonic force constants. 
    The inset is a blown up region between -80 and 140 cm$^{-1}$ in order to show phonon instabilities.} 
    \label{fig:phonons}
\end{figure}

Now that we have relaxed the structures within the SSCHA, we can check their dynamical stability by examining the phonon band structure. In Fig.~\ref{fig:phonons} (a) we show the phonon band structure of the $Immm$ phase at 25 GPa. Harmonic DFPT calculations show imaginary phonon modes at the $T$ point and on the $\Gamma - Y_0$ line. The softening of these modes is evidenced in the SSCHA Hessian phonons as well at 25 GPa and 0 K. These modes are mostly of hydrogen character despite being very low-frequency modes. Increasing temperature up to 100 K at 25 GPa stabilizes the phonon at $T$, which is in the commensurate grid of the SSCHA supercell. The instability on the $\Gamma - Y_0$ persists still in this case. 
However, this mode does not lie on a $\mathbf{q}$-point commensurate with the SSCHA supercell and may therefore arise from interpolation artifacts. To verify this, we performed calculations on larger SSCHA supercells using machine-learning interatomic potentials~\cite{yang2024mattersim} (see Supplementary Materials for details). In the larger supercell, all Hessian phonon frequencies—both at commensurate and interpolated $\mathbf{q}$-points—are found to be stable. This means that quantum and temperature effects push the dynamical stability of the RbH$_{12}$ down to 25 GPa, one of the lowest for binary hydride high-temperature superconductors. For the rest of the phonon modes, 100 K does not make a large difference. Finally, as observed in all hydrides, vibrons significantly soften in the SSCHA compared to the harmonic case. 

A similar discussion holds for the $P6_3/mmc$ phase at 10 GPa. Here, DFPT shows an instability in the $A$ high-symmetry point, which is commensurate with the SSCHA supercell and is stabilized by anharmonicity at 100 K (see the inset of Fig.~\ref{fig:phonons} (b)). SSCHA Hessian phonons show instabilities around the $\Gamma$ point and the $\Gamma - M$ high symmetry line similar to DFPT, but these are again interpolation issues since these instabilities are of similar size to the one in $A$, which is stabilized due to anharmonic and quantum effects (see Supplementary material, Section II). Similarly to other high-pressure hydrides, we see a large renormalization of the high-frequency hydrogen vibron modes. Worth noting here is that there are 12 high-frequency vibron modes, compared to only 6 for the $Immm$ phase. 
The phonon band structures of other phases at 50 GPa are shown in the Supplementary Material (Supplementary Figure 2), where in all cases a  similar impact of anharmonicity is observed, with a general tendency to stabilize dynamically the crystal structures.

\begin{figure}
    \centering
    \includegraphics[width=0.95\linewidth]{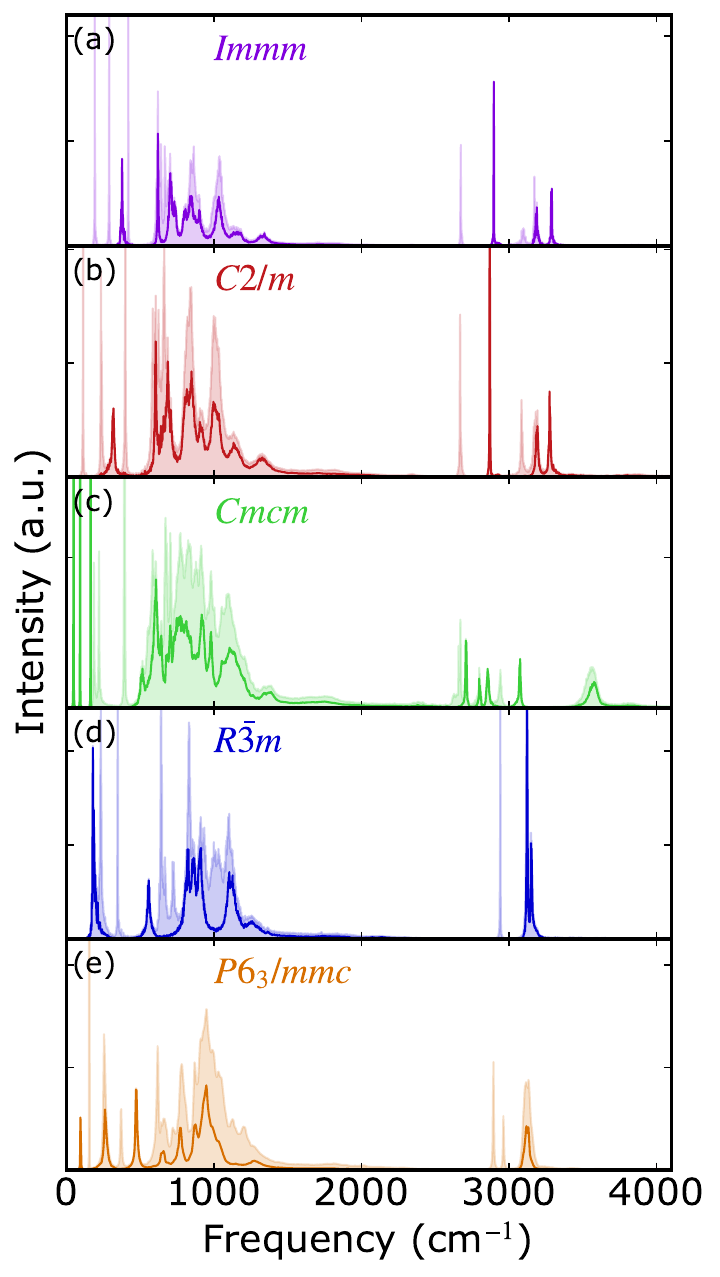}
    \caption{Phonon spectral function at $\Gamma$ for different phases of RbH$_{12}$ (shaded area) at 100 K and 50 GPa. The full line shows only modes that are Raman active.}
    \label{fig:raman}
\end{figure}

Fig.~\ref{fig:raman} shows the Raman active modes of the relevant structures at 50 GPa and 100 K. Since the calculation of the Raman tensor for metallic systems is not trivial~\cite{MILLS1970TheoryRaman}, we generated a random Raman tensor and symmetrized it according to the space group symmetries of a particular structure. This procedure was used only to identify modes with nonzero Raman activity. The resulting intensities were not used in constructing the phonon spectral functions; instead, all Raman-active phonon modes were assigned equal weight. Therefore, although the relative intensities shown in Fig.~\ref{fig:raman} should not be interpreted quantitatively, the positions and overall peak structure remain meaningful. Since we calculated phonon spectral functions within the dynamical bubble approximation at 100 K~\cite{SSCHA4}, the displayed broadenings of spectral lines are also realistic and accurate. While all relevant phases have similar global structures with three distinct segments (low-frequency Rb modes, middle-frequency H modes, and high-frequency H$_2$ vibron modes), they differ significantly in the Raman signatures. For example, high-frequency vibron H$_2$ modes are split into 4, 3, and 1 distinct bands for $Cmcm$, $Immm$ and $C2/m$, and $R\bar{3}m$ and $P6_3/mmc$ phases, respectively. $Immm$ and $C2/m$ phases, can not be properly differentiated in agreement with the findings from XRD patterns. The final two phases ($R\bar{3}m$ and $P6_3/mmc$) should also be easily distinguished by the structure of the middle-frequency hydrogen modes. While all phases show a somewhat large broadening of spectral lines due to phonon-phonon interaction even at 100 K for the middle-frequency phonon modes, the spectral functions do not significantly deviate from the Lorentzian lineshape. This is at odds with the findings for high pressure hydrogen~\cite{H,monacelli_black_2021}, where the large broadening of the hydrogen modes is accompanied with a significant deviation from the Lorentzian lineshape and appearance of satellite peaks. Also, the high-frequency vibron phonon modes do not show a large broadening possibly due to the large gap between these modes and the rest of the phonon spectra, which limits the phonon-phonon interaction because the energy conservation cannot be satisfied.  

\begin{figure}
    \centering
    \includegraphics[width=0.9\linewidth]{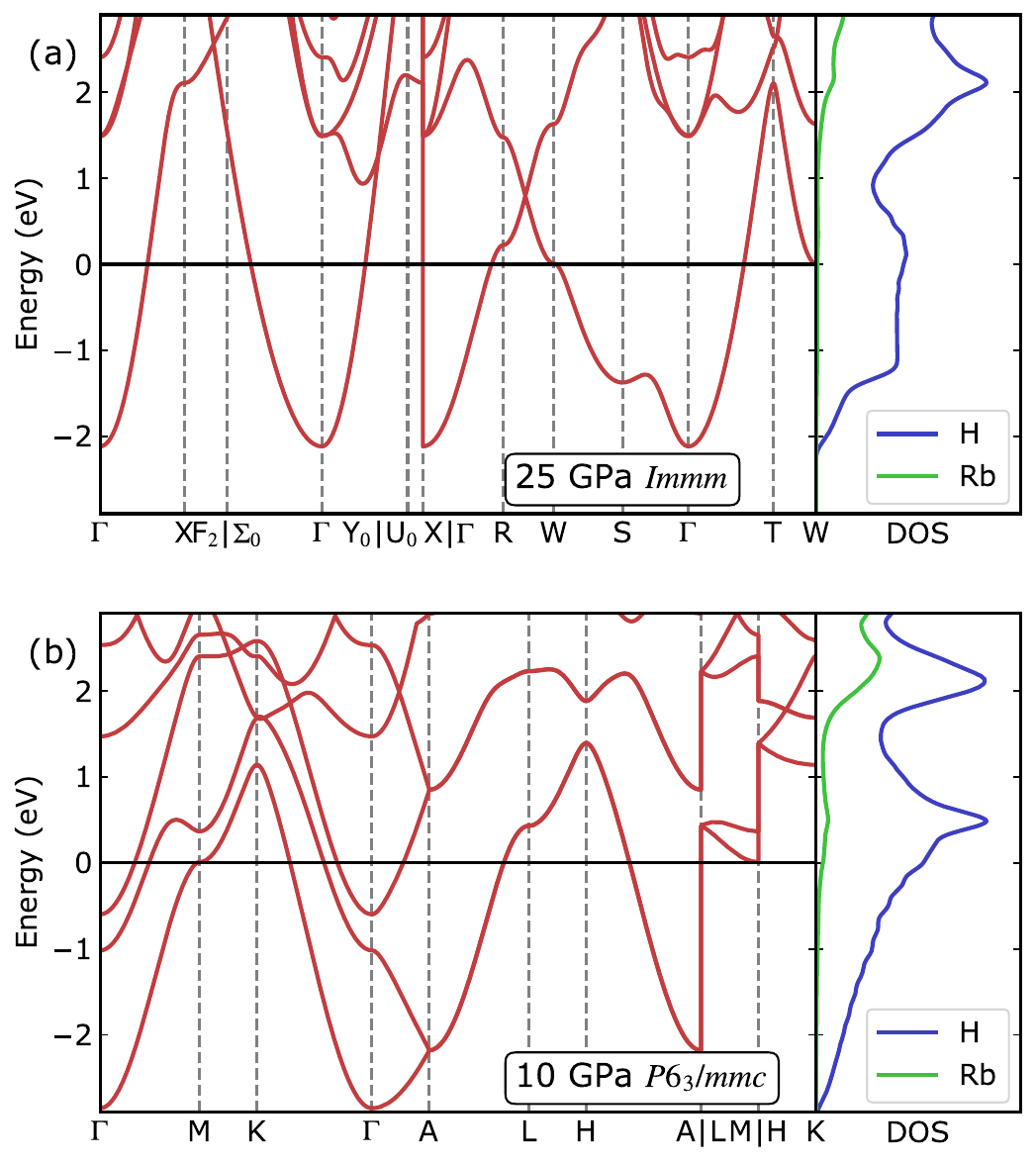}
    \caption{Electronic band structure and density of states of (a) $Immm$  RbH$_{12}$ at 25 GPa and (b) $P6_3/mmc$ RbH$_{12}$ at 10 GPa.}
    \label{fig:electronic_bands}
\end{figure}

Electronic structure calculations reveal that all of the competing phases are metallic and thus could host superconductivity. Fig.~\ref{fig:electronic_bands} (a) displays the electronic band structures of the $Immm$ phase at 25 and (b) the $P6_3/mmc$ phase at 10 GPa. Most of the states (>90\%) at the Fermi level are of hydrogen character, which is one of the most reliable estimators of possible high-temperature superconductivity~\cite{hydrides_at_ambient1, networkingvalue}. Despite the existence of Van Hove singularities near the Fermi level, the electronic density of states is more or less constant in this energy region for the $Immm$ structure, while the $P6_3/mmc$ phase has a noticeable peak at about 500 meV above the Fermi level. This is important because many of the approximations made for the calculation of the superconducting critical temperature in the Migdal-Eliashberg formalism rely on the assumption of a constant electronic density of states~\cite{Allen_Mitrovic, AME}. With increasing pressure, the electronic density of states at the Fermi level in the $Immm$ phase decreases. Other phases show similar trends, specifically, the electronic density of states does not vary a lot in the vicinity of the Fermi level and it is mostly composed of states with large hydrogen character. Electronic band structures of other phases at 50 GPa are shown in the Supplementary Material.

\begin{figure}
    \centering
    \includegraphics[width=0.9\linewidth]{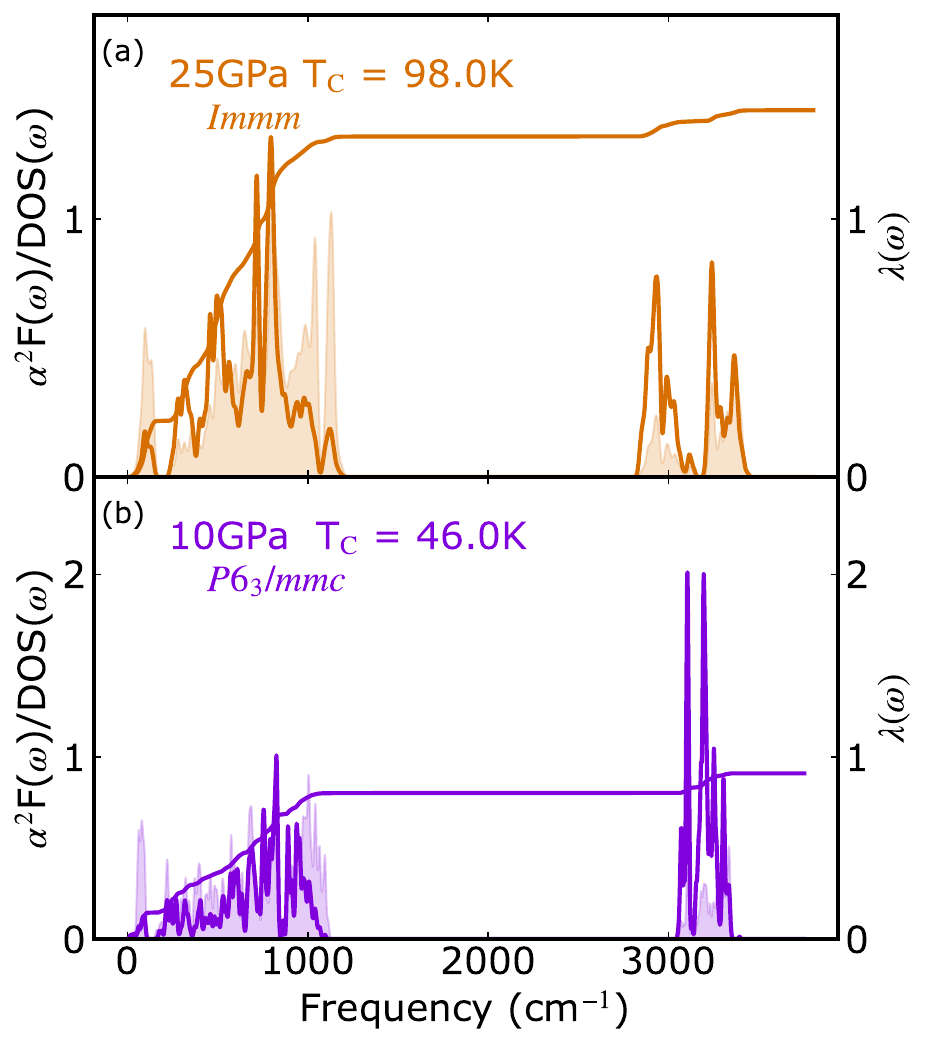}
    \caption{Eliashberg spectral function $\alpha ^2F(\omega)$ of RbH$_{12}$ in the (a) $Immm$ phase at 25 GPa and (b) the $P6_3/mmc$ phase at 10 GPa. The transparent filled line represents the phonon density of states scaled by the integral of $\alpha ^2F(\omega)$. The calculated critical temperature is marked in the figures. For the $Immm$ phase, we calculated the superconducting critical temperature using a Coulomb interaction obtained fully from first-principles calculations, whereas for the $P6_3/mmc$ phase we employed the value $\mu^* = 0.118$ estimated from the $Immm$ phase.}
    \label{fig:a2fs}
\end{figure}

The superconducting critical temperatures of the competing structures were estimated using the isotropic Migdal–Eliashberg equations. The commonly adopted value of the renormalized Coulomb interaction parameter, $\mu^* = 0.1$, frequently used for high-pressure hydrides, is likely underestimated~\cite{MgIrH6_ion, RbPH3}. Therefore, we calculated $\mu^*$ directly from first principles for the $Immm$ phase at 25 GPa. The details of this procedure are provided in the Supplementary Material. The resulting value, corresponding to a Matsubara frequency cutoff of ten times the Debye frequency, is $\mu^* = 0.118$, which we subsequently used in the superconducting temperature calculations for all other phases. In Fig.~\ref{fig:a2fs} we are showing the Eliashberg spectral function $\alpha ^2F(\omega)$ of two structures that are dynamically stable at low pressures and 100 K, $Immm$ at 25 GPa and $P6_3/mmc$ at 10 GPa. The calculated Eliashberg spectral function (see Fig.~\ref{fig:a2fs}) reveals that the electron-phonon coupling is fairly evenly distributed throughout the Brillouin zone, with the Eliashberg spectral function and phonon density of states closely following each other. This is not, however, true for the Rb-dominated modes and the high-frequency H phonon modes (around 1100 cm$^{-1}$), which show a lower electron-phonon coupling strength. The high-frequency vibron modes ($\sim$ 3000 cm$^{-1}$) in the $Immm$ phase on the other hand have a relatively higher electron-phonon coupling. The final electron-phonon coupling constant is fairly low in comparison to other superhydrides and is barely above one for the $Immm$ phase at 25 GPa, while it is below 1 for the $P6_3/mmc$ phase at 10 GPa.

All competing phases (see Supplementary Material for results for other phases) at 50 GPa have a critical temperature of between 59 and 111 K. The $Cmcm$ and $P6_3/mmc$ phases have a lower estimate of the critical temperature of 67 K and 59 K respectively. Both $Cmcm$ and $P6_3/mmc$ phases have two formula units per primitive cell. The lower electron-phonon coupling and critical temperature of the $Cmcm$ phase can be explained by its lower electronic density of states per atom at the Fermi level. On the other hand, the $P6_3/mmc$ phase actually has a higher density of states per atom at the Fermi level compared to the $Immm$. Additionally, the average phonon linewidth due to the electron-phonon interaction is similar in magnitude between $Immm$ and $P6_3/mmc$ phases. However, in the case of the $Immm$ phase lower frequency H phonon modes have larger phonon linewidth, which is then reflected in a higher electron-phonon coupling strength and critical temperature. $R\bar{3}m$ phase has an intermidiate estimate of T$_\textrm{C}$ of 78 K, while $Immm$ and $C2/m$ phases show the largest superconducting critical temperature of above 100 K at 50 GPa.

Pressure has a limited influence on the superconducting critical temperature. In the $Immm$ phase it increases with pressure from 98 K at 25 GPa to 113 K at 100 GPa. On the other hand, the electronic density of states at the Fermi level follows the opposite trend, decreasing with pressure, as well as the electron-phonon coupling constant $\lambda$. This is a consequence of the hardening of low-frequency phonon modes with applied pressure. The integrated Eliashberg spectral function $\alpha ^2F = \int \textrm{d}\omega \alpha ^2F(\omega)$ increases with pressure, which explains the increase of the critical temperature estimate~\cite{belli}. Finally, if we are using SSCHA Hessian frequencies instead of auxiliary phonon frequencies in the calculations for the $Immm$ phase, the estimate of the critical temperature increases up to 155 K at 50 GPa due to the softening of phonon modes due to higher order phonon-phonon interaction. The actual critical temperature should lie somewhere between these two values since Hessian phonons consistently overestimate the softening of phonon frequencies~\cite{H}.

\section{Conclusions}

In conclusion, we have investigated the thermodynamic and dynamical stability of the RbH$_{12}$ system at low-pressure conditions. The crystal structure prediction with \emph{ab initio} energies and volumes suggests five competing phases at 50 GPa. 
Including quantum and anharmonic effects in the estimation of the total free energy does not change the enthalpy hierarchy of phases and only slightly modifies the relative stabilities. At 50 GPa all of the studied systems appear to be dynamically stable, with $Immm$ and $P6_3/mmc$ retaining dynamical stability due to anharmonic effects even as low as 25 and 10 GPa, respectively. To aid the characterization of these materials in the experimental conditions, we simulate the XRD and Raman response of the competing phases and show that they should be easily distinguishable. All of the studied phases are metals with large hydrogen character of the electronic states at the Fermi level. Finally, we estimated the critical temperatures of the studied phases using isotropic Migdal-Eliashberg equations and found that they should show superconductivity between 59 K($P6_3/mmc$ phase) and 111 K ($Immm$ phase).
 
\section{Acknowledgments}

This work is supported by the European Research Council (ERC) under the European Unions Horizon 2020 research and innovation program (Grant Agreement No. 802533), the Spanish Ministry of Science and Innovation (Grant No. PID2022142861NA-I00), the Department of Education, Universities and Research of the Eusko Jaurlaritza and the University of the Basque Country UPV/EHU (Grant No. IT1527-22), and Simons Foundation through the Collaboration on New Frontiers in Superconductivity (Grant No. SFI-MPS-NFS-00006741-10). The project was also partially supported by the IKUR Strategy-High Performance Computing and Artificial Intelligence (HPC$\&$AI) 2025-2026 of the Department of Science, Universities and Innovation of the Basque Government. We acknowledge EuroHPC for granting us access to Lumi located in CSC’s data center in Kajaani, Finland, (Project ID EHPC-REG-2024R01-084) and to RES for giving us access to MareNostrum5, Spain, (Project ID FI-2024-2-0035). 

\bibliography{main}

\providecommand{\noopsort}[1]{}\providecommand{\singleletter}[1]{#1}%
\begin{thebibliography}{71}%
\makeatletter
\providecommand \@ifxundefined [1]{%
 \@ifx{#1\undefined}
}%
\providecommand \@ifnum [1]{%
 \ifnum #1\expandafter \@firstoftwo
 \else \expandafter \@secondoftwo
 \fi
}%
\providecommand \@ifx [1]{%
 \ifx #1\expandafter \@firstoftwo
 \else \expandafter \@secondoftwo
 \fi
}%
\providecommand \natexlab [1]{#1}%
\providecommand \enquote  [1]{``#1''}%
\providecommand \bibnamefont  [1]{#1}%
\providecommand \bibfnamefont [1]{#1}%
\providecommand \citenamefont [1]{#1}%
\providecommand \href@noop [0]{\@secondoftwo}%
\providecommand \href [0]{\begingroup \@sanitize@url \@href}%
\providecommand \@href[1]{\@@startlink{#1}\@@href}%
\providecommand \@@href[1]{\endgroup#1\@@endlink}%
\providecommand \@sanitize@url [0]{\catcode `\\12\catcode `\$12\catcode
  `\&12\catcode `\#12\catcode `\^12\catcode `\_12\catcode `\%12\relax}%
\providecommand \@@startlink[1]{}%
\providecommand \@@endlink[0]{}%
\providecommand \url  [0]{\begingroup\@sanitize@url \@url }%
\providecommand \@url [1]{\endgroup\@href {#1}{\urlprefix }}%
\providecommand \urlprefix  [0]{URL }%
\providecommand \Eprint [0]{\href }%
\providecommand \doibase [0]{https://doi.org/}%
\providecommand \selectlanguage [0]{\@gobble}%
\providecommand \bibinfo  [0]{\@secondoftwo}%
\providecommand \bibfield  [0]{\@secondoftwo}%
\providecommand \translation [1]{[#1]}%
\providecommand \BibitemOpen [0]{}%
\providecommand \bibitemStop [0]{}%
\providecommand \bibitemNoStop [0]{.\EOS\space}%
\providecommand \EOS [0]{\spacefactor3000\relax}%
\providecommand \BibitemShut  [1]{\csname bibitem#1\endcsname}%
\let\auto@bib@innerbib\@empty
\bibitem [{\citenamefont {Yao}\ and\ \citenamefont
  {Ma}(2021)}]{superconductivity_challenge}%
  \BibitemOpen
  \bibfield  {author} {\bibinfo {author} {\bibfnamefont {C.}~\bibnamefont
  {Yao}}\ and\ \bibinfo {author} {\bibfnamefont {Y.}~\bibnamefont {Ma}},\
  }\bibfield  {title} {\bibinfo {title} {Superconducting materials: Challenges
  and opportunities for large-scale applications},\ }\href
  {https://doi.org/https://doi.org/10.1016/j.isci.2021.102541} {\bibfield
  {journal} {\bibinfo  {journal} {iScience}\ }\textbf {\bibinfo {volume}
  {24}},\ \bibinfo {pages} {102541} (\bibinfo {year} {2021})}\BibitemShut
  {NoStop}%
\bibitem [{\citenamefont {Nagamatsu}\ \emph {et~al.}(2001)\citenamefont
  {Nagamatsu}, \citenamefont {Nakagawa}, \citenamefont {Muranaka},
  \citenamefont {Zenitani},\ and\ \citenamefont {Akimitsu}}]{mgb2}%
  \BibitemOpen
  \bibfield  {author} {\bibinfo {author} {\bibfnamefont {J.}~\bibnamefont
  {Nagamatsu}}, \bibinfo {author} {\bibfnamefont {N.}~\bibnamefont {Nakagawa}},
  \bibinfo {author} {\bibfnamefont {T.}~\bibnamefont {Muranaka}}, \bibinfo
  {author} {\bibfnamefont {Y.}~\bibnamefont {Zenitani}},\ and\ \bibinfo
  {author} {\bibfnamefont {J.}~\bibnamefont {Akimitsu}},\ }\bibfield  {title}
  {\bibinfo {title} {Superconductivity at 39{\thinspace}k in magnesium
  diboride},\ }\href {https://doi.org/10.1038/35065039} {\bibfield  {journal}
  {\bibinfo  {journal} {Nature}\ }\textbf {\bibinfo {volume} {410}},\ \bibinfo
  {pages} {63} (\bibinfo {year} {2001})}\BibitemShut {NoStop}%
\bibitem [{\citenamefont {Pickard}\ \emph {et~al.}(2020)\citenamefont
  {Pickard}, \citenamefont {Errea},\ and\ \citenamefont {Eremets}}]{hydrides1}%
  \BibitemOpen
  \bibfield  {author} {\bibinfo {author} {\bibfnamefont {C.~J.}\ \bibnamefont
  {Pickard}}, \bibinfo {author} {\bibfnamefont {I.}~\bibnamefont {Errea}},\
  and\ \bibinfo {author} {\bibfnamefont {M.~I.}\ \bibnamefont {Eremets}},\
  }\bibfield  {title} {\bibinfo {title} {Superconducting hydrides under
  pressure},\ }\href
  {https://doi.org/https://doi.org/10.1146/annurev-conmatphys-031218-013413}
  {\bibfield  {journal} {\bibinfo  {journal} {Annual Review of Condensed Matter
  Physics}\ }\textbf {\bibinfo {volume} {11}},\ \bibinfo {pages} {57} (\bibinfo
  {year} {2020})}\BibitemShut {NoStop}%
\bibitem [{\citenamefont {Eremets}\ \emph {et~al.}(2022)\citenamefont
  {Eremets}, \citenamefont {Minkov}, \citenamefont {Drozdov}, \citenamefont
  {Kong}, \citenamefont {Ksenofontov}, \citenamefont {Shylin}, \citenamefont
  {Bud'ko}, \citenamefont {Prozorov}, \citenamefont {Balakirev}, \citenamefont
  {Sun}, \citenamefont {Mozaffari},\ and\ \citenamefont {Balicas}}]{hydrides2}%
  \BibitemOpen
  \bibfield  {author} {\bibinfo {author} {\bibfnamefont {M.~I.}\ \bibnamefont
  {Eremets}}, \bibinfo {author} {\bibfnamefont {V.~S.}\ \bibnamefont {Minkov}},
  \bibinfo {author} {\bibfnamefont {A.~P.}\ \bibnamefont {Drozdov}}, \bibinfo
  {author} {\bibfnamefont {P.~P.}\ \bibnamefont {Kong}}, \bibinfo {author}
  {\bibfnamefont {V.}~\bibnamefont {Ksenofontov}}, \bibinfo {author}
  {\bibfnamefont {S.~I.}\ \bibnamefont {Shylin}}, \bibinfo {author}
  {\bibfnamefont {S.~L.}\ \bibnamefont {Bud'ko}}, \bibinfo {author}
  {\bibfnamefont {R.}~\bibnamefont {Prozorov}}, \bibinfo {author}
  {\bibfnamefont {F.~F.}\ \bibnamefont {Balakirev}}, \bibinfo {author}
  {\bibfnamefont {D.}~\bibnamefont {Sun}}, \bibinfo {author} {\bibfnamefont
  {S.}~\bibnamefont {Mozaffari}},\ and\ \bibinfo {author} {\bibfnamefont
  {L.}~\bibnamefont {Balicas}},\ }\bibfield  {title} {\bibinfo {title}
  {High-temperature superconductivity in hydrides: Experimental evidence and
  details},\ }\href {https://doi.org/10.1007/s10948-022-06148-1} {\bibfield
  {journal} {\bibinfo  {journal} {Journal of Superconductivity and Novel
  Magnetism}\ }\textbf {\bibinfo {volume} {35}},\ \bibinfo {pages} {965}
  (\bibinfo {year} {2022})}\BibitemShut {NoStop}%
\bibitem [{\citenamefont {Zhao}\ \emph {et~al.}(2023)\citenamefont {Zhao},
  \citenamefont {Huang}, \citenamefont {Zhang}, \citenamefont {Chen},
  \citenamefont {Du}, \citenamefont {Duan},\ and\ \citenamefont
  {Cui}}]{hydrides3}%
  \BibitemOpen
  \bibfield  {author} {\bibinfo {author} {\bibfnamefont {W.}~\bibnamefont
  {Zhao}}, \bibinfo {author} {\bibfnamefont {X.}~\bibnamefont {Huang}},
  \bibinfo {author} {\bibfnamefont {Z.}~\bibnamefont {Zhang}}, \bibinfo
  {author} {\bibfnamefont {S.}~\bibnamefont {Chen}}, \bibinfo {author}
  {\bibfnamefont {M.}~\bibnamefont {Du}}, \bibinfo {author} {\bibfnamefont
  {D.}~\bibnamefont {Duan}},\ and\ \bibinfo {author} {\bibfnamefont
  {T.}~\bibnamefont {Cui}},\ }\bibfield  {title} {\bibinfo {title}
  {{Superconducting ternary hydrides: progress and challenges}},\ }\href
  {https://doi.org/10.1093/nsr/nwad307} {\bibfield  {journal} {\bibinfo
  {journal} {National Science Review}\ }\textbf {\bibinfo {volume} {11}},\
  \bibinfo {pages} {nwad307} (\bibinfo {year} {2023})},\ \Eprint
  {https://arxiv.org/abs/https://academic.oup.com/nsr/article-pdf/11/7/nwad307/58227815/nwad307.pdf}
  {https://academic.oup.com/nsr/article-pdf/11/7/nwad307/58227815/nwad307.pdf}
  \BibitemShut {NoStop}%
\bibitem [{\citenamefont {Boeri}\ \emph {et~al.}(2022)\citenamefont {Boeri},
  \citenamefont {Hennig}, \citenamefont {Hirschfeld}, \citenamefont {Profeta},
  \citenamefont {Sanna}, \citenamefont {Zurek}, \citenamefont {Pickett},
  \citenamefont {Amsler}, \citenamefont {Dias}, \citenamefont {Eremets},
  \citenamefont {Heil}, \citenamefont {Hemley}, \citenamefont {Liu},
  \citenamefont {Ma}, \citenamefont {Pierleoni}, \citenamefont {Kolmogorov},
  \citenamefont {Rybin}, \citenamefont {Novoselov}, \citenamefont {Anisimov},
  \citenamefont {Oganov}, \citenamefont {Pickard}, \citenamefont {Bi},
  \citenamefont {Arita}, \citenamefont {Errea}, \citenamefont {Pellegrini},
  \citenamefont {Requist}, \citenamefont {Gross}, \citenamefont {Margine},
  \citenamefont {Xie}, \citenamefont {Quan}, \citenamefont {Hire},
  \citenamefont {Fanfarillo}, \citenamefont {Stewart}, \citenamefont {Hamlin},
  \citenamefont {Stanev}, \citenamefont {Gonnelli}, \citenamefont {Piatti},
  \citenamefont {Romanin}, \citenamefont {Daghero},\ and\ \citenamefont
  {Valenti}}]{hydrides4}%
  \BibitemOpen
  \bibfield  {author} {\bibinfo {author} {\bibfnamefont {L.}~\bibnamefont
  {Boeri}}, \bibinfo {author} {\bibfnamefont {R.}~\bibnamefont {Hennig}},
  \bibinfo {author} {\bibfnamefont {P.}~\bibnamefont {Hirschfeld}}, \bibinfo
  {author} {\bibfnamefont {G.}~\bibnamefont {Profeta}}, \bibinfo {author}
  {\bibfnamefont {A.}~\bibnamefont {Sanna}}, \bibinfo {author} {\bibfnamefont
  {E.}~\bibnamefont {Zurek}}, \bibinfo {author} {\bibfnamefont {W.~E.}\
  \bibnamefont {Pickett}}, \bibinfo {author} {\bibfnamefont {M.}~\bibnamefont
  {Amsler}}, \bibinfo {author} {\bibfnamefont {R.}~\bibnamefont {Dias}},
  \bibinfo {author} {\bibfnamefont {M.~I.}\ \bibnamefont {Eremets}}, \bibinfo
  {author} {\bibfnamefont {C.}~\bibnamefont {Heil}}, \bibinfo {author}
  {\bibfnamefont {R.~J.}\ \bibnamefont {Hemley}}, \bibinfo {author}
  {\bibfnamefont {H.}~\bibnamefont {Liu}}, \bibinfo {author} {\bibfnamefont
  {Y.}~\bibnamefont {Ma}}, \bibinfo {author} {\bibfnamefont {C.}~\bibnamefont
  {Pierleoni}}, \bibinfo {author} {\bibfnamefont {A.~N.}\ \bibnamefont
  {Kolmogorov}}, \bibinfo {author} {\bibfnamefont {N.}~\bibnamefont {Rybin}},
  \bibinfo {author} {\bibfnamefont {D.}~\bibnamefont {Novoselov}}, \bibinfo
  {author} {\bibfnamefont {V.}~\bibnamefont {Anisimov}}, \bibinfo {author}
  {\bibfnamefont {A.~R.}\ \bibnamefont {Oganov}}, \bibinfo {author}
  {\bibfnamefont {C.~J.}\ \bibnamefont {Pickard}}, \bibinfo {author}
  {\bibfnamefont {T.}~\bibnamefont {Bi}}, \bibinfo {author} {\bibfnamefont
  {R.}~\bibnamefont {Arita}}, \bibinfo {author} {\bibfnamefont
  {I.}~\bibnamefont {Errea}}, \bibinfo {author} {\bibfnamefont
  {C.}~\bibnamefont {Pellegrini}}, \bibinfo {author} {\bibfnamefont
  {R.}~\bibnamefont {Requist}}, \bibinfo {author} {\bibfnamefont {E.~K.~U.}\
  \bibnamefont {Gross}}, \bibinfo {author} {\bibfnamefont {E.~R.}\ \bibnamefont
  {Margine}}, \bibinfo {author} {\bibfnamefont {S.~R.}\ \bibnamefont {Xie}},
  \bibinfo {author} {\bibfnamefont {Y.}~\bibnamefont {Quan}}, \bibinfo {author}
  {\bibfnamefont {A.}~\bibnamefont {Hire}}, \bibinfo {author} {\bibfnamefont
  {L.}~\bibnamefont {Fanfarillo}}, \bibinfo {author} {\bibfnamefont {G.~R.}\
  \bibnamefont {Stewart}}, \bibinfo {author} {\bibfnamefont {J.~J.}\
  \bibnamefont {Hamlin}}, \bibinfo {author} {\bibfnamefont {V.}~\bibnamefont
  {Stanev}}, \bibinfo {author} {\bibfnamefont {R.~S.}\ \bibnamefont
  {Gonnelli}}, \bibinfo {author} {\bibfnamefont {E.}~\bibnamefont {Piatti}},
  \bibinfo {author} {\bibfnamefont {D.}~\bibnamefont {Romanin}}, \bibinfo
  {author} {\bibfnamefont {D.}~\bibnamefont {Daghero}},\ and\ \bibinfo {author}
  {\bibfnamefont {R.}~\bibnamefont {Valenti}},\ }\bibfield  {title} {\bibinfo
  {title} {The 2021 room-temperature superconductivity roadmap},\ }\href
  {https://doi.org/10.1088/1361-648X/ac2864} {\bibfield  {journal} {\bibinfo
  {journal} {Journal of Physics: Condensed Matter}\ }\textbf {\bibinfo {volume}
  {34}},\ \bibinfo {pages} {183002} (\bibinfo {year} {2022})}\BibitemShut
  {NoStop}%
\bibitem [{\citenamefont {Flores-Livas}\ \emph {et~al.}(2020)\citenamefont
  {Flores-Livas}, \citenamefont {Boeri}, \citenamefont {Sanna}, \citenamefont
  {Profeta}, \citenamefont {Arita},\ and\ \citenamefont
  {Eremets}}]{FLORESLIVAS20201}%
  \BibitemOpen
  \bibfield  {author} {\bibinfo {author} {\bibfnamefont {J.~A.}\ \bibnamefont
  {Flores-Livas}}, \bibinfo {author} {\bibfnamefont {L.}~\bibnamefont {Boeri}},
  \bibinfo {author} {\bibfnamefont {A.}~\bibnamefont {Sanna}}, \bibinfo
  {author} {\bibfnamefont {G.}~\bibnamefont {Profeta}}, \bibinfo {author}
  {\bibfnamefont {R.}~\bibnamefont {Arita}},\ and\ \bibinfo {author}
  {\bibfnamefont {M.}~\bibnamefont {Eremets}},\ }\bibfield  {title} {\bibinfo
  {title} {A perspective on conventional high-temperature superconductors at
  high pressure: Methods and materials},\ }\href
  {https://doi.org/https://doi.org/10.1016/j.physrep.2020.02.003} {\bibfield
  {journal} {\bibinfo  {journal} {Physics Reports}\ }\textbf {\bibinfo {volume}
  {856}},\ \bibinfo {pages} {1} (\bibinfo {year} {2020})},\ \bibinfo {note} {a
  perspective on conventional high-temperature superconductors at high
  pressure: Methods and materials}\BibitemShut {NoStop}%
\bibitem [{\citenamefont {Ashcroft}(1968)}]{Ashcroft1968}%
  \BibitemOpen
  \bibfield  {author} {\bibinfo {author} {\bibfnamefont {N.~W.}\ \bibnamefont
  {Ashcroft}},\ }\bibfield  {title} {\bibinfo {title} {{Metallic Hydrogen: A
  High-Temperature Superconductor?}},\ }\href
  {https://doi.org/10.1103/PhysRevLett.21.1748} {\bibfield  {journal} {\bibinfo
   {journal} {Phys. Rev. Lett.}\ }\textbf {\bibinfo {volume} {21}},\ \bibinfo
  {pages} {1748} (\bibinfo {year} {1968})}\BibitemShut {NoStop}%
\bibitem [{\citenamefont {Ashcroft}(2004)}]{Ashcroft2004}%
  \BibitemOpen
  \bibfield  {author} {\bibinfo {author} {\bibfnamefont {N.~W.}\ \bibnamefont
  {Ashcroft}},\ }\bibfield  {title} {\bibinfo {title} {{Hydrogen Dominant
  Metallic Alloys: High Temperature Superconductors?}},\ }\href
  {https://doi.org/10.1103/PhysRevLett.92.187002} {\bibfield  {journal}
  {\bibinfo  {journal} {Phys. Rev. Lett.}\ }\textbf {\bibinfo {volume} {92}},\
  \bibinfo {pages} {187002} (\bibinfo {year} {2004})}\BibitemShut {NoStop}%
\bibitem [{\citenamefont {Duan}\ \emph {et~al.}(2014)\citenamefont {Duan},
  \citenamefont {Liu}, \citenamefont {Tian}, \citenamefont {Li}, \citenamefont
  {Huang}, \citenamefont {Zhao}, \citenamefont {Yu}, \citenamefont {Liu},
  \citenamefont {Tian},\ and\ \citenamefont {Cui}}]{H3Spred}%
  \BibitemOpen
  \bibfield  {author} {\bibinfo {author} {\bibfnamefont {D.}~\bibnamefont
  {Duan}}, \bibinfo {author} {\bibfnamefont {Y.}~\bibnamefont {Liu}}, \bibinfo
  {author} {\bibfnamefont {F.}~\bibnamefont {Tian}}, \bibinfo {author}
  {\bibfnamefont {D.}~\bibnamefont {Li}}, \bibinfo {author} {\bibfnamefont
  {X.}~\bibnamefont {Huang}}, \bibinfo {author} {\bibfnamefont
  {Z.}~\bibnamefont {Zhao}}, \bibinfo {author} {\bibfnamefont {H.}~\bibnamefont
  {Yu}}, \bibinfo {author} {\bibfnamefont {B.}~\bibnamefont {Liu}}, \bibinfo
  {author} {\bibfnamefont {W.}~\bibnamefont {Tian}},\ and\ \bibinfo {author}
  {\bibfnamefont {T.}~\bibnamefont {Cui}},\ }\bibfield  {title} {\bibinfo
  {title} {Pressure-induced metallization of dense ({H2S}){2H2} with high-{T}c
  superconductivity},\ }\href {https://doi.org/10.1038/srep06968} {\bibfield
  {journal} {\bibinfo  {journal} {Scientific Reports}\ }\textbf {\bibinfo
  {volume} {4}},\ \bibinfo {pages} {6968} (\bibinfo {year} {2014})}\BibitemShut
  {NoStop}%
\bibitem [{\citenamefont {Drozdov}\ \emph {et~al.}(2015)\citenamefont
  {Drozdov}, \citenamefont {Eremets}, \citenamefont {Troyan}, \citenamefont
  {Ksenofontov},\ and\ \citenamefont {Shylin}}]{H3S}%
  \BibitemOpen
  \bibfield  {author} {\bibinfo {author} {\bibfnamefont {A.~P.}\ \bibnamefont
  {Drozdov}}, \bibinfo {author} {\bibfnamefont {M.~I.}\ \bibnamefont
  {Eremets}}, \bibinfo {author} {\bibfnamefont {I.~A.}\ \bibnamefont {Troyan}},
  \bibinfo {author} {\bibfnamefont {V.}~\bibnamefont {Ksenofontov}},\ and\
  \bibinfo {author} {\bibfnamefont {S.~I.}\ \bibnamefont {Shylin}},\ }\bibfield
   {title} {\bibinfo {title} {{Conventional superconductivity at 203 kelvin at
  high pressures in the sulfur hydride system}},\ }\href
  {https://doi.org/10.1038/nature14964} {\bibfield  {journal} {\bibinfo
  {journal} {Nature}\ }\textbf {\bibinfo {volume} {525}},\ \bibinfo {pages}
  {73} (\bibinfo {year} {2015})}\BibitemShut {NoStop}%
\bibitem [{\citenamefont {Somayazulu}\ \emph {et~al.}(2019)\citenamefont
  {Somayazulu}, \citenamefont {Ahart}, \citenamefont {Mishra}, \citenamefont
  {Geballe}, \citenamefont {Baldini}, \citenamefont {Meng}, \citenamefont
  {Struzhkin},\ and\ \citenamefont {Hemley}}]{LaH10}%
  \BibitemOpen
  \bibfield  {author} {\bibinfo {author} {\bibfnamefont {M.}~\bibnamefont
  {Somayazulu}}, \bibinfo {author} {\bibfnamefont {M.}~\bibnamefont {Ahart}},
  \bibinfo {author} {\bibfnamefont {A.~K.}\ \bibnamefont {Mishra}}, \bibinfo
  {author} {\bibfnamefont {Z.~M.}\ \bibnamefont {Geballe}}, \bibinfo {author}
  {\bibfnamefont {M.}~\bibnamefont {Baldini}}, \bibinfo {author} {\bibfnamefont
  {Y.}~\bibnamefont {Meng}}, \bibinfo {author} {\bibfnamefont {V.~V.}\
  \bibnamefont {Struzhkin}},\ and\ \bibinfo {author} {\bibfnamefont {R.~J.}\
  \bibnamefont {Hemley}},\ }\bibfield  {title} {\bibinfo {title} {Evidence for
  superconductivity above 260 k in lanthanum superhydride at megabar
  pressures},\ }\href {https://doi.org/10.1103/PhysRevLett.122.027001}
  {\bibfield  {journal} {\bibinfo  {journal} {Phys. Rev. Lett.}\ }\textbf
  {\bibinfo {volume} {122}},\ \bibinfo {pages} {027001} (\bibinfo {year}
  {2019})}\BibitemShut {NoStop}%
\bibitem [{\citenamefont {Drozdov}\ \emph {et~al.}(2019)\citenamefont
  {Drozdov}, \citenamefont {Kong}, \citenamefont {Minkov}, \citenamefont
  {Besedin}, \citenamefont {Kuzovnikov}, \citenamefont {Mozaffari},
  \citenamefont {Balicas}, \citenamefont {Balakirev}, \citenamefont {Graf},
  \citenamefont {Prakapenka}, \citenamefont {Greenberg}, \citenamefont
  {Knyazev}, \citenamefont {Tkacz},\ and\ \citenamefont {Eremets}}]{LaH10_2}%
  \BibitemOpen
  \bibfield  {author} {\bibinfo {author} {\bibfnamefont {A.~P.}\ \bibnamefont
  {Drozdov}}, \bibinfo {author} {\bibfnamefont {P.~P.}\ \bibnamefont {Kong}},
  \bibinfo {author} {\bibfnamefont {V.~S.}\ \bibnamefont {Minkov}}, \bibinfo
  {author} {\bibfnamefont {S.~P.}\ \bibnamefont {Besedin}}, \bibinfo {author}
  {\bibfnamefont {M.~A.}\ \bibnamefont {Kuzovnikov}}, \bibinfo {author}
  {\bibfnamefont {S.}~\bibnamefont {Mozaffari}}, \bibinfo {author}
  {\bibfnamefont {L.}~\bibnamefont {Balicas}}, \bibinfo {author} {\bibfnamefont
  {F.~F.}\ \bibnamefont {Balakirev}}, \bibinfo {author} {\bibfnamefont {D.~E.}\
  \bibnamefont {Graf}}, \bibinfo {author} {\bibfnamefont {V.~B.}\ \bibnamefont
  {Prakapenka}}, \bibinfo {author} {\bibfnamefont {E.}~\bibnamefont
  {Greenberg}}, \bibinfo {author} {\bibfnamefont {D.~A.}\ \bibnamefont
  {Knyazev}}, \bibinfo {author} {\bibfnamefont {M.}~\bibnamefont {Tkacz}},\
  and\ \bibinfo {author} {\bibfnamefont {M.~I.}\ \bibnamefont {Eremets}},\
  }\bibfield  {title} {\bibinfo {title} {Superconductivity at 250 {K} in
  lanthanum hydride under high pressures},\ }\href@noop {} {\bibfield
  {journal} {\bibinfo  {journal} {Nature}\ }\textbf {\bibinfo {volume} {569}},\
  \bibinfo {pages} {528} (\bibinfo {year} {2019})}\BibitemShut {NoStop}%
\bibitem [{\citenamefont {Kong}\ \emph {et~al.}(2021)\citenamefont {Kong},
  \citenamefont {Minkov}, \citenamefont {Kuzovnikov}, \citenamefont {Drozdov},
  \citenamefont {Besedin}, \citenamefont {Mozaffari}, \citenamefont {Balicas},
  \citenamefont {Balakirev}, \citenamefont {Prakapenka}, \citenamefont
  {Chariton}, \citenamefont {Knyazev}, \citenamefont {Greenberg},\ and\
  \citenamefont {Eremets}}]{YH96}%
  \BibitemOpen
  \bibfield  {author} {\bibinfo {author} {\bibfnamefont {P.}~\bibnamefont
  {Kong}}, \bibinfo {author} {\bibfnamefont {V.~S.}\ \bibnamefont {Minkov}},
  \bibinfo {author} {\bibfnamefont {M.~A.}\ \bibnamefont {Kuzovnikov}},
  \bibinfo {author} {\bibfnamefont {A.~P.}\ \bibnamefont {Drozdov}}, \bibinfo
  {author} {\bibfnamefont {S.~P.}\ \bibnamefont {Besedin}}, \bibinfo {author}
  {\bibfnamefont {S.}~\bibnamefont {Mozaffari}}, \bibinfo {author}
  {\bibfnamefont {L.}~\bibnamefont {Balicas}}, \bibinfo {author} {\bibfnamefont
  {F.~F.}\ \bibnamefont {Balakirev}}, \bibinfo {author} {\bibfnamefont {V.~B.}\
  \bibnamefont {Prakapenka}}, \bibinfo {author} {\bibfnamefont
  {S.}~\bibnamefont {Chariton}}, \bibinfo {author} {\bibfnamefont {D.~A.}\
  \bibnamefont {Knyazev}}, \bibinfo {author} {\bibfnamefont {E.}~\bibnamefont
  {Greenberg}},\ and\ \bibinfo {author} {\bibfnamefont {M.~I.}\ \bibnamefont
  {Eremets}},\ }\bibfield  {title} {\bibinfo {title} {Superconductivity up to
  243{\thinspace}k in the yttrium-hydrogen system under high pressure},\ }\href
  {https://doi.org/10.1038/s41467-021-25372-2} {\bibfield  {journal} {\bibinfo
  {journal} {Nature Communications}\ }\textbf {\bibinfo {volume} {12}},\
  \bibinfo {pages} {5075} (\bibinfo {year} {2021})}\BibitemShut {NoStop}%
\bibitem [{\citenamefont {Troyan}\ \emph {et~al.}(2021)\citenamefont {Troyan},
  \citenamefont {Semenok}, \citenamefont {Kvashnin}, \citenamefont {Sadakov},
  \citenamefont {Sobolevskiy}, \citenamefont {Pudalov}, \citenamefont
  {Ivanova}, \citenamefont {Prakapenka}, \citenamefont {Greenberg},
  \citenamefont {Gavriliuk}, \citenamefont {Lyubutin}, \citenamefont
  {Struzhkin}, \citenamefont {Bergara}, \citenamefont {Errea}, \citenamefont
  {Bianco}, \citenamefont {Calandra}, \citenamefont {Mauri}, \citenamefont
  {Monacelli}, \citenamefont {Akashi},\ and\ \citenamefont {Oganov}}]{YH6}%
  \BibitemOpen
  \bibfield  {author} {\bibinfo {author} {\bibfnamefont {I.~A.}\ \bibnamefont
  {Troyan}}, \bibinfo {author} {\bibfnamefont {D.~V.}\ \bibnamefont {Semenok}},
  \bibinfo {author} {\bibfnamefont {A.~G.}\ \bibnamefont {Kvashnin}}, \bibinfo
  {author} {\bibfnamefont {A.~V.}\ \bibnamefont {Sadakov}}, \bibinfo {author}
  {\bibfnamefont {O.~A.}\ \bibnamefont {Sobolevskiy}}, \bibinfo {author}
  {\bibfnamefont {V.~M.}\ \bibnamefont {Pudalov}}, \bibinfo {author}
  {\bibfnamefont {A.~G.}\ \bibnamefont {Ivanova}}, \bibinfo {author}
  {\bibfnamefont {V.~B.}\ \bibnamefont {Prakapenka}}, \bibinfo {author}
  {\bibfnamefont {E.}~\bibnamefont {Greenberg}}, \bibinfo {author}
  {\bibfnamefont {A.~G.}\ \bibnamefont {Gavriliuk}}, \bibinfo {author}
  {\bibfnamefont {I.~S.}\ \bibnamefont {Lyubutin}}, \bibinfo {author}
  {\bibfnamefont {V.~V.}\ \bibnamefont {Struzhkin}}, \bibinfo {author}
  {\bibfnamefont {A.}~\bibnamefont {Bergara}}, \bibinfo {author} {\bibfnamefont
  {I.}~\bibnamefont {Errea}}, \bibinfo {author} {\bibfnamefont
  {R.}~\bibnamefont {Bianco}}, \bibinfo {author} {\bibfnamefont
  {M.}~\bibnamefont {Calandra}}, \bibinfo {author} {\bibfnamefont
  {F.}~\bibnamefont {Mauri}}, \bibinfo {author} {\bibfnamefont
  {L.}~\bibnamefont {Monacelli}}, \bibinfo {author} {\bibfnamefont
  {R.}~\bibnamefont {Akashi}},\ and\ \bibinfo {author} {\bibfnamefont {A.~R.}\
  \bibnamefont {Oganov}},\ }\bibfield  {title} {\bibinfo {title} {Anomalous
  high-temperature superconductivity in yh6},\ }\href
  {https://doi.org/https://doi.org/10.1002/adma.202006832} {\bibfield
  {journal} {\bibinfo  {journal} {Advanced Materials}\ }\textbf {\bibinfo
  {volume} {33}},\ \bibinfo {pages} {2006832} (\bibinfo {year} {2021})},\
  \Eprint
  {https://arxiv.org/abs/https://onlinelibrary.wiley.com/doi/pdf/10.1002/adma.202006832}
  {https://onlinelibrary.wiley.com/doi/pdf/10.1002/adma.202006832} \BibitemShut
  {NoStop}%
\bibitem [{\citenamefont {Ma}\ \emph {et~al.}(2022)\citenamefont {Ma},
  \citenamefont {Wang}, \citenamefont {Xie}, \citenamefont {Yang},
  \citenamefont {Wang}, \citenamefont {Zhou}, \citenamefont {Liu},
  \citenamefont {Yu}, \citenamefont {Zhao}, \citenamefont {Wang}, \citenamefont
  {Liu},\ and\ \citenamefont {Ma}}]{CaH6}%
  \BibitemOpen
  \bibfield  {author} {\bibinfo {author} {\bibfnamefont {L.}~\bibnamefont
  {Ma}}, \bibinfo {author} {\bibfnamefont {K.}~\bibnamefont {Wang}}, \bibinfo
  {author} {\bibfnamefont {Y.}~\bibnamefont {Xie}}, \bibinfo {author}
  {\bibfnamefont {X.}~\bibnamefont {Yang}}, \bibinfo {author} {\bibfnamefont
  {Y.}~\bibnamefont {Wang}}, \bibinfo {author} {\bibfnamefont {M.}~\bibnamefont
  {Zhou}}, \bibinfo {author} {\bibfnamefont {H.}~\bibnamefont {Liu}}, \bibinfo
  {author} {\bibfnamefont {X.}~\bibnamefont {Yu}}, \bibinfo {author}
  {\bibfnamefont {Y.}~\bibnamefont {Zhao}}, \bibinfo {author} {\bibfnamefont
  {H.}~\bibnamefont {Wang}}, \bibinfo {author} {\bibfnamefont {G.}~\bibnamefont
  {Liu}},\ and\ \bibinfo {author} {\bibfnamefont {Y.}~\bibnamefont {Ma}},\
  }\bibfield  {title} {\bibinfo {title} {High-temperature superconducting phase
  in clathrate calcium hydride ${\mathrm{cah}}_{6}$ up to 215 k at a pressure
  of 172 gpa},\ }\href {https://doi.org/10.1103/PhysRevLett.128.167001}
  {\bibfield  {journal} {\bibinfo  {journal} {Phys. Rev. Lett.}\ }\textbf
  {\bibinfo {volume} {128}},\ \bibinfo {pages} {167001} (\bibinfo {year}
  {2022})}\BibitemShut {NoStop}%
\bibitem [{\citenamefont {Li}\ \emph {et~al.}(2022{\natexlab{a}})\citenamefont
  {Li}, \citenamefont {He}, \citenamefont {Zhang}, \citenamefont {Wang},
  \citenamefont {Zhang}, \citenamefont {Jia}, \citenamefont {Feng},
  \citenamefont {Lu}, \citenamefont {Zhao}, \citenamefont {Zhang},
  \citenamefont {Min}, \citenamefont {Long}, \citenamefont {Yu}, \citenamefont
  {Wang}, \citenamefont {Ye}, \citenamefont {Zhang}, \citenamefont
  {Prakapenka}, \citenamefont {Chariton}, \citenamefont {Ginsberg},
  \citenamefont {Bass}, \citenamefont {Yuan}, \citenamefont {Liu},\ and\
  \citenamefont {Jin}}]{CaH6_2}%
  \BibitemOpen
  \bibfield  {author} {\bibinfo {author} {\bibfnamefont {Z.}~\bibnamefont
  {Li}}, \bibinfo {author} {\bibfnamefont {X.}~\bibnamefont {He}}, \bibinfo
  {author} {\bibfnamefont {C.}~\bibnamefont {Zhang}}, \bibinfo {author}
  {\bibfnamefont {X.}~\bibnamefont {Wang}}, \bibinfo {author} {\bibfnamefont
  {S.}~\bibnamefont {Zhang}}, \bibinfo {author} {\bibfnamefont
  {Y.}~\bibnamefont {Jia}}, \bibinfo {author} {\bibfnamefont {S.}~\bibnamefont
  {Feng}}, \bibinfo {author} {\bibfnamefont {K.}~\bibnamefont {Lu}}, \bibinfo
  {author} {\bibfnamefont {J.}~\bibnamefont {Zhao}}, \bibinfo {author}
  {\bibfnamefont {J.}~\bibnamefont {Zhang}}, \bibinfo {author} {\bibfnamefont
  {B.}~\bibnamefont {Min}}, \bibinfo {author} {\bibfnamefont {Y.}~\bibnamefont
  {Long}}, \bibinfo {author} {\bibfnamefont {R.}~\bibnamefont {Yu}}, \bibinfo
  {author} {\bibfnamefont {L.}~\bibnamefont {Wang}}, \bibinfo {author}
  {\bibfnamefont {M.}~\bibnamefont {Ye}}, \bibinfo {author} {\bibfnamefont
  {Z.}~\bibnamefont {Zhang}}, \bibinfo {author} {\bibfnamefont
  {V.}~\bibnamefont {Prakapenka}}, \bibinfo {author} {\bibfnamefont
  {S.}~\bibnamefont {Chariton}}, \bibinfo {author} {\bibfnamefont {P.~A.}\
  \bibnamefont {Ginsberg}}, \bibinfo {author} {\bibfnamefont {J.}~\bibnamefont
  {Bass}}, \bibinfo {author} {\bibfnamefont {S.}~\bibnamefont {Yuan}}, \bibinfo
  {author} {\bibfnamefont {H.}~\bibnamefont {Liu}},\ and\ \bibinfo {author}
  {\bibfnamefont {C.}~\bibnamefont {Jin}},\ }\bibfield  {title} {\bibinfo
  {title} {Superconductivity above 200 k discovered in superhydrides of
  calcium},\ }\href {https://doi.org/10.1038/s41467-022-30454-w} {\bibfield
  {journal} {\bibinfo  {journal} {Nature Communications}\ }\textbf {\bibinfo
  {volume} {13}},\ \bibinfo {pages} {2863} (\bibinfo {year}
  {2022}{\natexlab{a}})}\BibitemShut {NoStop}%
\bibitem [{\citenamefont {Chen}\ \emph {et~al.}(2021)\citenamefont {Chen},
  \citenamefont {Semenok}, \citenamefont {Huang}, \citenamefont {Shu},
  \citenamefont {Li}, \citenamefont {Duan}, \citenamefont {Cui},\ and\
  \citenamefont {Oganov}}]{CeH10}%
  \BibitemOpen
  \bibfield  {author} {\bibinfo {author} {\bibfnamefont {W.}~\bibnamefont
  {Chen}}, \bibinfo {author} {\bibfnamefont {D.~V.}\ \bibnamefont {Semenok}},
  \bibinfo {author} {\bibfnamefont {X.}~\bibnamefont {Huang}}, \bibinfo
  {author} {\bibfnamefont {H.}~\bibnamefont {Shu}}, \bibinfo {author}
  {\bibfnamefont {X.}~\bibnamefont {Li}}, \bibinfo {author} {\bibfnamefont
  {D.}~\bibnamefont {Duan}}, \bibinfo {author} {\bibfnamefont {T.}~\bibnamefont
  {Cui}},\ and\ \bibinfo {author} {\bibfnamefont {A.~R.}\ \bibnamefont
  {Oganov}},\ }\bibfield  {title} {\bibinfo {title} {High-temperature
  superconducting phases in cerium superhydride with a ${T}_{c}$ up to 115 k
  below a pressure of 1 megabar},\ }\href
  {https://doi.org/10.1103/PhysRevLett.127.117001} {\bibfield  {journal}
  {\bibinfo  {journal} {Phys. Rev. Lett.}\ }\textbf {\bibinfo {volume} {127}},\
  \bibinfo {pages} {117001} (\bibinfo {year} {2021})}\BibitemShut {NoStop}%
\bibitem [{\citenamefont {Song}\ \emph {et~al.}(2023)\citenamefont {Song},
  \citenamefont {Bi}, \citenamefont {Nakamoto}, \citenamefont {Shimizu},
  \citenamefont {Liu}, \citenamefont {Zou}, \citenamefont {Liu}, \citenamefont
  {Wang},\ and\ \citenamefont {Ma}}]{LaBeH8}%
  \BibitemOpen
  \bibfield  {author} {\bibinfo {author} {\bibfnamefont {Y.}~\bibnamefont
  {Song}}, \bibinfo {author} {\bibfnamefont {J.}~\bibnamefont {Bi}}, \bibinfo
  {author} {\bibfnamefont {Y.}~\bibnamefont {Nakamoto}}, \bibinfo {author}
  {\bibfnamefont {K.}~\bibnamefont {Shimizu}}, \bibinfo {author} {\bibfnamefont
  {H.}~\bibnamefont {Liu}}, \bibinfo {author} {\bibfnamefont {B.}~\bibnamefont
  {Zou}}, \bibinfo {author} {\bibfnamefont {G.}~\bibnamefont {Liu}}, \bibinfo
  {author} {\bibfnamefont {H.}~\bibnamefont {Wang}},\ and\ \bibinfo {author}
  {\bibfnamefont {Y.}~\bibnamefont {Ma}},\ }\bibfield  {title} {\bibinfo
  {title} {Stoichiometric ternary superhydride ${\mathrm{labeh}}_{8}$ as a new
  template for high-temperature superconductivity at 110 k under 80 gpa},\
  }\href {https://doi.org/10.1103/PhysRevLett.130.266001} {\bibfield  {journal}
  {\bibinfo  {journal} {Phys. Rev. Lett.}\ }\textbf {\bibinfo {volume} {130}},\
  \bibinfo {pages} {266001} (\bibinfo {year} {2023})}\BibitemShut {NoStop}%
\bibitem [{\citenamefont {Cerqueira}\ \emph {et~al.}()\citenamefont
  {Cerqueira}, \citenamefont {Fang}, \citenamefont {Errea}, \citenamefont
  {Sanna},\ and\ \citenamefont {Marques}}]{hydrides_at_ambient1}%
  \BibitemOpen
  \bibfield  {author} {\bibinfo {author} {\bibfnamefont {T.~F.~T.}\
  \bibnamefont {Cerqueira}}, \bibinfo {author} {\bibfnamefont {Y.-W.}\
  \bibnamefont {Fang}}, \bibinfo {author} {\bibfnamefont {I.}~\bibnamefont
  {Errea}}, \bibinfo {author} {\bibfnamefont {A.}~\bibnamefont {Sanna}},\ and\
  \bibinfo {author} {\bibfnamefont {M.~A.~L.}\ \bibnamefont {Marques}},\
  }\bibfield  {title} {\bibinfo {title} {Searching materials space for hydride
  superconductors at ambient pressure},\ }\href
  {https://doi.org/https://doi.org/10.1002/adfm.202404043} {\bibfield
  {journal} {\bibinfo  {journal} {Advanced Functional Materials}\ }\textbf
  {\bibinfo {volume} {n/a}},\ \bibinfo {pages} {2404043}},\ \Eprint
  {https://arxiv.org/abs/https://onlinelibrary.wiley.com/doi/pdf/10.1002/adfm.202404043}
  {https://onlinelibrary.wiley.com/doi/pdf/10.1002/adfm.202404043} \BibitemShut
  {NoStop}%
\bibitem [{\citenamefont {Gao}\ \emph {et~al.}(2025)\citenamefont {Gao},
  \citenamefont {Cerqueira}, \citenamefont {Sanna}, \citenamefont {Fang},
  \citenamefont {Dangi{\'{c}}}, \citenamefont {Errea}, \citenamefont {Wang},
  \citenamefont {Botti},\ and\ \citenamefont
  {Marques}}]{gao2025maximumtcconventionalsuperconductors}%
  \BibitemOpen
  \bibfield  {author} {\bibinfo {author} {\bibfnamefont {K.}~\bibnamefont
  {Gao}}, \bibinfo {author} {\bibfnamefont {T.~F.~T.}\ \bibnamefont
  {Cerqueira}}, \bibinfo {author} {\bibfnamefont {A.}~\bibnamefont {Sanna}},
  \bibinfo {author} {\bibfnamefont {Y.-W.}\ \bibnamefont {Fang}}, \bibinfo
  {author} {\bibfnamefont {{\DJ}.}~\bibnamefont {Dangi{\'{c}}}}, \bibinfo
  {author} {\bibfnamefont {I.}~\bibnamefont {Errea}}, \bibinfo {author}
  {\bibfnamefont {H.-C.}\ \bibnamefont {Wang}}, \bibinfo {author}
  {\bibfnamefont {S.}~\bibnamefont {Botti}},\ and\ \bibinfo {author}
  {\bibfnamefont {M.~A.~L.}\ \bibnamefont {Marques}},\ }\bibfield  {title}
  {\bibinfo {title} {The maximum tc of conventional superconductors at ambient
  pressure},\ }\href {https://doi.org/10.1038/s41467-025-63702-w} {\bibfield
  {journal} {\bibinfo  {journal} {Nature Communications}\ }\textbf {\bibinfo
  {volume} {16}},\ \bibinfo {pages} {8253} (\bibinfo {year}
  {2025})}\BibitemShut {NoStop}%
\bibitem [{\citenamefont {Dolui}\ \emph {et~al.}(2024)\citenamefont {Dolui},
  \citenamefont {Conway}, \citenamefont {Heil}, \citenamefont {Strobel},
  \citenamefont {Prasankumar},\ and\ \citenamefont {Pickard}}]{MgIrH6_chris}%
  \BibitemOpen
  \bibfield  {author} {\bibinfo {author} {\bibfnamefont {K.}~\bibnamefont
  {Dolui}}, \bibinfo {author} {\bibfnamefont {L.~J.}\ \bibnamefont {Conway}},
  \bibinfo {author} {\bibfnamefont {C.}~\bibnamefont {Heil}}, \bibinfo {author}
  {\bibfnamefont {T.~A.}\ \bibnamefont {Strobel}}, \bibinfo {author}
  {\bibfnamefont {R.~P.}\ \bibnamefont {Prasankumar}},\ and\ \bibinfo {author}
  {\bibfnamefont {C.~J.}\ \bibnamefont {Pickard}},\ }\bibfield  {title}
  {\bibinfo {title} {Feasible route to high-temperature ambient-pressure
  hydride superconductivity},\ }\href
  {https://doi.org/10.1103/PhysRevLett.132.166001} {\bibfield  {journal}
  {\bibinfo  {journal} {Phys. Rev. Lett.}\ }\textbf {\bibinfo {volume} {132}},\
  \bibinfo {pages} {166001} (\bibinfo {year} {2024})}\BibitemShut {NoStop}%
\bibitem [{\citenamefont {Sanna}\ \emph {et~al.}(2024)\citenamefont {Sanna},
  \citenamefont {Cerqueira}, \citenamefont {Fang}, \citenamefont {Errea},
  \citenamefont {Ludwig},\ and\ \citenamefont {Marques}}]{MgIrH6_ion}%
  \BibitemOpen
  \bibfield  {author} {\bibinfo {author} {\bibfnamefont {A.}~\bibnamefont
  {Sanna}}, \bibinfo {author} {\bibfnamefont {T.~F.~T.}\ \bibnamefont
  {Cerqueira}}, \bibinfo {author} {\bibfnamefont {Y.-W.}\ \bibnamefont {Fang}},
  \bibinfo {author} {\bibfnamefont {I.}~\bibnamefont {Errea}}, \bibinfo
  {author} {\bibfnamefont {A.}~\bibnamefont {Ludwig}},\ and\ \bibinfo {author}
  {\bibfnamefont {M.~A.~L.}\ \bibnamefont {Marques}},\ }\bibfield  {title}
  {\bibinfo {title} {Prediction of ambient pressure conventional
  superconductivity above 80 k in hydride compounds},\ }\href
  {https://doi.org/10.1038/s41524-024-01214-9} {\bibfield  {journal} {\bibinfo
  {journal} {npj Computational Materials}\ }\textbf {\bibinfo {volume} {10}},\
  \bibinfo {pages} {44} (\bibinfo {year} {2024})}\BibitemShut {NoStop}%
\bibitem [{\citenamefont {Đorđe Dangić}\ \emph {et~al.}(2025)\citenamefont
  {Đorđe Dangić}, \citenamefont {Fang}, \citenamefont {Cerqueira},
  \citenamefont {Sanna}, \citenamefont {Novoa}, \citenamefont {Gao},
  \citenamefont {Marques},\ and\ \citenamefont {Errea}}]{RbPH3}%
  \BibitemOpen
  \bibfield  {author} {\bibinfo {author} {\bibnamefont {Đorđe Dangić}},
  \bibinfo {author} {\bibfnamefont {Y.-W.}\ \bibnamefont {Fang}}, \bibinfo
  {author} {\bibfnamefont {T.~F.}\ \bibnamefont {Cerqueira}}, \bibinfo {author}
  {\bibfnamefont {A.}~\bibnamefont {Sanna}}, \bibinfo {author} {\bibfnamefont
  {T.}~\bibnamefont {Novoa}}, \bibinfo {author} {\bibfnamefont
  {H.}~\bibnamefont {Gao}}, \bibinfo {author} {\bibfnamefont {M.~A.}\
  \bibnamefont {Marques}},\ and\ \bibinfo {author} {\bibfnamefont
  {I.}~\bibnamefont {Errea}},\ }\bibfield  {title} {\bibinfo {title} {Ambient
  pressure high temperature superconductivity in rbph3 facilitated by ionic
  anharmonicity},\ }\href
  {https://doi.org/https://doi.org/10.1016/j.commt.2025.100043} {\bibfield
  {journal} {\bibinfo  {journal} {Computational Materials Today}\ }\textbf
  {\bibinfo {volume} {8}},\ \bibinfo {pages} {100043} (\bibinfo {year}
  {2025})}\BibitemShut {NoStop}%
\bibitem [{\citenamefont {Monacelli}\ \emph
  {et~al.}(2021{\natexlab{a}})\citenamefont {Monacelli}, \citenamefont
  {Bianco}, \citenamefont {Cherubini}, \citenamefont {Calandra}, \citenamefont
  {Errea},\ and\ \citenamefont {Mauri}}]{SSCHA1}%
  \BibitemOpen
  \bibfield  {author} {\bibinfo {author} {\bibfnamefont {L.}~\bibnamefont
  {Monacelli}}, \bibinfo {author} {\bibfnamefont {R.}~\bibnamefont {Bianco}},
  \bibinfo {author} {\bibfnamefont {M.}~\bibnamefont {Cherubini}}, \bibinfo
  {author} {\bibfnamefont {M.}~\bibnamefont {Calandra}}, \bibinfo {author}
  {\bibfnamefont {I.}~\bibnamefont {Errea}},\ and\ \bibinfo {author}
  {\bibfnamefont {F.}~\bibnamefont {Mauri}},\ }\bibfield  {title} {\bibinfo
  {title} {The stochastic self-consistent harmonic approximation: calculating
  vibrational properties of materials with full quantum and anharmonic
  effects},\ }\href {https://doi.org/10.1088/1361-648X/ac066b} {\bibfield
  {journal} {\bibinfo  {journal} {Journal of Physics: Condensed Matter}\
  }\textbf {\bibinfo {volume} {33}},\ \bibinfo {pages} {363001} (\bibinfo
  {year} {2021}{\natexlab{a}})}\BibitemShut {NoStop}%
\bibitem [{\citenamefont {Errea}\ \emph
  {et~al.}(2013{\natexlab{a}})\citenamefont {Errea}, \citenamefont {Calandra},\
  and\ \citenamefont {Mauri}}]{SSCHA2}%
  \BibitemOpen
  \bibfield  {author} {\bibinfo {author} {\bibfnamefont {I.}~\bibnamefont
  {Errea}}, \bibinfo {author} {\bibfnamefont {M.}~\bibnamefont {Calandra}},\
  and\ \bibinfo {author} {\bibfnamefont {F.}~\bibnamefont {Mauri}},\ }\bibfield
   {title} {\bibinfo {title} {First-principles theory of anharmonicity and the
  inverse isotope effect in superconducting palladium-hydride compounds},\
  }\href {https://doi.org/10.1103/PhysRevLett.111.177002} {\bibfield  {journal}
  {\bibinfo  {journal} {Phys. Rev. Lett.}\ }\textbf {\bibinfo {volume} {111}},\
  \bibinfo {pages} {177002} (\bibinfo {year} {2013}{\natexlab{a}})}\BibitemShut
  {NoStop}%
\bibitem [{\citenamefont {Errea}\ \emph {et~al.}(2014)\citenamefont {Errea},
  \citenamefont {Calandra},\ and\ \citenamefont {Mauri}}]{SSCHA3}%
  \BibitemOpen
  \bibfield  {author} {\bibinfo {author} {\bibfnamefont {I.}~\bibnamefont
  {Errea}}, \bibinfo {author} {\bibfnamefont {M.}~\bibnamefont {Calandra}},\
  and\ \bibinfo {author} {\bibfnamefont {F.}~\bibnamefont {Mauri}},\ }\bibfield
   {title} {\bibinfo {title} {Anharmonic free energies and phonon dispersions
  from the stochastic self-consistent harmonic approximation: Application to
  platinum and palladium hydrides},\ }\href
  {https://doi.org/10.1103/PhysRevB.89.064302} {\bibfield  {journal} {\bibinfo
  {journal} {Phys. Rev. B}\ }\textbf {\bibinfo {volume} {89}},\ \bibinfo
  {pages} {064302} (\bibinfo {year} {2014})}\BibitemShut {NoStop}%
\bibitem [{\citenamefont {Bianco}\ \emph {et~al.}(2017)\citenamefont {Bianco},
  \citenamefont {Errea}, \citenamefont {Paulatto}, \citenamefont {Calandra},\
  and\ \citenamefont {Mauri}}]{SSCHA4}%
  \BibitemOpen
  \bibfield  {author} {\bibinfo {author} {\bibfnamefont {R.}~\bibnamefont
  {Bianco}}, \bibinfo {author} {\bibfnamefont {I.}~\bibnamefont {Errea}},
  \bibinfo {author} {\bibfnamefont {L.}~\bibnamefont {Paulatto}}, \bibinfo
  {author} {\bibfnamefont {M.}~\bibnamefont {Calandra}},\ and\ \bibinfo
  {author} {\bibfnamefont {F.}~\bibnamefont {Mauri}},\ }\bibfield  {title}
  {\bibinfo {title} {Second-order structural phase transitions, free energy
  curvature, and temperature-dependent anharmonic phonons in the
  self-consistent harmonic approximation: Theory and stochastic
  implementation},\ }\href {https://doi.org/10.1103/PhysRevB.96.014111}
  {\bibfield  {journal} {\bibinfo  {journal} {Phys. Rev. B}\ }\textbf {\bibinfo
  {volume} {96}},\ \bibinfo {pages} {014111} (\bibinfo {year}
  {2017})}\BibitemShut {NoStop}%
\bibitem [{\citenamefont {Monacelli}\ \emph {et~al.}(2018)\citenamefont
  {Monacelli}, \citenamefont {Errea}, \citenamefont {Calandra},\ and\
  \citenamefont {Mauri}}]{SSCHA5}%
  \BibitemOpen
  \bibfield  {author} {\bibinfo {author} {\bibfnamefont {L.}~\bibnamefont
  {Monacelli}}, \bibinfo {author} {\bibfnamefont {I.}~\bibnamefont {Errea}},
  \bibinfo {author} {\bibfnamefont {M.}~\bibnamefont {Calandra}},\ and\
  \bibinfo {author} {\bibfnamefont {F.}~\bibnamefont {Mauri}},\ }\bibfield
  {title} {\bibinfo {title} {Pressure and stress tensor of complex anharmonic
  crystals within the stochastic self-consistent harmonic approximation},\
  }\href {https://doi.org/10.1103/PhysRevB.98.024106} {\bibfield  {journal}
  {\bibinfo  {journal} {Phys. Rev. B}\ }\textbf {\bibinfo {volume} {98}},\
  \bibinfo {pages} {024106} (\bibinfo {year} {2018})}\BibitemShut {NoStop}%
\bibitem [{\citenamefont {Dangi{\'{c}}}\ \emph {et~al.}(2024)\citenamefont
  {Dangi{\'{c}}}, \citenamefont {Monacelli}, \citenamefont {Bianco},
  \citenamefont {Mauri},\ and\ \citenamefont {Errea}}]{H}%
  \BibitemOpen
  \bibfield  {author} {\bibinfo {author} {\bibfnamefont {{\DJ}.}~\bibnamefont
  {Dangi{\'{c}}}}, \bibinfo {author} {\bibfnamefont {L.}~\bibnamefont
  {Monacelli}}, \bibinfo {author} {\bibfnamefont {R.}~\bibnamefont {Bianco}},
  \bibinfo {author} {\bibfnamefont {F.}~\bibnamefont {Mauri}},\ and\ \bibinfo
  {author} {\bibfnamefont {I.}~\bibnamefont {Errea}},\ }\bibfield  {title}
  {\bibinfo {title} {Large impact of phonon lineshapes on the superconductivity
  of solid hydrogen},\ }\href {https://doi.org/10.1038/s42005-024-01643-4}
  {\bibfield  {journal} {\bibinfo  {journal} {Communications Physics}\ }\textbf
  {\bibinfo {volume} {7}},\ \bibinfo {pages} {150} (\bibinfo {year}
  {2024})}\BibitemShut {NoStop}%
\bibitem [{\citenamefont {Dangi\ifmmode~\acute{c}\else \'{c}\fi{}}\ \emph
  {et~al.}(2023)\citenamefont {Dangi\ifmmode~\acute{c}\else \'{c}\fi{}},
  \citenamefont {Garcia-Goiricelaya}, \citenamefont {Fang}, \citenamefont
  {Iba\~nez Azpiroz},\ and\ \citenamefont {Errea}}]{LuNH1}%
  \BibitemOpen
  \bibfield  {author} {\bibinfo {author} {\bibfnamefont {D.}~\bibnamefont
  {Dangi\ifmmode~\acute{c}\else \'{c}\fi{}}}, \bibinfo {author} {\bibfnamefont
  {P.}~\bibnamefont {Garcia-Goiricelaya}}, \bibinfo {author} {\bibfnamefont
  {Y.-W.}\ \bibnamefont {Fang}}, \bibinfo {author} {\bibfnamefont
  {J.}~\bibnamefont {Iba\~nez Azpiroz}},\ and\ \bibinfo {author} {\bibfnamefont
  {I.}~\bibnamefont {Errea}},\ }\bibfield  {title} {\bibinfo {title} {Ab initio
  study of the structural, vibrational, and optical properties of potential
  parent structures of nitrogen-doped lutetium hydride},\ }\href
  {https://doi.org/10.1103/PhysRevB.108.064517} {\bibfield  {journal} {\bibinfo
   {journal} {Phys. Rev. B}\ }\textbf {\bibinfo {volume} {108}},\ \bibinfo
  {pages} {064517} (\bibinfo {year} {2023})}\BibitemShut {NoStop}%
\bibitem [{\citenamefont {Fang}\ \emph {et~al.}(2024)\citenamefont {Fang},
  \citenamefont {Dangi{\'{c}}},\ and\ \citenamefont {Errea}}]{Fang2024}%
  \BibitemOpen
  \bibfield  {author} {\bibinfo {author} {\bibfnamefont {Y.-W.}\ \bibnamefont
  {Fang}}, \bibinfo {author} {\bibfnamefont {{\DJ}.}~\bibnamefont
  {Dangi{\'{c}}}},\ and\ \bibinfo {author} {\bibfnamefont {I.}~\bibnamefont
  {Errea}},\ }\bibfield  {title} {\bibinfo {title} {Assessing the feasibility
  of near-ambient conditions superconductivity in the lu-n-h system},\ }\href
  {https://doi.org/10.1038/s43246-024-00500-9} {\bibfield  {journal} {\bibinfo
  {journal} {Communications Materials}\ }\textbf {\bibinfo {volume} {5}},\
  \bibinfo {pages} {61} (\bibinfo {year} {2024})}\BibitemShut {NoStop}%
\bibitem [{\citenamefont {Errea}\ \emph {et~al.}(2016)\citenamefont {Errea},
  \citenamefont {Calandra}, \citenamefont {Pickard}, \citenamefont {Nelson},
  \citenamefont {Needs}, \citenamefont {Li}, \citenamefont {Liu}, \citenamefont
  {Zhang}, \citenamefont {Ma},\ and\ \citenamefont {Mauri}}]{H3S_ion2}%
  \BibitemOpen
  \bibfield  {author} {\bibinfo {author} {\bibfnamefont {I.}~\bibnamefont
  {Errea}}, \bibinfo {author} {\bibfnamefont {M.}~\bibnamefont {Calandra}},
  \bibinfo {author} {\bibfnamefont {C.~J.}\ \bibnamefont {Pickard}}, \bibinfo
  {author} {\bibfnamefont {J.~R.}\ \bibnamefont {Nelson}}, \bibinfo {author}
  {\bibfnamefont {R.~J.}\ \bibnamefont {Needs}}, \bibinfo {author}
  {\bibfnamefont {Y.}~\bibnamefont {Li}}, \bibinfo {author} {\bibfnamefont
  {H.}~\bibnamefont {Liu}}, \bibinfo {author} {\bibfnamefont {Y.}~\bibnamefont
  {Zhang}}, \bibinfo {author} {\bibfnamefont {Y.}~\bibnamefont {Ma}},\ and\
  \bibinfo {author} {\bibfnamefont {F.}~\bibnamefont {Mauri}},\ }\bibfield
  {title} {\bibinfo {title} {Quantum hydrogen-bond symmetrization in the
  superconducting hydrogen sulfide system},\ }\href
  {https://doi.org/10.1038/nature17175} {\bibfield  {journal} {\bibinfo
  {journal} {Nature}\ }\textbf {\bibinfo {volume} {532}},\ \bibinfo {pages}
  {81} (\bibinfo {year} {2016})}\BibitemShut {NoStop}%
\bibitem [{\citenamefont {Errea}\ \emph {et~al.}(2020)\citenamefont {Errea},
  \citenamefont {Belli}, \citenamefont {Monacelli}, \citenamefont {Sanna},
  \citenamefont {Koretsune}, \citenamefont {Tadano}, \citenamefont {Bianco},
  \citenamefont {Calandra}, \citenamefont {Arita}, \citenamefont {Mauri},\ and\
  \citenamefont {Flores-Livas}}]{LaH10_ion}%
  \BibitemOpen
  \bibfield  {author} {\bibinfo {author} {\bibfnamefont {I.}~\bibnamefont
  {Errea}}, \bibinfo {author} {\bibfnamefont {F.}~\bibnamefont {Belli}},
  \bibinfo {author} {\bibfnamefont {L.}~\bibnamefont {Monacelli}}, \bibinfo
  {author} {\bibfnamefont {A.}~\bibnamefont {Sanna}}, \bibinfo {author}
  {\bibfnamefont {T.}~\bibnamefont {Koretsune}}, \bibinfo {author}
  {\bibfnamefont {T.}~\bibnamefont {Tadano}}, \bibinfo {author} {\bibfnamefont
  {R.}~\bibnamefont {Bianco}}, \bibinfo {author} {\bibfnamefont
  {M.}~\bibnamefont {Calandra}}, \bibinfo {author} {\bibfnamefont
  {R.}~\bibnamefont {Arita}}, \bibinfo {author} {\bibfnamefont
  {F.}~\bibnamefont {Mauri}},\ and\ \bibinfo {author} {\bibfnamefont {J.~A.}\
  \bibnamefont {Flores-Livas}},\ }\bibfield  {title} {\bibinfo {title} {Quantum
  crystal structure in the 250-kelvin superconducting lanthanum hydride},\
  }\href {https://doi.org/10.1038/s41586-020-1955-z} {\bibfield  {journal}
  {\bibinfo  {journal} {Nature}\ }\textbf {\bibinfo {volume} {578}},\ \bibinfo
  {pages} {66} (\bibinfo {year} {2020})}\BibitemShut {NoStop}%
\bibitem [{\citenamefont {Errea}\ \emph
  {et~al.}(2013{\natexlab{b}})\citenamefont {Errea}, \citenamefont {Calandra},\
  and\ \citenamefont {Mauri}}]{PdH}%
  \BibitemOpen
  \bibfield  {author} {\bibinfo {author} {\bibfnamefont {I.}~\bibnamefont
  {Errea}}, \bibinfo {author} {\bibfnamefont {M.}~\bibnamefont {Calandra}},\
  and\ \bibinfo {author} {\bibfnamefont {F.}~\bibnamefont {Mauri}},\ }\bibfield
   {title} {\bibinfo {title} {First-principles theory of anharmonicity and the
  inverse isotope effect in superconducting palladium-hydride compounds},\
  }\href {https://doi.org/10.1103/PhysRevLett.111.177002} {\bibfield  {journal}
  {\bibinfo  {journal} {Phys. Rev. Lett.}\ }\textbf {\bibinfo {volume} {111}},\
  \bibinfo {pages} {177002} (\bibinfo {year} {2013}{\natexlab{b}})}\BibitemShut
  {NoStop}%
\bibitem [{\citenamefont {Lucrezi}\ \emph {et~al.}(2023)\citenamefont
  {Lucrezi}, \citenamefont {Kogler}, \citenamefont {Di~Cataldo}, \citenamefont
  {Aichhorn}, \citenamefont {Boeri},\ and\ \citenamefont {Heil}}]{Lucrezi2023}%
  \BibitemOpen
  \bibfield  {author} {\bibinfo {author} {\bibfnamefont {R.}~\bibnamefont
  {Lucrezi}}, \bibinfo {author} {\bibfnamefont {E.}~\bibnamefont {Kogler}},
  \bibinfo {author} {\bibfnamefont {S.}~\bibnamefont {Di~Cataldo}}, \bibinfo
  {author} {\bibfnamefont {M.}~\bibnamefont {Aichhorn}}, \bibinfo {author}
  {\bibfnamefont {L.}~\bibnamefont {Boeri}},\ and\ \bibinfo {author}
  {\bibfnamefont {C.}~\bibnamefont {Heil}},\ }\bibfield  {title} {\bibinfo
  {title} {Quantum lattice dynamics and their importance in ternary
  superhydride clathrates},\ }\href
  {https://doi.org/10.1038/s42005-023-01413-8} {\bibfield  {journal} {\bibinfo
  {journal} {Communications Physics}\ }\textbf {\bibinfo {volume} {6}},\
  \bibinfo {pages} {298} (\bibinfo {year} {2023})}\BibitemShut {NoStop}%
\bibitem [{\citenamefont {Lucrezi}\ \emph {et~al.}(2024)\citenamefont
  {Lucrezi}, \citenamefont {Ferreira}, \citenamefont {Aichhorn},\ and\
  \citenamefont {Heil}}]{Lucrezi2024}%
  \BibitemOpen
  \bibfield  {author} {\bibinfo {author} {\bibfnamefont {R.}~\bibnamefont
  {Lucrezi}}, \bibinfo {author} {\bibfnamefont {P.~P.}\ \bibnamefont
  {Ferreira}}, \bibinfo {author} {\bibfnamefont {M.}~\bibnamefont {Aichhorn}},\
  and\ \bibinfo {author} {\bibfnamefont {C.}~\bibnamefont {Heil}},\ }\bibfield
  {title} {\bibinfo {title} {Temperature and quantum anharmonic lattice effects
  on stability and superconductivity in lutetium trihydride},\ }\href
  {https://doi.org/10.1038/s41467-023-44326-4} {\bibfield  {journal} {\bibinfo
  {journal} {Nature Communications}\ }\textbf {\bibinfo {volume} {15}},\
  \bibinfo {pages} {441} (\bibinfo {year} {2024})}\BibitemShut {NoStop}%
\bibitem [{\citenamefont {Belli}\ and\ \citenamefont
  {Zurek}(2025)}]{belli2025}%
  \BibitemOpen
  \bibfield  {author} {\bibinfo {author} {\bibfnamefont {F.}~\bibnamefont
  {Belli}}\ and\ \bibinfo {author} {\bibfnamefont {E.}~\bibnamefont {Zurek}},\
  }\bibfield  {title} {\bibinfo {title} {Efficient modelling of anharmonicity
  and quantum effects in pdcuh2 with machine learning potentials},\ }\href
  {https://doi.org/10.1038/s41524-025-01553-1} {\bibfield  {journal} {\bibinfo
  {journal} {npj Computational Materials}\ }\textbf {\bibinfo {volume} {11}},\
  \bibinfo {pages} {87} (\bibinfo {year} {2025})}\BibitemShut {NoStop}%
\bibitem [{\citenamefont {He}\ \emph {et~al.}(2024)\citenamefont {He},
  \citenamefont {Zhao}, \citenamefont {Xie}, \citenamefont {Hermann},
  \citenamefont {Hemley}, \citenamefont {Liu},\ and\ \citenamefont
  {Ma}}]{LaScH}%
  \BibitemOpen
  \bibfield  {author} {\bibinfo {author} {\bibfnamefont {X.-L.}\ \bibnamefont
  {He}}, \bibinfo {author} {\bibfnamefont {W.}~\bibnamefont {Zhao}}, \bibinfo
  {author} {\bibfnamefont {Y.}~\bibnamefont {Xie}}, \bibinfo {author}
  {\bibfnamefont {A.}~\bibnamefont {Hermann}}, \bibinfo {author} {\bibfnamefont
  {R.~J.}\ \bibnamefont {Hemley}}, \bibinfo {author} {\bibfnamefont
  {H.}~\bibnamefont {Liu}},\ and\ \bibinfo {author} {\bibfnamefont
  {Y.}~\bibnamefont {Ma}},\ }\bibfield  {title} {\bibinfo {title} {Predicted
  hot superconductivity in lasc<sub>2</sub>h<sub>24</sub> under pressure},\
  }\href {https://doi.org/10.1073/pnas.2401840121} {\bibfield  {journal}
  {\bibinfo  {journal} {Proceedings of the National Academy of Sciences}\
  }\textbf {\bibinfo {volume} {121}},\ \bibinfo {pages} {e2401840121} (\bibinfo
  {year} {2024})},\ \Eprint
  {https://arxiv.org/abs/https://www.pnas.org/doi/pdf/10.1073/pnas.2401840121}
  {https://www.pnas.org/doi/pdf/10.1073/pnas.2401840121} \BibitemShut {NoStop}%
\bibitem [{\citenamefont {Jiang}\ \emph {et~al.}(2024)\citenamefont {Jiang},
  \citenamefont {Zhang}, \citenamefont {Song}, \citenamefont {Ma},
  \citenamefont {Sun}, \citenamefont {Miao}, \citenamefont {Cui},\ and\
  \citenamefont {Duan}}]{ThCeBeH}%
  \BibitemOpen
  \bibfield  {author} {\bibinfo {author} {\bibfnamefont {Q.}~\bibnamefont
  {Jiang}}, \bibinfo {author} {\bibfnamefont {Z.}~\bibnamefont {Zhang}},
  \bibinfo {author} {\bibfnamefont {H.}~\bibnamefont {Song}}, \bibinfo {author}
  {\bibfnamefont {Y.}~\bibnamefont {Ma}}, \bibinfo {author} {\bibfnamefont
  {Y.}~\bibnamefont {Sun}}, \bibinfo {author} {\bibfnamefont {M.}~\bibnamefont
  {Miao}}, \bibinfo {author} {\bibfnamefont {T.}~\bibnamefont {Cui}},\ and\
  \bibinfo {author} {\bibfnamefont {D.}~\bibnamefont {Duan}},\ }\bibfield
  {title} {\bibinfo {title} {Ternary superconducting hydrides stabilized via th
  and ce elements at mild pressures},\ }\href
  {https://doi.org/https://doi.org/10.1016/j.fmre.2022.11.010} {\bibfield
  {journal} {\bibinfo  {journal} {Fundamental Research}\ }\textbf {\bibinfo
  {volume} {4}},\ \bibinfo {pages} {550} (\bibinfo {year} {2024})}\BibitemShut
  {NoStop}%
\bibitem [{\citenamefont {Sun}\ \emph {et~al.}(2022)\citenamefont {Sun},
  \citenamefont {Sun}, \citenamefont {Zhong},\ and\ \citenamefont
  {Liu}}]{ThCeBeH2}%
  \BibitemOpen
  \bibfield  {author} {\bibinfo {author} {\bibfnamefont {Y.}~\bibnamefont
  {Sun}}, \bibinfo {author} {\bibfnamefont {S.}~\bibnamefont {Sun}}, \bibinfo
  {author} {\bibfnamefont {X.}~\bibnamefont {Zhong}},\ and\ \bibinfo {author}
  {\bibfnamefont {H.}~\bibnamefont {Liu}},\ }\bibfield  {title} {\bibinfo
  {title} {Prediction for high superconducting ternary hydrides below megabar
  pressure},\ }\href {https://doi.org/10.1088/1361-648X/ac9bba} {\bibfield
  {journal} {\bibinfo  {journal} {Journal of Physics: Condensed Matter}\
  }\textbf {\bibinfo {volume} {34}},\ \bibinfo {pages} {505404} (\bibinfo
  {year} {2022})}\BibitemShut {NoStop}%
\bibitem [{\citenamefont {Di~Cataldo}\ \emph {et~al.}(2021)\citenamefont
  {Di~Cataldo}, \citenamefont {Heil}, \citenamefont {von~der Linden},\ and\
  \citenamefont {Boeri}}]{LaBH8}%
  \BibitemOpen
  \bibfield  {author} {\bibinfo {author} {\bibfnamefont {S.}~\bibnamefont
  {Di~Cataldo}}, \bibinfo {author} {\bibfnamefont {C.}~\bibnamefont {Heil}},
  \bibinfo {author} {\bibfnamefont {W.}~\bibnamefont {von~der Linden}},\ and\
  \bibinfo {author} {\bibfnamefont {L.}~\bibnamefont {Boeri}},\ }\bibfield
  {title} {\bibinfo {title} {$\mathrm{La}{\mathrm{bh}}_{8}$: Towards
  high-${T}_{c}$ low-pressure superconductivity in ternary superhydrides},\
  }\href {https://doi.org/10.1103/PhysRevB.104.L020511} {\bibfield  {journal}
  {\bibinfo  {journal} {Phys. Rev. B}\ }\textbf {\bibinfo {volume} {104}},\
  \bibinfo {pages} {L020511} (\bibinfo {year} {2021})}\BibitemShut {NoStop}%
\bibitem [{\citenamefont {Li}\ \emph {et~al.}(2022{\natexlab{b}})\citenamefont
  {Li}, \citenamefont {Wang}, \citenamefont {Sun}, \citenamefont {Lu},\ and\
  \citenamefont {Peng}}]{RbBH8}%
  \BibitemOpen
  \bibfield  {author} {\bibinfo {author} {\bibfnamefont {S.}~\bibnamefont
  {Li}}, \bibinfo {author} {\bibfnamefont {H.}~\bibnamefont {Wang}}, \bibinfo
  {author} {\bibfnamefont {W.}~\bibnamefont {Sun}}, \bibinfo {author}
  {\bibfnamefont {C.}~\bibnamefont {Lu}},\ and\ \bibinfo {author}
  {\bibfnamefont {F.}~\bibnamefont {Peng}},\ }\bibfield  {title} {\bibinfo
  {title} {Superconductivity in compressed ternary alkaline boron hydrides},\
  }\href {https://doi.org/10.1103/PhysRevB.105.224107} {\bibfield  {journal}
  {\bibinfo  {journal} {Phys. Rev. B}\ }\textbf {\bibinfo {volume} {105}},\
  \bibinfo {pages} {224107} (\bibinfo {year} {2022}{\natexlab{b}})}\BibitemShut
  {NoStop}%
\bibitem [{\citenamefont {Lucrezi}\ \emph {et~al.}(2022)\citenamefont
  {Lucrezi}, \citenamefont {Di~Cataldo}, \citenamefont {von~der Linden},
  \citenamefont {Boeri},\ and\ \citenamefont {Heil}}]{BaSiH}%
  \BibitemOpen
  \bibfield  {author} {\bibinfo {author} {\bibfnamefont {R.}~\bibnamefont
  {Lucrezi}}, \bibinfo {author} {\bibfnamefont {S.}~\bibnamefont {Di~Cataldo}},
  \bibinfo {author} {\bibfnamefont {W.}~\bibnamefont {von~der Linden}},
  \bibinfo {author} {\bibfnamefont {L.}~\bibnamefont {Boeri}},\ and\ \bibinfo
  {author} {\bibfnamefont {C.}~\bibnamefont {Heil}},\ }\bibfield  {title}
  {\bibinfo {title} {In-silico synthesis of lowest-pressure high-tc ternary
  superhydrides},\ }\href {https://doi.org/10.1038/s41524-022-00801-y}
  {\bibfield  {journal} {\bibinfo  {journal} {npj Computational Materials}\
  }\textbf {\bibinfo {volume} {8}},\ \bibinfo {pages} {119} (\bibinfo {year}
  {2022})}\BibitemShut {NoStop}%
\bibitem [{\citenamefont {Kuzovnikov}\ \emph {et~al.}(2025)\citenamefont
  {Kuzovnikov}, \citenamefont {Wang}, \citenamefont {Wang}, \citenamefont
  {Marque\~no}, \citenamefont {Shuttleworth}, \citenamefont {Strain},
  \citenamefont {Gregoryanz}, \citenamefont {Zurek}, \citenamefont {Pe\~na
  Alvarez},\ and\ \citenamefont {Howie}}]{RbH_exp}%
  \BibitemOpen
  \bibfield  {author} {\bibinfo {author} {\bibfnamefont {M.~A.}\ \bibnamefont
  {Kuzovnikov}}, \bibinfo {author} {\bibfnamefont {B.}~\bibnamefont {Wang}},
  \bibinfo {author} {\bibfnamefont {X.}~\bibnamefont {Wang}}, \bibinfo {author}
  {\bibfnamefont {T.}~\bibnamefont {Marque\~no}}, \bibinfo {author}
  {\bibfnamefont {H.~A.}\ \bibnamefont {Shuttleworth}}, \bibinfo {author}
  {\bibfnamefont {C.}~\bibnamefont {Strain}}, \bibinfo {author} {\bibfnamefont
  {E.}~\bibnamefont {Gregoryanz}}, \bibinfo {author} {\bibfnamefont
  {E.}~\bibnamefont {Zurek}}, \bibinfo {author} {\bibfnamefont
  {M.}~\bibnamefont {Pe\~na Alvarez}},\ and\ \bibinfo {author} {\bibfnamefont
  {R.~T.}\ \bibnamefont {Howie}},\ }\bibfield  {title} {\bibinfo {title} {High
  pressure synthesis of rubidium superhydrides},\ }\href
  {https://doi.org/10.1103/PhysRevLett.134.196102} {\bibfield  {journal}
  {\bibinfo  {journal} {Phys. Rev. Lett.}\ }\textbf {\bibinfo {volume} {134}},\
  \bibinfo {pages} {196102} (\bibinfo {year} {2025})}\BibitemShut {NoStop}%
\bibitem [{\citenamefont {Zhou}\ \emph {et~al.}(2024)\citenamefont {Zhou},
  \citenamefont {Semenok}, \citenamefont {Galasso}, \citenamefont {Alabarse},
  \citenamefont {Sannikov}, \citenamefont {Troyan}, \citenamefont {Nakamoto},
  \citenamefont {Shimizu},\ and\ \citenamefont {Oganov}}]{SemenokRbH}%
  \BibitemOpen
  \bibfield  {author} {\bibinfo {author} {\bibfnamefont {D.}~\bibnamefont
  {Zhou}}, \bibinfo {author} {\bibfnamefont {D.}~\bibnamefont {Semenok}},
  \bibinfo {author} {\bibfnamefont {M.}~\bibnamefont {Galasso}}, \bibinfo
  {author} {\bibfnamefont {F.~G.}\ \bibnamefont {Alabarse}}, \bibinfo {author}
  {\bibfnamefont {D.}~\bibnamefont {Sannikov}}, \bibinfo {author}
  {\bibfnamefont {I.~A.}\ \bibnamefont {Troyan}}, \bibinfo {author}
  {\bibfnamefont {Y.}~\bibnamefont {Nakamoto}}, \bibinfo {author}
  {\bibfnamefont {K.}~\bibnamefont {Shimizu}},\ and\ \bibinfo {author}
  {\bibfnamefont {A.~R.}\ \bibnamefont {Oganov}},\ }\bibfield  {title}
  {\bibinfo {title} {Raisins in a hydrogen pie: Ultrastable cesium and rubidium
  polyhydrides},\ }\href
  {https://doi.org/https://doi.org/10.1002/aenm.202400077} {\bibfield
  {journal} {\bibinfo  {journal} {Advanced Energy Materials}\ }\textbf
  {\bibinfo {volume} {14}},\ \bibinfo {pages} {2400077} (\bibinfo {year}
  {2024})},\ \Eprint
  {https://arxiv.org/abs/https://advanced.onlinelibrary.wiley.com/doi/pdf/10.1002/aenm.202400077}
  {https://advanced.onlinelibrary.wiley.com/doi/pdf/10.1002/aenm.202400077}
  \BibitemShut {NoStop}%
\bibitem [{\citenamefont {Hutcheon}\ \emph {et~al.}(2020)\citenamefont
  {Hutcheon}, \citenamefont {Shipley},\ and\ \citenamefont {Needs}}]{RbH}%
  \BibitemOpen
  \bibfield  {author} {\bibinfo {author} {\bibfnamefont {M.~J.}\ \bibnamefont
  {Hutcheon}}, \bibinfo {author} {\bibfnamefont {A.~M.}\ \bibnamefont
  {Shipley}},\ and\ \bibinfo {author} {\bibfnamefont {R.~J.}\ \bibnamefont
  {Needs}},\ }\bibfield  {title} {\bibinfo {title} {Predicting novel
  superconducting hydrides using machine learning approaches},\ }\href
  {https://doi.org/10.1103/PhysRevB.101.144505} {\bibfield  {journal} {\bibinfo
   {journal} {Phys. Rev. B}\ }\textbf {\bibinfo {volume} {101}},\ \bibinfo
  {pages} {144505} (\bibinfo {year} {2020})}\BibitemShut {NoStop}%
\bibitem [{\citenamefont {Hooper}\ and\ \citenamefont
  {Zurek}(2012)}]{Hooper2012Rubidium}%
  \BibitemOpen
  \bibfield  {author} {\bibinfo {author} {\bibfnamefont {J.}~\bibnamefont
  {Hooper}}\ and\ \bibinfo {author} {\bibfnamefont {E.}~\bibnamefont {Zurek}},\
  }\bibfield  {title} {\bibinfo {title} {Rubidium polyhydrides under pressure:
  Emergence of the linear h3$^-$ species},\ }\href
  {https://doi.org/https://doi.org/10.1002/chem.201103205} {\bibfield
  {journal} {\bibinfo  {journal} {Chemistry – A European Journal}\ }\textbf
  {\bibinfo {volume} {18}},\ \bibinfo {pages} {5013} (\bibinfo {year}
  {2012})},\ \Eprint
  {https://arxiv.org/abs/https://chemistry-europe.onlinelibrary.wiley.com/doi/pdf/10.1002/chem.201103205}
  {https://chemistry-europe.onlinelibrary.wiley.com/doi/pdf/10.1002/chem.201103205}
  \BibitemShut {NoStop}%
\bibitem [{\citenamefont {Perdew}\ \emph {et~al.}(1996)\citenamefont {Perdew},
  \citenamefont {Burke},\ and\ \citenamefont {Ernzerhof}}]{PBE}%
  \BibitemOpen
  \bibfield  {author} {\bibinfo {author} {\bibfnamefont {J.~P.}\ \bibnamefont
  {Perdew}}, \bibinfo {author} {\bibfnamefont {K.}~\bibnamefont {Burke}},\ and\
  \bibinfo {author} {\bibfnamefont {M.}~\bibnamefont {Ernzerhof}},\ }\bibfield
  {title} {\bibinfo {title} {Generalized gradient approximation made simple},\
  }\href {https://doi.org/10.1103/PhysRevLett.77.3865} {\bibfield  {journal}
  {\bibinfo  {journal} {Phys. Rev. Lett.}\ }\textbf {\bibinfo {volume} {77}},\
  \bibinfo {pages} {3865} (\bibinfo {year} {1996})}\BibitemShut {NoStop}%
\bibitem [{\citenamefont {Giannozzi}\ \emph {et~al.}(2020)\citenamefont
  {Giannozzi}, \citenamefont {Baseggio}, \citenamefont {Bonfà}, \citenamefont
  {Brunato}, \citenamefont {Car}, \citenamefont {Carnimeo}, \citenamefont
  {Cavazzoni}, \citenamefont {de~Gironcoli}, \citenamefont {Delugas},
  \citenamefont {Ferrari~Ruffino}, \citenamefont {Ferretti}, \citenamefont
  {Marzari}, \citenamefont {Timrov}, \citenamefont {Urru},\ and\ \citenamefont
  {Baroni}}]{QE1}%
  \BibitemOpen
  \bibfield  {author} {\bibinfo {author} {\bibfnamefont {P.}~\bibnamefont
  {Giannozzi}}, \bibinfo {author} {\bibfnamefont {O.}~\bibnamefont {Baseggio}},
  \bibinfo {author} {\bibfnamefont {P.}~\bibnamefont {Bonfà}}, \bibinfo
  {author} {\bibfnamefont {D.}~\bibnamefont {Brunato}}, \bibinfo {author}
  {\bibfnamefont {R.}~\bibnamefont {Car}}, \bibinfo {author} {\bibfnamefont
  {I.}~\bibnamefont {Carnimeo}}, \bibinfo {author} {\bibfnamefont
  {C.}~\bibnamefont {Cavazzoni}}, \bibinfo {author} {\bibfnamefont
  {S.}~\bibnamefont {de~Gironcoli}}, \bibinfo {author} {\bibfnamefont
  {P.}~\bibnamefont {Delugas}}, \bibinfo {author} {\bibfnamefont
  {F.}~\bibnamefont {Ferrari~Ruffino}}, \bibinfo {author} {\bibfnamefont
  {A.}~\bibnamefont {Ferretti}}, \bibinfo {author} {\bibfnamefont
  {N.}~\bibnamefont {Marzari}}, \bibinfo {author} {\bibfnamefont
  {I.}~\bibnamefont {Timrov}}, \bibinfo {author} {\bibfnamefont
  {A.}~\bibnamefont {Urru}},\ and\ \bibinfo {author} {\bibfnamefont
  {S.}~\bibnamefont {Baroni}},\ }\bibfield  {title} {\bibinfo {title} {{Quantum
  ESPRESSO toward the exascale}},\ }\href {https://doi.org/10.1063/5.0005082}
  {\bibfield  {journal} {\bibinfo  {journal} {The Journal of Chemical Physics}\
  }\textbf {\bibinfo {volume} {152}},\ \bibinfo {pages} {154105} (\bibinfo
  {year} {2020})},\ \Eprint
  {https://arxiv.org/abs/https://pubs.aip.org/aip/jcp/article-pdf/doi/10.1063/5.0005082/16721881/154105\_1\_online.pdf}
  {https://pubs.aip.org/aip/jcp/article-pdf/doi/10.1063/5.0005082/16721881/154105\_1\_online.pdf}
  \BibitemShut {NoStop}%
\bibitem [{\citenamefont {Giannozzi}\ \emph {et~al.}(2017)\citenamefont
  {Giannozzi}, \citenamefont {Andreussi}, \citenamefont {Brumme}, \citenamefont
  {Bunau}, \citenamefont {Nardelli}, \citenamefont {Calandra}, \citenamefont
  {Car}, \citenamefont {Cavazzoni}, \citenamefont {Ceresoli}, \citenamefont
  {Cococcioni}, \citenamefont {Colonna}, \citenamefont {Carnimeo},
  \citenamefont {Corso}, \citenamefont {de~Gironcoli}, \citenamefont {Delugas},
  \citenamefont {DiStasio}, \citenamefont {Ferretti}, \citenamefont {Floris},
  \citenamefont {Fratesi}, \citenamefont {Fugallo}, \citenamefont {Gebauer},
  \citenamefont {Gerstmann}, \citenamefont {Giustino}, \citenamefont {Gorni},
  \citenamefont {Jia}, \citenamefont {Kawamura}, \citenamefont {Ko},
  \citenamefont {Kokalj}, \citenamefont {Küçükbenli}, \citenamefont
  {Lazzeri}, \citenamefont {Marsili}, \citenamefont {Marzari}, \citenamefont
  {Mauri}, \citenamefont {Nguyen}, \citenamefont {Nguyen}, \citenamefont {de-la
  Roza}, \citenamefont {Paulatto}, \citenamefont {Poncé}, \citenamefont
  {Rocca}, \citenamefont {Sabatini}, \citenamefont {Santra}, \citenamefont
  {Schlipf}, \citenamefont {Seitsonen}, \citenamefont {Smogunov}, \citenamefont
  {Timrov}, \citenamefont {Thonhauser}, \citenamefont {Umari}, \citenamefont
  {Vast}, \citenamefont {Wu},\ and\ \citenamefont {Baroni}}]{QE2}%
  \BibitemOpen
  \bibfield  {author} {\bibinfo {author} {\bibfnamefont {P.}~\bibnamefont
  {Giannozzi}}, \bibinfo {author} {\bibfnamefont {O.}~\bibnamefont
  {Andreussi}}, \bibinfo {author} {\bibfnamefont {T.}~\bibnamefont {Brumme}},
  \bibinfo {author} {\bibfnamefont {O.}~\bibnamefont {Bunau}}, \bibinfo
  {author} {\bibfnamefont {M.~B.}\ \bibnamefont {Nardelli}}, \bibinfo {author}
  {\bibfnamefont {M.}~\bibnamefont {Calandra}}, \bibinfo {author}
  {\bibfnamefont {R.}~\bibnamefont {Car}}, \bibinfo {author} {\bibfnamefont
  {C.}~\bibnamefont {Cavazzoni}}, \bibinfo {author} {\bibfnamefont
  {D.}~\bibnamefont {Ceresoli}}, \bibinfo {author} {\bibfnamefont
  {M.}~\bibnamefont {Cococcioni}}, \bibinfo {author} {\bibfnamefont
  {N.}~\bibnamefont {Colonna}}, \bibinfo {author} {\bibfnamefont
  {I.}~\bibnamefont {Carnimeo}}, \bibinfo {author} {\bibfnamefont {A.~D.}\
  \bibnamefont {Corso}}, \bibinfo {author} {\bibfnamefont {S.}~\bibnamefont
  {de~Gironcoli}}, \bibinfo {author} {\bibfnamefont {P.}~\bibnamefont
  {Delugas}}, \bibinfo {author} {\bibfnamefont {R.~A.}\ \bibnamefont
  {DiStasio}}, \bibinfo {author} {\bibfnamefont {A.}~\bibnamefont {Ferretti}},
  \bibinfo {author} {\bibfnamefont {A.}~\bibnamefont {Floris}}, \bibinfo
  {author} {\bibfnamefont {G.}~\bibnamefont {Fratesi}}, \bibinfo {author}
  {\bibfnamefont {G.}~\bibnamefont {Fugallo}}, \bibinfo {author} {\bibfnamefont
  {R.}~\bibnamefont {Gebauer}}, \bibinfo {author} {\bibfnamefont
  {U.}~\bibnamefont {Gerstmann}}, \bibinfo {author} {\bibfnamefont
  {F.}~\bibnamefont {Giustino}}, \bibinfo {author} {\bibfnamefont
  {T.}~\bibnamefont {Gorni}}, \bibinfo {author} {\bibfnamefont
  {J.}~\bibnamefont {Jia}}, \bibinfo {author} {\bibfnamefont {M.}~\bibnamefont
  {Kawamura}}, \bibinfo {author} {\bibfnamefont {H.-Y.}\ \bibnamefont {Ko}},
  \bibinfo {author} {\bibfnamefont {A.}~\bibnamefont {Kokalj}}, \bibinfo
  {author} {\bibfnamefont {E.}~\bibnamefont {Küçükbenli}}, \bibinfo {author}
  {\bibfnamefont {M.}~\bibnamefont {Lazzeri}}, \bibinfo {author} {\bibfnamefont
  {M.}~\bibnamefont {Marsili}}, \bibinfo {author} {\bibfnamefont
  {N.}~\bibnamefont {Marzari}}, \bibinfo {author} {\bibfnamefont
  {F.}~\bibnamefont {Mauri}}, \bibinfo {author} {\bibfnamefont {N.~L.}\
  \bibnamefont {Nguyen}}, \bibinfo {author} {\bibfnamefont {H.-V.}\
  \bibnamefont {Nguyen}}, \bibinfo {author} {\bibfnamefont {A.~O.}\
  \bibnamefont {de-la Roza}}, \bibinfo {author} {\bibfnamefont
  {L.}~\bibnamefont {Paulatto}}, \bibinfo {author} {\bibfnamefont
  {S.}~\bibnamefont {Poncé}}, \bibinfo {author} {\bibfnamefont
  {D.}~\bibnamefont {Rocca}}, \bibinfo {author} {\bibfnamefont
  {R.}~\bibnamefont {Sabatini}}, \bibinfo {author} {\bibfnamefont
  {B.}~\bibnamefont {Santra}}, \bibinfo {author} {\bibfnamefont
  {M.}~\bibnamefont {Schlipf}}, \bibinfo {author} {\bibfnamefont {A.~P.}\
  \bibnamefont {Seitsonen}}, \bibinfo {author} {\bibfnamefont {A.}~\bibnamefont
  {Smogunov}}, \bibinfo {author} {\bibfnamefont {I.}~\bibnamefont {Timrov}},
  \bibinfo {author} {\bibfnamefont {T.}~\bibnamefont {Thonhauser}}, \bibinfo
  {author} {\bibfnamefont {P.}~\bibnamefont {Umari}}, \bibinfo {author}
  {\bibfnamefont {N.}~\bibnamefont {Vast}}, \bibinfo {author} {\bibfnamefont
  {X.}~\bibnamefont {Wu}},\ and\ \bibinfo {author} {\bibfnamefont
  {S.}~\bibnamefont {Baroni}},\ }\bibfield  {title} {\bibinfo {title} {Advanced
  capabilities for materials modelling with quantum espresso},\ }\href
  {https://doi.org/10.1088/1361-648X/aa8f79} {\bibfield  {journal} {\bibinfo
  {journal} {Journal of Physics: Condensed Matter}\ }\textbf {\bibinfo {volume}
  {29}},\ \bibinfo {pages} {465901} (\bibinfo {year} {2017})}\BibitemShut
  {NoStop}%
\bibitem [{\citenamefont {Giannozzi}\ \emph {et~al.}(2009)\citenamefont
  {Giannozzi}, \citenamefont {Baroni}, \citenamefont {Bonini}, \citenamefont
  {Calandra}, \citenamefont {Car}, \citenamefont {Cavazzoni}, \citenamefont
  {Ceresoli}, \citenamefont {Chiarotti}, \citenamefont {Cococcioni},
  \citenamefont {Dabo}, \citenamefont {Corso}, \citenamefont {de~Gironcoli},
  \citenamefont {Fabris}, \citenamefont {Fratesi}, \citenamefont {Gebauer},
  \citenamefont {Gerstmann}, \citenamefont {Gougoussis}, \citenamefont
  {Kokalj}, \citenamefont {Lazzeri}, \citenamefont {Martin-Samos},
  \citenamefont {Marzari}, \citenamefont {Mauri}, \citenamefont {Mazzarello},
  \citenamefont {Paolini}, \citenamefont {Pasquarello}, \citenamefont
  {Paulatto}, \citenamefont {Sbraccia}, \citenamefont {Scandolo}, \citenamefont
  {Sclauzero}, \citenamefont {Seitsonen}, \citenamefont {Smogunov},
  \citenamefont {Umari},\ and\ \citenamefont {Wentzcovitch}}]{QE3}%
  \BibitemOpen
  \bibfield  {author} {\bibinfo {author} {\bibfnamefont {P.}~\bibnamefont
  {Giannozzi}}, \bibinfo {author} {\bibfnamefont {S.}~\bibnamefont {Baroni}},
  \bibinfo {author} {\bibfnamefont {N.}~\bibnamefont {Bonini}}, \bibinfo
  {author} {\bibfnamefont {M.}~\bibnamefont {Calandra}}, \bibinfo {author}
  {\bibfnamefont {R.}~\bibnamefont {Car}}, \bibinfo {author} {\bibfnamefont
  {C.}~\bibnamefont {Cavazzoni}}, \bibinfo {author} {\bibfnamefont
  {D.}~\bibnamefont {Ceresoli}}, \bibinfo {author} {\bibfnamefont {G.~L.}\
  \bibnamefont {Chiarotti}}, \bibinfo {author} {\bibfnamefont {M.}~\bibnamefont
  {Cococcioni}}, \bibinfo {author} {\bibfnamefont {I.}~\bibnamefont {Dabo}},
  \bibinfo {author} {\bibfnamefont {A.~D.}\ \bibnamefont {Corso}}, \bibinfo
  {author} {\bibfnamefont {S.}~\bibnamefont {de~Gironcoli}}, \bibinfo {author}
  {\bibfnamefont {S.}~\bibnamefont {Fabris}}, \bibinfo {author} {\bibfnamefont
  {G.}~\bibnamefont {Fratesi}}, \bibinfo {author} {\bibfnamefont
  {R.}~\bibnamefont {Gebauer}}, \bibinfo {author} {\bibfnamefont
  {U.}~\bibnamefont {Gerstmann}}, \bibinfo {author} {\bibfnamefont
  {C.}~\bibnamefont {Gougoussis}}, \bibinfo {author} {\bibfnamefont
  {A.}~\bibnamefont {Kokalj}}, \bibinfo {author} {\bibfnamefont
  {M.}~\bibnamefont {Lazzeri}}, \bibinfo {author} {\bibfnamefont
  {L.}~\bibnamefont {Martin-Samos}}, \bibinfo {author} {\bibfnamefont
  {N.}~\bibnamefont {Marzari}}, \bibinfo {author} {\bibfnamefont
  {F.}~\bibnamefont {Mauri}}, \bibinfo {author} {\bibfnamefont
  {R.}~\bibnamefont {Mazzarello}}, \bibinfo {author} {\bibfnamefont
  {S.}~\bibnamefont {Paolini}}, \bibinfo {author} {\bibfnamefont
  {A.}~\bibnamefont {Pasquarello}}, \bibinfo {author} {\bibfnamefont
  {L.}~\bibnamefont {Paulatto}}, \bibinfo {author} {\bibfnamefont
  {C.}~\bibnamefont {Sbraccia}}, \bibinfo {author} {\bibfnamefont
  {S.}~\bibnamefont {Scandolo}}, \bibinfo {author} {\bibfnamefont
  {G.}~\bibnamefont {Sclauzero}}, \bibinfo {author} {\bibfnamefont {A.~P.}\
  \bibnamefont {Seitsonen}}, \bibinfo {author} {\bibfnamefont {A.}~\bibnamefont
  {Smogunov}}, \bibinfo {author} {\bibfnamefont {P.}~\bibnamefont {Umari}},\
  and\ \bibinfo {author} {\bibfnamefont {R.~M.}\ \bibnamefont {Wentzcovitch}},\
  }\bibfield  {title} {\bibinfo {title} {Quantum espresso: a modular and
  open-source software project for quantum simulations of materials},\ }\href
  {https://doi.org/10.1088/0953-8984/21/39/395502} {\bibfield  {journal}
  {\bibinfo  {journal} {Journal of Physics: Condensed Matter}\ }\textbf
  {\bibinfo {volume} {21}},\ \bibinfo {pages} {395502} (\bibinfo {year}
  {2009})}\BibitemShut {NoStop}%
\bibitem [{\citenamefont {Marzari}\ \emph {et~al.}(1999)\citenamefont
  {Marzari}, \citenamefont {Vanderbilt}, \citenamefont {De~Vita},\ and\
  \citenamefont {Payne}}]{mv}%
  \BibitemOpen
  \bibfield  {author} {\bibinfo {author} {\bibfnamefont {N.}~\bibnamefont
  {Marzari}}, \bibinfo {author} {\bibfnamefont {D.}~\bibnamefont {Vanderbilt}},
  \bibinfo {author} {\bibfnamefont {A.}~\bibnamefont {De~Vita}},\ and\ \bibinfo
  {author} {\bibfnamefont {M.~C.}\ \bibnamefont {Payne}},\ }\bibfield  {title}
  {\bibinfo {title} {Thermal contraction and disordering of the al(110)
  surface},\ }\href {https://doi.org/10.1103/PhysRevLett.82.3296} {\bibfield
  {journal} {\bibinfo  {journal} {Phys. Rev. Lett.}\ }\textbf {\bibinfo
  {volume} {82}},\ \bibinfo {pages} {3296} (\bibinfo {year}
  {1999})}\BibitemShut {NoStop}%
\bibitem [{\citenamefont {Yamashita}\ \emph {et~al.}(2021)\citenamefont
  {Yamashita}, \citenamefont {Kanehira}, \citenamefont {Sato}, \citenamefont
  {Kino}, \citenamefont {Terayama}, \citenamefont {Sawahata}, \citenamefont
  {Sato}, \citenamefont {Utsuno}, \citenamefont {Tsuda}, \citenamefont
  {Miyake},\ and\ \citenamefont {Oguchi}}]{cryspy}%
  \BibitemOpen
  \bibfield  {author} {\bibinfo {author} {\bibfnamefont {T.}~\bibnamefont
  {Yamashita}}, \bibinfo {author} {\bibfnamefont {S.}~\bibnamefont {Kanehira}},
  \bibinfo {author} {\bibfnamefont {N.}~\bibnamefont {Sato}}, \bibinfo {author}
  {\bibfnamefont {H.}~\bibnamefont {Kino}}, \bibinfo {author} {\bibfnamefont
  {K.}~\bibnamefont {Terayama}}, \bibinfo {author} {\bibfnamefont
  {H.}~\bibnamefont {Sawahata}}, \bibinfo {author} {\bibfnamefont
  {T.}~\bibnamefont {Sato}}, \bibinfo {author} {\bibfnamefont {F.}~\bibnamefont
  {Utsuno}}, \bibinfo {author} {\bibfnamefont {K.}~\bibnamefont {Tsuda}},
  \bibinfo {author} {\bibfnamefont {T.}~\bibnamefont {Miyake}},\ and\ \bibinfo
  {author} {\bibfnamefont {T.}~\bibnamefont {Oguchi}},\ }\bibfield  {title}
  {\bibinfo {title} {Cryspy: a crystal structure prediction tool accelerated by
  machine learning},\ }\href {https://doi.org/10.1080/27660400.2021.1943171}
  {\bibfield  {journal} {\bibinfo  {journal} {Science and Technology of
  Advanced Materials: Methods}\ }\textbf {\bibinfo {volume} {1}},\ \bibinfo
  {pages} {87} (\bibinfo {year} {2021})},\ \Eprint
  {https://arxiv.org/abs/https://doi.org/10.1080/27660400.2021.1943171}
  {https://doi.org/10.1080/27660400.2021.1943171} \BibitemShut {NoStop}%
\bibitem [{\citenamefont {Yang}\ \emph {et~al.}(2024)\citenamefont {Yang},
  \citenamefont {Hu}, \citenamefont {Zhou}, \citenamefont {Liu}, \citenamefont
  {Shi}, \citenamefont {Li}, \citenamefont {Li}, \citenamefont {Chen},
  \citenamefont {Chen}, \citenamefont {Zeni}, \citenamefont {Horton},
  \citenamefont {Pinsler}, \citenamefont {Fowler}, \citenamefont {Zügner},
  \citenamefont {Xie}, \citenamefont {Smith}, \citenamefont {Sun},
  \citenamefont {Wang}, \citenamefont {Kong}, \citenamefont {Liu},
  \citenamefont {Hao},\ and\ \citenamefont {Lu}}]{yang2024mattersim}%
  \BibitemOpen
  \bibfield  {author} {\bibinfo {author} {\bibfnamefont {H.}~\bibnamefont
  {Yang}}, \bibinfo {author} {\bibfnamefont {C.}~\bibnamefont {Hu}}, \bibinfo
  {author} {\bibfnamefont {Y.}~\bibnamefont {Zhou}}, \bibinfo {author}
  {\bibfnamefont {X.}~\bibnamefont {Liu}}, \bibinfo {author} {\bibfnamefont
  {Y.}~\bibnamefont {Shi}}, \bibinfo {author} {\bibfnamefont {J.}~\bibnamefont
  {Li}}, \bibinfo {author} {\bibfnamefont {G.}~\bibnamefont {Li}}, \bibinfo
  {author} {\bibfnamefont {Z.}~\bibnamefont {Chen}}, \bibinfo {author}
  {\bibfnamefont {S.}~\bibnamefont {Chen}}, \bibinfo {author} {\bibfnamefont
  {C.}~\bibnamefont {Zeni}}, \bibinfo {author} {\bibfnamefont {M.}~\bibnamefont
  {Horton}}, \bibinfo {author} {\bibfnamefont {R.}~\bibnamefont {Pinsler}},
  \bibinfo {author} {\bibfnamefont {A.}~\bibnamefont {Fowler}}, \bibinfo
  {author} {\bibfnamefont {D.}~\bibnamefont {Zügner}}, \bibinfo {author}
  {\bibfnamefont {T.}~\bibnamefont {Xie}}, \bibinfo {author} {\bibfnamefont
  {J.}~\bibnamefont {Smith}}, \bibinfo {author} {\bibfnamefont
  {L.}~\bibnamefont {Sun}}, \bibinfo {author} {\bibfnamefont {Q.}~\bibnamefont
  {Wang}}, \bibinfo {author} {\bibfnamefont {L.}~\bibnamefont {Kong}}, \bibinfo
  {author} {\bibfnamefont {C.}~\bibnamefont {Liu}}, \bibinfo {author}
  {\bibfnamefont {H.}~\bibnamefont {Hao}},\ and\ \bibinfo {author}
  {\bibfnamefont {Z.}~\bibnamefont {Lu}},\ }\bibfield  {title} {\bibinfo
  {title} {Mattersim: A deep learning atomistic model across elements,
  temperatures and pressures},\ }\href {https://arxiv.org/abs/2405.04967}
  {\bibfield  {journal} {\bibinfo  {journal} {arXiv preprint arXiv:2405.04967}\
  } (\bibinfo {year} {2024})},\ \Eprint {https://arxiv.org/abs/2405.04967}
  {arXiv:2405.04967 [cond-mat.mtrl-sci]} \BibitemShut {NoStop}%
\bibitem [{\citenamefont {Pellegrini}\ \emph {et~al.}(2022)\citenamefont
  {Pellegrini}, \citenamefont {Heid},\ and\ \citenamefont
  {Sanna}}]{Pellegrini_2022}%
  \BibitemOpen
  \bibfield  {author} {\bibinfo {author} {\bibfnamefont {C.}~\bibnamefont
  {Pellegrini}}, \bibinfo {author} {\bibfnamefont {R.}~\bibnamefont {Heid}},\
  and\ \bibinfo {author} {\bibfnamefont {A.}~\bibnamefont {Sanna}},\ }\bibfield
   {title} {\bibinfo {title} {Eliashberg theory with ab-initio coulomb
  interactions: a minimal numerical scheme applied to layered
  superconductors},\ }\href {https://doi.org/10.1088/2515-7639/ac6041}
  {\bibfield  {journal} {\bibinfo  {journal} {Journal of Physics: Materials}\
  }\textbf {\bibinfo {volume} {5}},\ \bibinfo {pages} {024007} (\bibinfo {year}
  {2022})}\BibitemShut {NoStop}%
\bibitem [{\citenamefont {Kogler}\ \emph {et~al.}(2025)\citenamefont {Kogler},
  \citenamefont {Spath}, \citenamefont {Lucrezi}, \citenamefont {Mori},
  \citenamefont {Zhu}, \citenamefont {Li}, \citenamefont {Margine},\ and\
  \citenamefont {Heil}}]{IsoME}%
  \BibitemOpen
  \bibfield  {author} {\bibinfo {author} {\bibfnamefont {E.}~\bibnamefont
  {Kogler}}, \bibinfo {author} {\bibfnamefont {D.}~\bibnamefont {Spath}},
  \bibinfo {author} {\bibfnamefont {R.}~\bibnamefont {Lucrezi}}, \bibinfo
  {author} {\bibfnamefont {H.}~\bibnamefont {Mori}}, \bibinfo {author}
  {\bibfnamefont {Z.}~\bibnamefont {Zhu}}, \bibinfo {author} {\bibfnamefont
  {Z.}~\bibnamefont {Li}}, \bibinfo {author} {\bibfnamefont {E.~R.}\
  \bibnamefont {Margine}},\ and\ \bibinfo {author} {\bibfnamefont
  {C.}~\bibnamefont {Heil}},\ }\bibfield  {title} {\bibinfo {title} {Isome:
  Streamlining high-precision eliashberg calculations},\ }\href
  {https://doi.org/https://doi.org/10.1016/j.cpc.2025.109720} {\bibfield
  {journal} {\bibinfo  {journal} {Computer Physics Communications}\ }\textbf
  {\bibinfo {volume} {315}},\ \bibinfo {pages} {109720} (\bibinfo {year}
  {2025})}\BibitemShut {NoStop}%
\bibitem [{\citenamefont {Marzari}\ and\ \citenamefont
  {Vanderbilt}(1997)}]{MLWF}%
  \BibitemOpen
  \bibfield  {author} {\bibinfo {author} {\bibfnamefont {N.}~\bibnamefont
  {Marzari}}\ and\ \bibinfo {author} {\bibfnamefont {D.}~\bibnamefont
  {Vanderbilt}},\ }\bibfield  {title} {\bibinfo {title} {Maximally localized
  generalized wannier functions for composite energy bands},\ }\href
  {https://doi.org/10.1103/PhysRevB.56.12847} {\bibfield  {journal} {\bibinfo
  {journal} {Phys. Rev. B}\ }\textbf {\bibinfo {volume} {56}},\ \bibinfo
  {pages} {12847} (\bibinfo {year} {1997})}\BibitemShut {NoStop}%
\bibitem [{\citenamefont {Marzari}\ \emph {et~al.}(2012)\citenamefont
  {Marzari}, \citenamefont {Mostofi}, \citenamefont {Yates}, \citenamefont
  {Souza},\ and\ \citenamefont {Vanderbilt}}]{MLWF2}%
  \BibitemOpen
  \bibfield  {author} {\bibinfo {author} {\bibfnamefont {N.}~\bibnamefont
  {Marzari}}, \bibinfo {author} {\bibfnamefont {A.~A.}\ \bibnamefont
  {Mostofi}}, \bibinfo {author} {\bibfnamefont {J.~R.}\ \bibnamefont {Yates}},
  \bibinfo {author} {\bibfnamefont {I.}~\bibnamefont {Souza}},\ and\ \bibinfo
  {author} {\bibfnamefont {D.}~\bibnamefont {Vanderbilt}},\ }\bibfield  {title}
  {\bibinfo {title} {Maximally localized wannier functions: Theory and
  applications},\ }\href {https://doi.org/10.1103/RevModPhys.84.1419}
  {\bibfield  {journal} {\bibinfo  {journal} {Rev. Mod. Phys.}\ }\textbf
  {\bibinfo {volume} {84}},\ \bibinfo {pages} {1419} (\bibinfo {year}
  {2012})}\BibitemShut {NoStop}%
\bibitem [{\citenamefont {Pizzi}\ \emph {et~al.}(2020)\citenamefont {Pizzi},
  \citenamefont {Vitale}, \citenamefont {Arita}, \citenamefont {Blügel},
  \citenamefont {Freimuth}, \citenamefont {Géranton}, \citenamefont
  {Gibertini}, \citenamefont {Gresch}, \citenamefont {Johnson}, \citenamefont
  {Koretsune}, \citenamefont {Ibañez-Azpiroz}, \citenamefont {Lee},
  \citenamefont {Lihm}, \citenamefont {Marchand}, \citenamefont {Marrazzo},
  \citenamefont {Mokrousov}, \citenamefont {Mustafa}, \citenamefont {Nohara},
  \citenamefont {Nomura}, \citenamefont {Paulatto}, \citenamefont {Poncé},
  \citenamefont {Ponweiser}, \citenamefont {Qiao}, \citenamefont {Thöle},
  \citenamefont {Tsirkin}, \citenamefont {Wierzbowska}, \citenamefont
  {Marzari}, \citenamefont {Vanderbilt}, \citenamefont {Souza}, \citenamefont
  {Mostofi},\ and\ \citenamefont {Yates}}]{wan90}%
  \BibitemOpen
  \bibfield  {author} {\bibinfo {author} {\bibfnamefont {G.}~\bibnamefont
  {Pizzi}}, \bibinfo {author} {\bibfnamefont {V.}~\bibnamefont {Vitale}},
  \bibinfo {author} {\bibfnamefont {R.}~\bibnamefont {Arita}}, \bibinfo
  {author} {\bibfnamefont {S.}~\bibnamefont {Blügel}}, \bibinfo {author}
  {\bibfnamefont {F.}~\bibnamefont {Freimuth}}, \bibinfo {author}
  {\bibfnamefont {G.}~\bibnamefont {Géranton}}, \bibinfo {author}
  {\bibfnamefont {M.}~\bibnamefont {Gibertini}}, \bibinfo {author}
  {\bibfnamefont {D.}~\bibnamefont {Gresch}}, \bibinfo {author} {\bibfnamefont
  {C.}~\bibnamefont {Johnson}}, \bibinfo {author} {\bibfnamefont
  {T.}~\bibnamefont {Koretsune}}, \bibinfo {author} {\bibfnamefont
  {J.}~\bibnamefont {Ibañez-Azpiroz}}, \bibinfo {author} {\bibfnamefont
  {H.}~\bibnamefont {Lee}}, \bibinfo {author} {\bibfnamefont {J.-M.}\
  \bibnamefont {Lihm}}, \bibinfo {author} {\bibfnamefont {D.}~\bibnamefont
  {Marchand}}, \bibinfo {author} {\bibfnamefont {A.}~\bibnamefont {Marrazzo}},
  \bibinfo {author} {\bibfnamefont {Y.}~\bibnamefont {Mokrousov}}, \bibinfo
  {author} {\bibfnamefont {J.~I.}\ \bibnamefont {Mustafa}}, \bibinfo {author}
  {\bibfnamefont {Y.}~\bibnamefont {Nohara}}, \bibinfo {author} {\bibfnamefont
  {Y.}~\bibnamefont {Nomura}}, \bibinfo {author} {\bibfnamefont
  {L.}~\bibnamefont {Paulatto}}, \bibinfo {author} {\bibfnamefont
  {S.}~\bibnamefont {Poncé}}, \bibinfo {author} {\bibfnamefont
  {T.}~\bibnamefont {Ponweiser}}, \bibinfo {author} {\bibfnamefont
  {J.}~\bibnamefont {Qiao}}, \bibinfo {author} {\bibfnamefont {F.}~\bibnamefont
  {Thöle}}, \bibinfo {author} {\bibfnamefont {S.~S.}\ \bibnamefont {Tsirkin}},
  \bibinfo {author} {\bibfnamefont {M.}~\bibnamefont {Wierzbowska}}, \bibinfo
  {author} {\bibfnamefont {N.}~\bibnamefont {Marzari}}, \bibinfo {author}
  {\bibfnamefont {D.}~\bibnamefont {Vanderbilt}}, \bibinfo {author}
  {\bibfnamefont {I.}~\bibnamefont {Souza}}, \bibinfo {author} {\bibfnamefont
  {A.~A.}\ \bibnamefont {Mostofi}},\ and\ \bibinfo {author} {\bibfnamefont
  {J.~R.}\ \bibnamefont {Yates}},\ }\bibfield  {title} {\bibinfo {title}
  {Wannier90 as a community code: new features and applications},\ }\href
  {https://doi.org/10.1088/1361-648X/ab51ff} {\bibfield  {journal} {\bibinfo
  {journal} {Journal of Physics: Condensed Matter}\ }\textbf {\bibinfo {volume}
  {32}},\ \bibinfo {pages} {165902} (\bibinfo {year} {2020})}\BibitemShut
  {NoStop}%
\bibitem [{\citenamefont {Souza}\ \emph {et~al.}(2001)\citenamefont {Souza},
  \citenamefont {Marzari},\ and\ \citenamefont {Vanderbilt}}]{disentangle}%
  \BibitemOpen
  \bibfield  {author} {\bibinfo {author} {\bibfnamefont {I.}~\bibnamefont
  {Souza}}, \bibinfo {author} {\bibfnamefont {N.}~\bibnamefont {Marzari}},\
  and\ \bibinfo {author} {\bibfnamefont {D.}~\bibnamefont {Vanderbilt}},\
  }\bibfield  {title} {\bibinfo {title} {Maximally localized wannier functions
  for entangled energy bands},\ }\href
  {https://doi.org/10.1103/PhysRevB.65.035109} {\bibfield  {journal} {\bibinfo
  {journal} {Phys. Rev. B}\ }\textbf {\bibinfo {volume} {65}},\ \bibinfo
  {pages} {035109} (\bibinfo {year} {2001})}\BibitemShut {NoStop}%
\bibitem [{\citenamefont {Yates}\ \emph {et~al.}(2007)\citenamefont {Yates},
  \citenamefont {Wang}, \citenamefont {Vanderbilt},\ and\ \citenamefont
  {Souza}}]{adaptive_smearing}%
  \BibitemOpen
  \bibfield  {author} {\bibinfo {author} {\bibfnamefont {J.~R.}\ \bibnamefont
  {Yates}}, \bibinfo {author} {\bibfnamefont {X.}~\bibnamefont {Wang}},
  \bibinfo {author} {\bibfnamefont {D.}~\bibnamefont {Vanderbilt}},\ and\
  \bibinfo {author} {\bibfnamefont {I.}~\bibnamefont {Souza}},\ }\bibfield
  {title} {\bibinfo {title} {Spectral and fermi surface properties from wannier
  interpolation},\ }\href {https://doi.org/10.1103/PhysRevB.75.195121}
  {\bibfield  {journal} {\bibinfo  {journal} {Phys. Rev. B}\ }\textbf {\bibinfo
  {volume} {75}},\ \bibinfo {pages} {195121} (\bibinfo {year}
  {2007})}\BibitemShut {NoStop}%
\bibitem [{\citenamefont {Momma}\ and\ \citenamefont {Izumi}(2011)}]{VESTA}%
  \BibitemOpen
  \bibfield  {author} {\bibinfo {author} {\bibfnamefont {K.}~\bibnamefont
  {Momma}}\ and\ \bibinfo {author} {\bibfnamefont {F.}~\bibnamefont {Izumi}},\
  }\bibfield  {title} {\bibinfo {title} {{{\it VESTA3} for three-dimensional
  visualization of crystal, volumetric and morphology data}},\ }\href
  {https://doi.org/10.1107/S0021889811038970} {\bibfield  {journal} {\bibinfo
  {journal} {Journal of Applied Crystallography}\ }\textbf {\bibinfo {volume}
  {44}},\ \bibinfo {pages} {1272} (\bibinfo {year} {2011})}\BibitemShut
  {NoStop}%
\bibitem [{\citenamefont {Mills}\ \emph {et~al.}(1970)\citenamefont {Mills},
  \citenamefont {Maradudin},\ and\ \citenamefont
  {Burstein}}]{MILLS1970TheoryRaman}%
  \BibitemOpen
  \bibfield  {author} {\bibinfo {author} {\bibfnamefont {D.}~\bibnamefont
  {Mills}}, \bibinfo {author} {\bibfnamefont {A.}~\bibnamefont {Maradudin}},\
  and\ \bibinfo {author} {\bibfnamefont {E.}~\bibnamefont {Burstein}},\
  }\bibfield  {title} {\bibinfo {title} {Theory of the raman effect in
  metals},\ }\href
  {https://doi.org/https://doi.org/10.1016/0003-4916(70)90028-X} {\bibfield
  {journal} {\bibinfo  {journal} {Annals of Physics}\ }\textbf {\bibinfo
  {volume} {56}},\ \bibinfo {pages} {504} (\bibinfo {year} {1970})}\BibitemShut
  {NoStop}%
\bibitem [{\citenamefont {Monacelli}\ \emph
  {et~al.}(2021{\natexlab{b}})\citenamefont {Monacelli}, \citenamefont {Errea},
  \citenamefont {Calandra},\ and\ \citenamefont
  {Mauri}}]{monacelli_black_2021}%
  \BibitemOpen
  \bibfield  {author} {\bibinfo {author} {\bibfnamefont {L.}~\bibnamefont
  {Monacelli}}, \bibinfo {author} {\bibfnamefont {I.}~\bibnamefont {Errea}},
  \bibinfo {author} {\bibfnamefont {M.}~\bibnamefont {Calandra}},\ and\
  \bibinfo {author} {\bibfnamefont {F.}~\bibnamefont {Mauri}},\ }\bibfield
  {title} {\bibinfo {title} {Black metal hydrogen above 360 {GPa} driven by
  proton quantum fluctuations},\ }\href
  {https://doi.org/10.1038/s41567-020-1009-3} {\bibfield  {journal} {\bibinfo
  {journal} {Nature Physics}\ }\textbf {\bibinfo {volume} {17}},\ \bibinfo
  {pages} {63} (\bibinfo {year} {2021}{\natexlab{b}})}\BibitemShut {NoStop}%
\bibitem [{\citenamefont {Belli}\ \emph {et~al.}(2021)\citenamefont {Belli},
  \citenamefont {Novoa}, \citenamefont {Contreras-Garc{\'i}a},\ and\
  \citenamefont {Errea}}]{networkingvalue}%
  \BibitemOpen
  \bibfield  {author} {\bibinfo {author} {\bibfnamefont {F.}~\bibnamefont
  {Belli}}, \bibinfo {author} {\bibfnamefont {T.}~\bibnamefont {Novoa}},
  \bibinfo {author} {\bibfnamefont {J.}~\bibnamefont {Contreras-Garc{\'i}a}},\
  and\ \bibinfo {author} {\bibfnamefont {I.}~\bibnamefont {Errea}},\ }\bibfield
   {title} {\bibinfo {title} {Strong correlation between electronic bonding
  network and critical temperature in hydrogen-based superconductors},\ }\href
  {https://doi.org/10.1038/s41467-021-25687-0} {\bibfield  {journal} {\bibinfo
  {journal} {Nature Communications}\ }\textbf {\bibinfo {volume} {12}},\
  \bibinfo {pages} {5381} (\bibinfo {year} {2021})}\BibitemShut {NoStop}%
\bibitem [{\citenamefont {Allen}\ and\ \citenamefont
  {Mitrović}(1983)}]{Allen_Mitrovic}%
  \BibitemOpen
  \bibfield  {author} {\bibinfo {author} {\bibfnamefont {P.~B.}\ \bibnamefont
  {Allen}}\ and\ \bibinfo {author} {\bibfnamefont {B.}~\bibnamefont
  {Mitrović}},\ }\bibfield  {title} {\bibinfo {title} {Theory of
  superconducting tc}\ }(\bibinfo  {publisher} {Academic Press},\ \bibinfo
  {year} {1983})\ pp.\ \bibinfo {pages} {1--92}\BibitemShut {NoStop}%
\bibitem [{\citenamefont {Margine}\ and\ \citenamefont {Giustino}(2013)}]{AME}%
  \BibitemOpen
  \bibfield  {author} {\bibinfo {author} {\bibfnamefont {E.~R.}\ \bibnamefont
  {Margine}}\ and\ \bibinfo {author} {\bibfnamefont {F.}~\bibnamefont
  {Giustino}},\ }\bibfield  {title} {\bibinfo {title} {Anisotropic
  migdal-eliashberg theory using wannier functions},\ }\href
  {https://doi.org/10.1103/PhysRevB.87.024505} {\bibfield  {journal} {\bibinfo
  {journal} {Phys. Rev. B}\ }\textbf {\bibinfo {volume} {87}},\ \bibinfo
  {pages} {024505} (\bibinfo {year} {2013})}\BibitemShut {NoStop}%
\bibitem [{\citenamefont {Belli}\ \emph {et~al.}(2025)\citenamefont {Belli},
  \citenamefont {Zurek},\ and\ \citenamefont {Errea}}]{belli}%
  \BibitemOpen
  \bibfield  {author} {\bibinfo {author} {\bibfnamefont {F.}~\bibnamefont
  {Belli}}, \bibinfo {author} {\bibfnamefont {E.}~\bibnamefont {Zurek}},\ and\
  \bibinfo {author} {\bibfnamefont {I.}~\bibnamefont {Errea}},\ }\href
  {https://arxiv.org/abs/2501.14420} {\bibinfo {title} {A chemical bonding
  based descriptor for predicting the impact of quantum nuclear and anharmonic
  effects on hydrogen-based superconductors}} (\bibinfo {year} {2025}),\
  \Eprint {https://arxiv.org/abs/2501.14420} {arXiv:2501.14420
  [cond-mat.supr-con]} \BibitemShut {NoStop}%
\bibitem [{\citenamefont {Errea}\ \emph {et~al.}(2012)\citenamefont {Errea},
  \citenamefont {Rousseau}, \citenamefont {Eiguren},\ and\ \citenamefont
  {Bergara}}]{IonBruno}%
  \BibitemOpen
  \bibfield  {author} {\bibinfo {author} {\bibfnamefont {I.}~\bibnamefont
  {Errea}}, \bibinfo {author} {\bibfnamefont {B.}~\bibnamefont {Rousseau}},
  \bibinfo {author} {\bibfnamefont {A.}~\bibnamefont {Eiguren}},\ and\ \bibinfo
  {author} {\bibfnamefont {A.}~\bibnamefont {Bergara}},\ }\bibfield  {title}
  {\bibinfo {title} {Optical properties of calcium under pressure from
  first-principles calculations},\ }\href
  {https://doi.org/10.1103/PhysRevB.86.085106} {\bibfield  {journal} {\bibinfo
  {journal} {Phys. Rev. B}\ }\textbf {\bibinfo {volume} {86}},\ \bibinfo
  {pages} {085106} (\bibinfo {year} {2012})}\BibitemShut {NoStop}%
\bibitem [{\citenamefont {Iba\~nez Azpiroz}\ \emph {et~al.}(2014)\citenamefont
  {Iba\~nez Azpiroz}, \citenamefont {Rousseau}, \citenamefont {Eiguren},\ and\
  \citenamefont {Bergara}}]{JulenBruno}%
  \BibitemOpen
  \bibfield  {author} {\bibinfo {author} {\bibfnamefont {J.}~\bibnamefont
  {Iba\~nez Azpiroz}}, \bibinfo {author} {\bibfnamefont {B.}~\bibnamefont
  {Rousseau}}, \bibinfo {author} {\bibfnamefont {A.}~\bibnamefont {Eiguren}},\
  and\ \bibinfo {author} {\bibfnamefont {A.}~\bibnamefont {Bergara}},\
  }\bibfield  {title} {\bibinfo {title} {Ab initio analysis of plasmon
  dispersion in sodium under pressure},\ }\href
  {https://doi.org/10.1103/PhysRevB.89.085102} {\bibfield  {journal} {\bibinfo
  {journal} {Phys. Rev. B}\ }\textbf {\bibinfo {volume} {89}},\ \bibinfo
  {pages} {085102} (\bibinfo {year} {2014})}\BibitemShut {NoStop}%
\end{thebibliography}%

\onecolumngrid
\renewcommand{\figurename}{Supplementary Figure}
\renewcommand{\tablename}{Supplementary Table}
\setcounter{figure}{0}
\setcounter{table}{0}
\setcounter{section}{0}
\newpage
\newpage
\newpage

\section*{Supplementary material for: Superconductivity in RbH$_{12}$ at low pressures: an \emph{ab initio} study}

\section{Superconducting critical temperature of RbH$_{12}$ at 50 GPa}

\begin{figure}[h!]
	\centering
	\includegraphics[height=0.18\textheight]{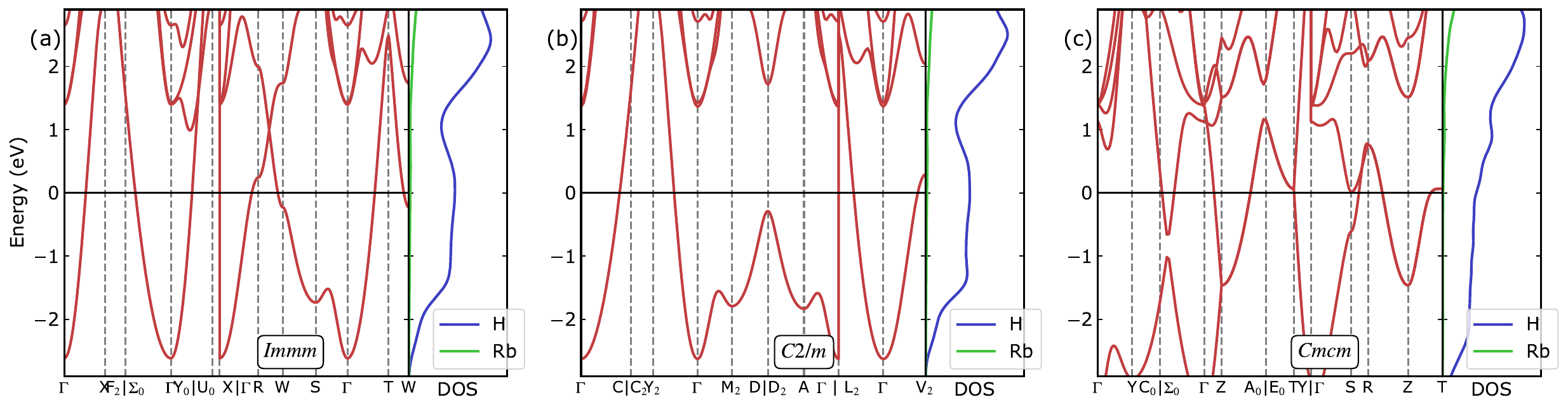}
	\includegraphics[height=0.18\textheight]{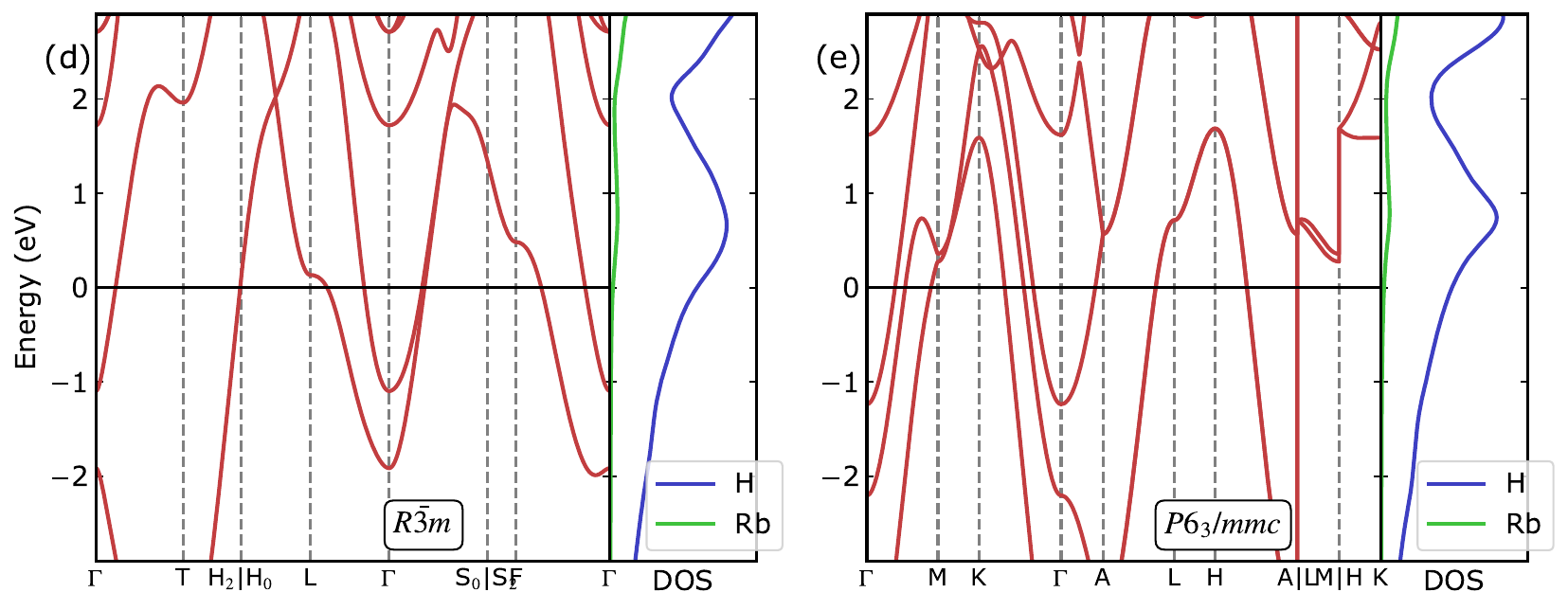}
	\caption{Electronic band structure and density of states for RbH$_{12}$ at 50 GPa in (a) $Immm$, (b) $C/2m$, (c) $Cmcm$, (d) $R\bar{3}m$ and (e) $P6_3/mmc$ phase.}
	\label{fig:el_bands_50GPa}
\end{figure}

In Supp. Figure~\ref{fig:el_bands_50GPa} we show electronic band structure and density of states for RbH$_{12}$ for different phases at 50 GPa. All phases are metalic with large majority of states having H character.

\begin{figure}[h!]
	\centering
	\includegraphics[height=0.18\textheight]{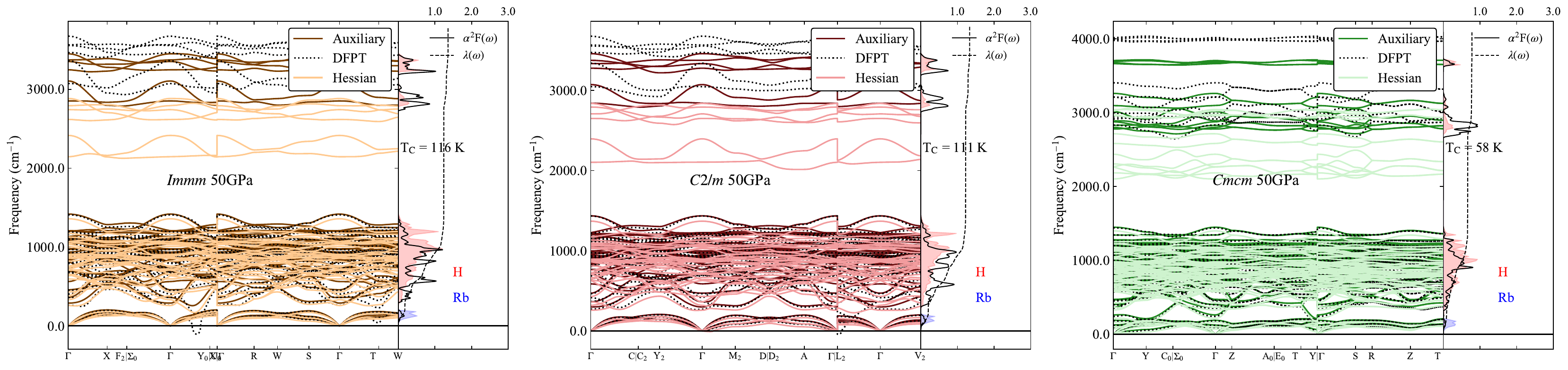}
	\includegraphics[height=0.18\textheight]{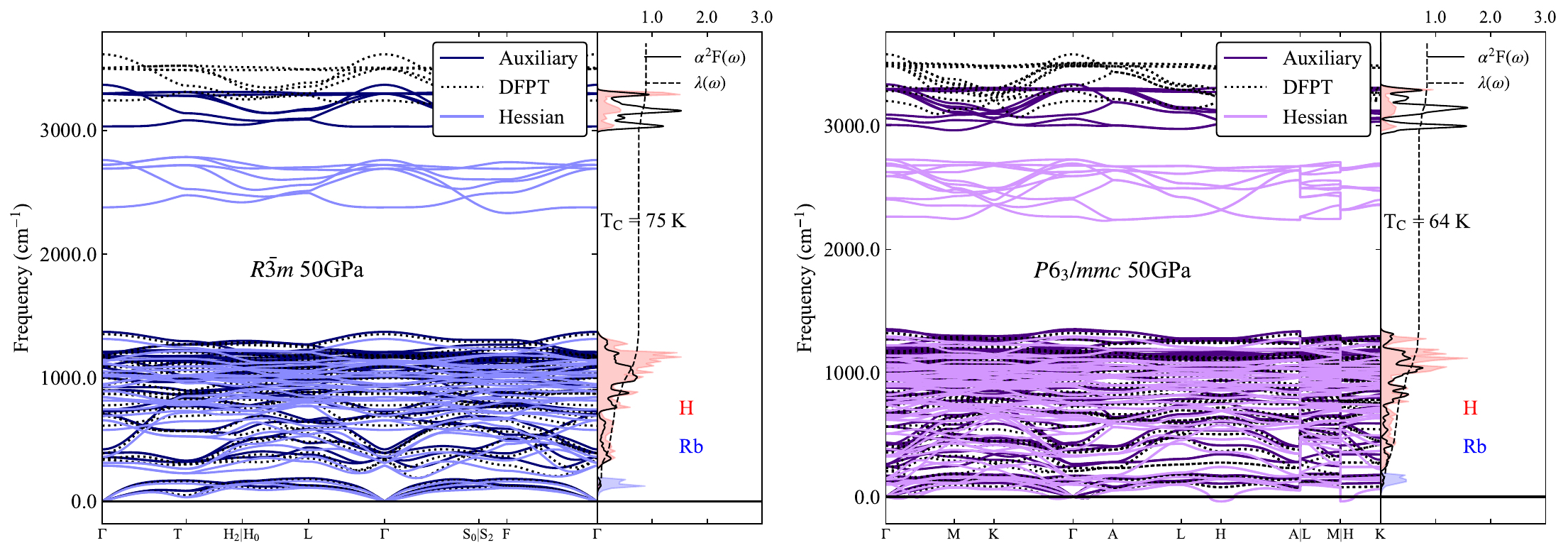}
	\caption{Phonon band structure and Eliashberg spectral function for RbH$_{12}$ at 50 GPa in (a) $Immm$, (b) $C/2m$, (c) $Cmcm$, (d) $R\bar{3}m$ and (e) $P6_3/mmc$ phase. The shaded regions in the side panel represent atom resolved auxiliary SSCHA phonon density of states.}
	\label{fig:a2fs_50gpa}
\end{figure}

Supp. Figure~\ref{fig:a2fs_50gpa} reports phonon band structure of all phases of RbH$_{12}$ at 50 GPa. Most of the phases are dynamically stable even in harmonic approximation. Some instabilities that can be seen are consequence of the fact that these DFPT phonon dispersions were calculated for structures that are minima of total free energy and not BOES. In all cases there is a strong softening of high-frequency optical modes. The side plots show calculated Eliashberg spectral function $\alpha ^2F$ and phonon density of states calculated for SSCHA auxiliary force constants. Most of the electron-phonon coupling comes from hydrogen modes. The electrons are interacting more strongly with the high frequency optical modes, however this does not have a large influence on $\lambda$.

\begin{table}[]
	\begin{tabular}{|c|c|c|c|c|c|}
		\hline
		50 GPa                                         & $Immm$ & $C2/m$ & $Cmcm$ & $R\bar{3}m$ & $P6_3/mmc$ \\ \hline
		$\lambda$                                      & 1.36   & 1.33   & 0.78   & 0.89        & 0.84       \\ \hline
		$\omega _\textrm{log}$ (K)                     & 879    & 848    & 1103   & 1134        & 1078       \\ \hline
		$\omega_2$ (K)                                 & 1600   & 1587   & 1930   & 2065        & 2141       \\ \hline
		$\int \textrm{d}\omega \alpha ^2F(\omega)$ (K) & 818    & 792    & 591    & 702         & 679        \\ \hline
		DOS(E$_\textrm{F}$) (1/meV/atom)               & 20.8  & 20.7   & 16.1   & 19.6        & 19.4       \\ \hline
		T$^\textrm{AD} _\textrm{C}$ (K)                & 95     & 90     & 59     & 72          & 53         \\ \hline
		T$^\textrm{E} _\textrm{C}$ (K)                 & 111    & 106    & 67     & 78          & 59         \\ \hline
	\end{tabular}
	\caption{Superconducting properties of different phases of RbH$_{12}$ at 50 GPa. The definitions of each property is given in the text below. T$^\textrm{AD} _\textrm{C}$ and T$^\textrm{E} _\textrm{C}$ are estimates of superconducting critical temperature using Allen-Dynes formula and solution of Migdal-Eliashberg equations~\ref{eq:mustar} with $\mu^* = 0.118$.}
	\label{tb:1}
\end{table}

\section{Interpolation issues of Hessian of free energy}

To further substantiate our interpretation that the imaginary frequencies of total free energy Hessian in Fig. 3 of the main text are due to the interpolation issues, we have performed additional calculations beyond those presented in the main part.

Given the high computational cost associated with evaluating the required number of atomic forces, energies, and stresses, we employed the recently developed MatterSim machine learning potential~\cite{yang2024mattersim}. Specifically, we fine-tuned the foundation model (“mattersim-v1.0.0-1M”) using our previously computed DFT data for smaller supercells to accurately reproduce the relevant portions of the potential energy surface.

\begin{figure}
	\centering
	\includegraphics[width=0.85\linewidth]{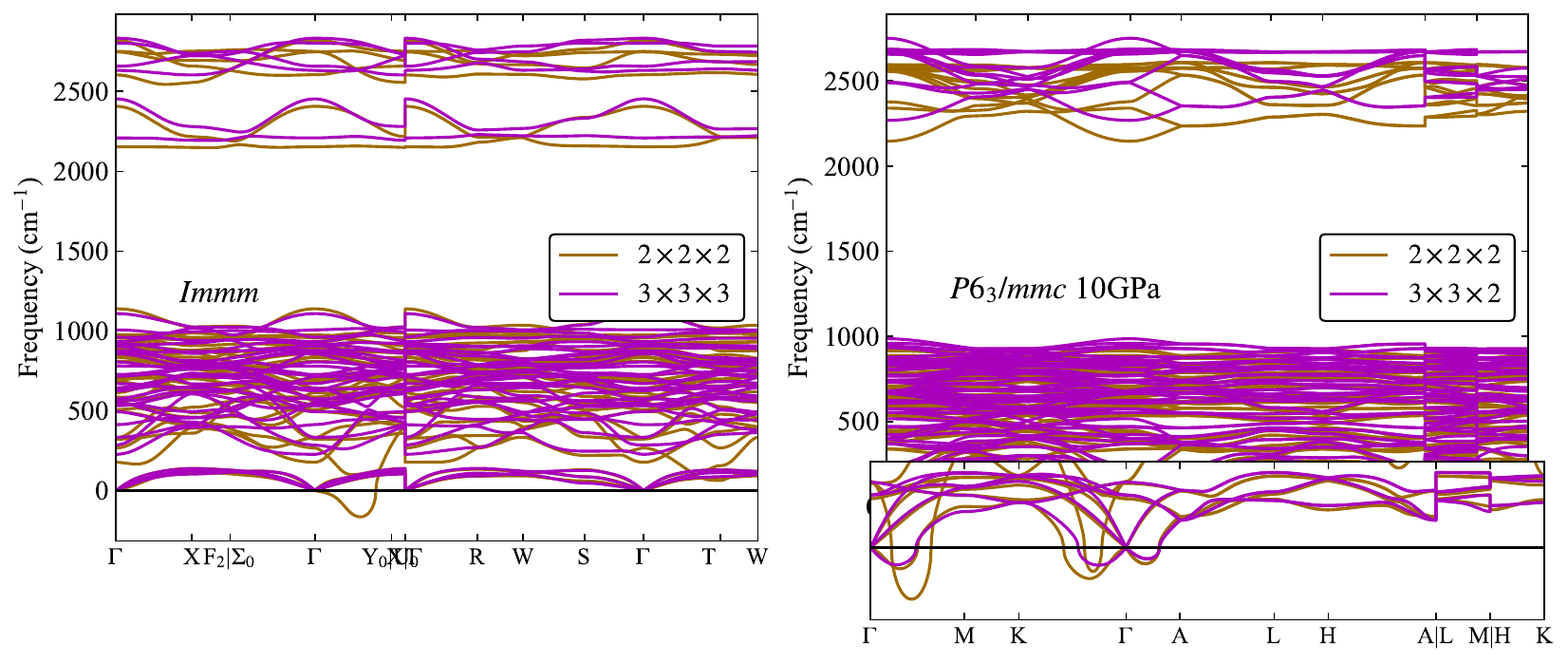}
	\caption{Phonon band structures of $Immm$ (10 GPa) and $P6_3/mmc$ phases calculated from the Hessian of the total free energy.}
	\label{fig:phonon_larger_grid}
\end{figure}

We then carried out SSCHA relaxations for the $Immm$ phase of RbH$_{12}$ at 25 GPa using a $3\times3\times3$ supercell (see Fig.~\ref{fig:phonon_larger_grid}). In this case, all Hessian phonon frequencies—both directly computed and interpolated—are positive. This confirms our earlier interpretation that the negative frequencies reported previously originated from interpolation artifacts rather than true dynamical instabilities.

We also repeated the analysis for the $P6_3/mmc$ phase. Since this structure contains two formula units per primitive cell, the largest feasible supercell corresponds to $3\times3\times2$ (468 atoms). In this system, we observe a clear renormalization and hardening of the Hessian phonon frequencies. The residual small imaginary modes are confined near the $\Gamma$ point, consistent with remaining interpolation inaccuracies rather than genuine instabilities. We further improved SSCHA algorithm in order to allow us to use larger cells and calculated free energy Hessian of this phase for $3\times 3\times 3$ and $4\times 4\times 2$ supercells. In both cases, the
interpolated phonon dispersions still exhibit irregular behavior in the vicinity of the $\Gamma$ point. However, importantly, none of the commensurate $\mathbf{q}$-points show dynamical instability. This observation supports our interpretation that the apparent instabilities near $\Gamma$ originate from interpolation artifacts rather than true physical soft modes. Extending the calculation to a $4\times 4\times 3$ supercell (1248 atoms) is currently computationally prohibitive, even with the improved algorithm.

These additional results strengthen our conclusion that the previously reported imaginary frequencies were numerical artifacts and that both $Immm$ and $P6_3/mmc$ phases are dynamically stable at the corresponding pressures. 

\section{Migdal-Eliashberg equations with the explict Coulomb interaction}

To estimate the superconducting critical temperature, we employed the isotropic approximation of the Migdal–Eliashberg equations. The set of equations to be solved is~\cite{Pellegrini_2022,IsoME}:
\begin{align*}
Z (i\omega_n) &= 1 + \frac{k_BT}{N_F\omega_n}\sum _{m}\int \textrm{d}\varepsilon' N(\varepsilon')\frac{\omega_{m}Z(i\omega _{m})}{\Theta(\varepsilon',i\omega _{m})} \lambda(i\omega_n,i\omega_m), \\
\chi (i\omega_n) &= -\frac{k_BT}{N_F}\int\textrm{d}\varepsilon' N(\varepsilon ')\sum _m\frac{\varepsilon ' - \varepsilon _F + \chi(i\omega_m)}{\Theta(\varepsilon',i\omega_m)} \lambda(i\omega_n,i\omega_m), \\
\phi ^{ph}(i\omega_n) &= \frac{k_BT}{N_F}\int\textrm{d}\varepsilon' N(\varepsilon ')\sum _m \frac{\phi(\varepsilon',i\omega _m)}{\Theta(\varepsilon',i\omega _m)}\lambda(i\omega_n,i\omega_m), \numberthis \label{eq:ME} \\
\phi ^c(\varepsilon) &= -k_BT\int\textrm{d}\varepsilon' N(\varepsilon ')W(\varepsilon,\varepsilon')\sum _m \frac{\phi(\varepsilon',i\omega _m)} {\Theta(\varepsilon',i\omega _m)}, \\
N_e &= \int\textrm{d}\varepsilon' N(\varepsilon ')\left(1 - k_BT\sum _m\frac{\varepsilon ' - \varepsilon _F + \chi(i\omega_m)}{\Theta(\varepsilon',i\omega_m)}\right).
\end{align*}
In these equations, $i\omega_n$ denotes the $n$-th Matsubara frequency, $Z(i\omega_n)$ is the electronic mass-renormalization function, and $\chi(i\omega_n)$ represents the Fermi-level shift. The denominator entering all equations is defined as:
\[
\Theta(\varepsilon, i\omega_n) = [\omega_n Z(i\omega_n)]^2 + [\varepsilon - \varepsilon_F + \chi(i\omega_n)]^2 + \phi^2(\varepsilon,i\omega_n).
\]
The superconducting order parameter $\phi(\varepsilon,i\omega_n)$ is written as the sum of a Coulombic part $\phi^c(\varepsilon)$, which depends on the electronic energy, and a phononic part $\phi^{ph}(i\omega_n)$, which depends on the Matsubara frequency. This separation reflects the fact that the electron–phonon interaction is only relevant near the Fermi surface, while the Coulomb term extends over a broader energy range.  

To make the numerical solution tractable, we introduce a cutoff $\omega_c$ in the sum over Matsubara frequencies. For $|\omega_n| > \omega_c$, we assume $Z(i\omega_n) - 1 = \phi^{ph}(i\omega_n) = \chi(i\omega_n) = 0$, which allows analytic evaluation of the Matsubara sums in the equations for $\phi^c$ and $N_e$. The inclusion of $N_e$ guarantees electron-number conservation by adjusting the Fermi level self-consistently at each iteration.

The electron-phonon coupling is described via $\lambda(i\omega _n, i\omega _m)$:
\begin{align*}
\lambda (i\omega _n, i\omega _m) = 2\int _0 ^{\infty}\frac{\Omega}{\Omega ^2 + (\omega _n - \omega _m)^2}\alpha ^2F (\Omega)\mathrm{d}\Omega .
\end{align*}
Elishberg spectral function $\alpha ^2F (\Omega)$ is given by:
\begin{align*}
\alpha ^2F (\Omega) = \frac{1}{N_\mathbf{q}}\sum_{a,b,\mathbf{q}} \Delta ^{ab} (\mathbf{q})B^{ab}(\mathbf{q},\Omega),
\end{align*}
where $a,b$ are compact Cartesian and atom indices, $\mathbf{q}$ is the phonon wave vector, $B^{ab}(\mathbf{q},\Omega)$ is the phonon spectral function. Phonon spectral function $B^{ab}(\mathbf{q},\Omega)$ is calculated using SSCHA auxiliary phonon frequencies and eigenvectors~\cite{H}:

\begin{align*}
B^{ab}(\mathbf{q},\Omega) = \sum _{\nu} e^a_{\nu}(\mathbf{q})e^{b*}_{\nu}(\mathbf{q})\delta(\Omega - \omega_\nu(\mathbf{q})).
\end{align*}

$\Delta ^{ab}  (\mathbf{q})$ is the electron-phonon matrix element averaged over the Fermi surface:
\begin{align*}
\Delta ^{ab}  (\mathbf{q}) =& \frac{1}{N _FN_\mathbf{k}}\sum _\mathbf{k} d^a _{n\mathbf{k},n'\mathbf{k}+\mathbf{q}}d^b _{n'\mathbf{k}+\mathbf{q}, n\mathbf{k}}\times \\
&\times\delta (\varepsilon _{n\mathbf{k}} - \varepsilon _F)\delta (\varepsilon _{n'\mathbf{k} + \mathbf{q}} - \varepsilon _F).
\end{align*}
Here $\mathbf{k}$ is the wave vector of electronic state, $N _F$ is the electronic density of states at the Fermi level, $\varepsilon _{n\mathbf{k}}$ is the energy of electronic state of wave vector $\mathbf{k}$ and band index $n$, and $\varepsilon _F$ is the Fermi level as before. $d^a _{n\mathbf{k},n'\mathbf{k}+\mathbf{q}}$ is the deformation potential: $d^a _{n\mathbf{k},n'\mathbf{k}+\mathbf{q}} = \langle n\mathbf{k}|\frac{\delta V}{\delta u^a(\mathbf{q})}|n'\mathbf{k}+\mathbf{q}\rangle$.

The energy averaged Coulomb interaction (it is purely real since we assume it is taken for $\Omega = 0$, see below) is given by:
\begin{align*}
W(\varepsilon',\varepsilon) = \frac{1}{N(\varepsilon')N(\varepsilon)}\sum_{n\mathbf{k},n'\mathbf{q}}W_{n\mathbf{k},n'\mathbf{k} +\mathbf{q}}(\Omega = 0)\delta(\varepsilon - \varepsilon_{n\mathbf{k}})\delta(\varepsilon' - \varepsilon_{n'\mathbf{k} + \mathbf{q}}).
\end{align*}
Coulomb matrix elements $W_{n\mathbf{k},n'\mathbf{k} + \mathbf{q}}(\Omega)$ are then given by:
\begin{align*}
W_{n\mathbf{k},n'\mathbf{k} + \mathbf{q}}(\Omega) = 4\pi\sum_{\mathbf{G}, \mathbf{G}'}\epsilon ^{-1}_{\mathbf{G},\mathbf{G'}} (\mathbf{q}, \Omega) \frac{\langle n'\mathbf{k}+\mathbf{q}|e^{-i(\mathbf{q} + \mathbf{G})\mathbf{r}}|n\mathbf{k}\rangle\langle n\mathbf{k}|e^{-i(\mathbf{q} + \mathbf{G}')\mathbf{r}}|n'\mathbf{k}+\mathbf{q}\rangle}{|\mathbf{q} + \mathbf{G}||\mathbf{q} + \mathbf{G}'|}.
\end{align*}
Here dielectric function $\epsilon ^{-1}_{\mathbf{G},\mathbf{G'}}(\mathbf{q}, \Omega)$ is calculated in the RPA approximation:
\begin{align*}
\epsilon ^{-1}_{\mathbf{G},\mathbf{G'}}(\mathbf{q}, \Omega) &= \delta _{\mathbf{G},\mathbf{G}'} + \frac{4\pi}{|\mathbf{q} + \mathbf{G}|^2}\chi_{\mathbf{G},\mathbf{G}'}(\mathbf{q}, \Omega), \\
\chi &= (1 - \chi ^0K)^{-1}\chi^0, 
\end{align*}
where:
\begin{align*}
K_{\mathbf{G},\mathbf{G'}}(\mathbf{q}) &= \frac{4\pi}{|\mathbf{q} + \mathbf{G}|^2}\delta_{\mathbf{G},\mathbf{G}'} \quad \textrm{and} \\
\chi ^0 _{\mathbf{G},\mathbf{G'}}(\mathbf{q}, \Omega) = \frac{1}{V}\sum _{\mathbf{k},n,n'} \langle n'\mathbf{k}+\mathbf{q}|e^{-i(\mathbf{q} + \mathbf{G})\mathbf{r}}|n\mathbf{k}\rangle&\langle n\mathbf{k}|e^{-i(\mathbf{q} + \mathbf{G}')\mathbf{r}}|n'\mathbf{k}+\mathbf{q}\rangle\frac{f_{n\mathbf{k}} - f_{n'\mathbf{k} + \mathbf{q}}}{\varepsilon _{n\mathbf{k}} - \varepsilon_{n'\mathbf{k}+\mathbf{q}} + \Omega + i\eta}.
\end{align*}
The denominator in $\chi ^0 _{\mathbf{G},\mathbf{G'}}(\mathbf{q}, \Omega)$ is expressed using identity $\lim_{\eta\rightarrow 0}\frac{1}{x \pm i\eta} = \mp i\pi\delta(x) + \mathcal{P}(\frac{1}{x})$.

To calculate the dielectric function and Coulomb matrix elements, we employed a workflow based on maximally localized Wannier functions~\cite{MLWF, MLWF2, IonBruno,JulenBruno}. First, we wannierized the electronic states of the $Immm$ phase of RbH$_{12}$ using an $8\times8\times8$ coarse $\mathbf{k}$-point grid~\cite{wan90}. The imaginary part of $\chi^0_{\mathbf{G},\mathbf{G}'}(\mathbf{q},\Omega)$ was computed by approximating the Dirac $\delta$-function with a Gaussian of adaptive width~\cite{adaptive_smearing}, chosen according to the local density of the $\mathbf{k}$-point sampling involved in the summation. The real part of $\chi^0_{\mathbf{G},\mathbf{G}'}(\mathbf{q},\Omega)$ was then obtained via the Kramers–Kronig relation. For this reason, it is necessary to perform Wannierization over a wide energy window (here, $\pm30$~eV around the Fermi level). In this particular case, the disentanglement procedure~\cite{disentangle} yielded 58 Wannier functions from a total of 100 calculated DFT bands. When solving the Migdal–Eliashberg equations [Eq.~\ref{eq:ME}], we subsequently restricted the energy window to the range of $-20$ to $20$~eV.

Electronic energies on dense grids were evaluated using Wannier interpolation, which is computationally efficient. However, direct Fourier interpolation of the plane-wave matrix elements 
\[
\rho_{n,n'}(\mathbf{k},\mathbf{q},\mathbf{G}) = \langle n'\mathbf{k}+\mathbf{q} | e^{-i(\mathbf{q}+\mathbf{G})\mathbf{r}} | n\mathbf{k} \rangle
\]
is not possible because these quantities are not periodic with respect to $\mathbf{q}$ and exhibit phase discontinuities arising from the arbitrary phases of DFT wavefunctions~\cite{IonBruno}. To overcome this, we rotated the matrix elements to the Wannier basis,
\[
\rho^W_{mm'}(\mathbf{k},\mathbf{q},\mathbf{G}) = \sum_{n,n'} V^*_{mn}\, \rho_{nn'}(\mathbf{k},\mathbf{q},\mathbf{G})\, V_{n'm'},
\]
where they become smooth functions of $\mathbf{q}$ and can thus be interpolated efficiently using a linear interpolation scheme. This interpolation is first applied over the $\mathbf{q}$-dependence of the matrix elements, and subsequently, the Wannier-interpolated quantities are used to evaluate the matrix elements on a dense $\mathbf{k}$-point grid.

\begin{figure}
	\centering
	\includegraphics[width=0.85\linewidth]{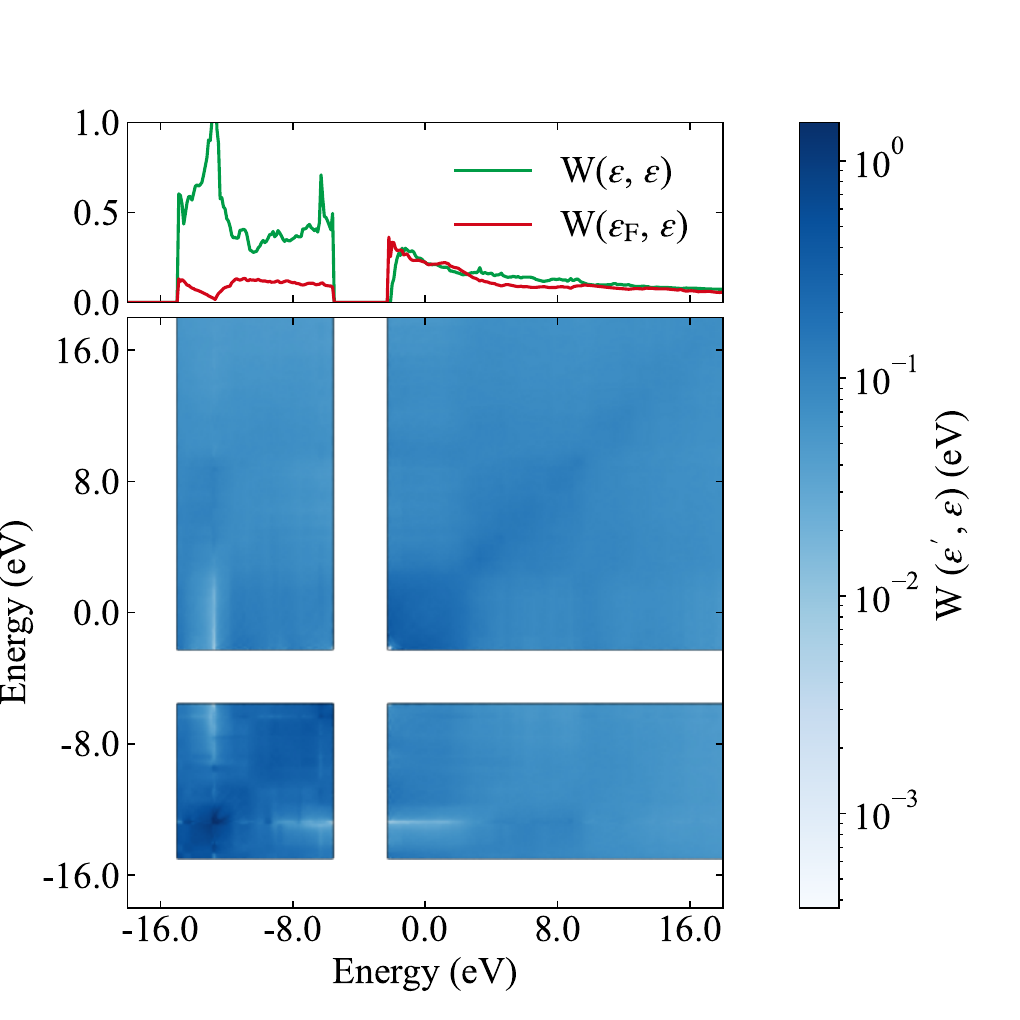}
	\caption{Electronic energy averaged Coulomb interaction in $Immm$ phase of RbH$_{12}$ at 25 GPa. The top panel shows a line plot for the $W(\varepsilon', \varepsilon')$ and $W(\varepsilon _F, \varepsilon)$.}
	\label{fig:wee}
\end{figure}

The energy-averaged Coulomb potential $W(\varepsilon',\varepsilon)$ was calculated on a $36\times36\times36$ $\mathbf{k}$-point and $6\times6\times6$ $\mathbf{q}$-point grid, including up to the ninth shell of $\mathbf{G}$-vectors (37 in total). The result is shown in Fig.~\ref{fig:wee}. As expected, $W(\varepsilon,\varepsilon)$ exhibits a pronounced peak for the core states and gradually decreases for higher-energy conduction states. The largest values are observed along and near the diagonal $\varepsilon' = \varepsilon$.

There are several further approximations one can make to the set of equation ~\ref{eq:ME}. Firstly, one could assume that the density of states $N(\varepsilon)$ is constant $N(\varepsilon _F)$ for Matsubara frequency dependent quantities. This means that $\chi(i\omega _n) = 0$ and thus we do not need equations for $\chi$ and $N_e$. Next the integration over energy in equations for $Z$ and $\phi ^{ph}$ can be done analytically:
\begin{align*}
Z(i\omega _n) &= 1 +\pi\frac{k_BT}{\omega _n}\sum_m \frac{\omega_mZ(i\omega_m)}{\sqrt{\Theta(\varepsilon _F,i\omega_m)}}\lambda(i\omega_n,i\omega_m), \\
\phi ^{ph}(i\omega_n) &=\pi k_BT\sum_m \frac{\phi(\varepsilon _F,i\omega_m)}{\sqrt{\Theta(\varepsilon_F,i\omega_m)}}\lambda(i\omega_n,i\omega_m). \numberthis \label{eq:constdos}
\end{align*}
Equation for the Coulombic part of the order parameter stays the same. 

Finally, the Coulomb contribution can be incorporated into a single effective parameter, $\mu^*$, which modifies the equation for $\phi^{ph}$ as:
\begin{align*}
\phi^{ph}(i\omega _n) = \pi k_BT\sum _m \frac{\phi^{ph}(i\omega_m)}{\sqrt{\Theta(\varepsilon _F,i\omega_m)}}\left(\lambda(i\omega_n,i\omega_m) - \mu^*\right).~\numberthis \label{eq:mustar}
\end{align*}
The parameter $\mu^*$ is commonly interpreted as the renormalized, Fermi-surface-averaged Coulomb interaction, $\mu = N(\varepsilon_F)W(\varepsilon _F,\varepsilon _F)$. We determine $\mu^*$ by solving the constant-DOS Eliashberg equations with the full Coulomb interaction [Eq.~\ref{eq:constdos}] and matching the resulting critical temperature to that obtained from Eq.~\ref{eq:mustar}. For a Matsubara frequency cutoff of $10\omega_D$ ($\omega _D$ is the largest phonon frequency), this procedure yields $\mu^* = 0.118$. In the case of constant DOS approximation~\ref{eq:constdos}, the critical temperature converges rapidly with increasing Matsubara frequency cutoff, and identical values are obtained for cutoffs between $4$ and $12$~$\omega_D$.

\begin{figure}
	\centering
	\includegraphics[width=0.65\linewidth]{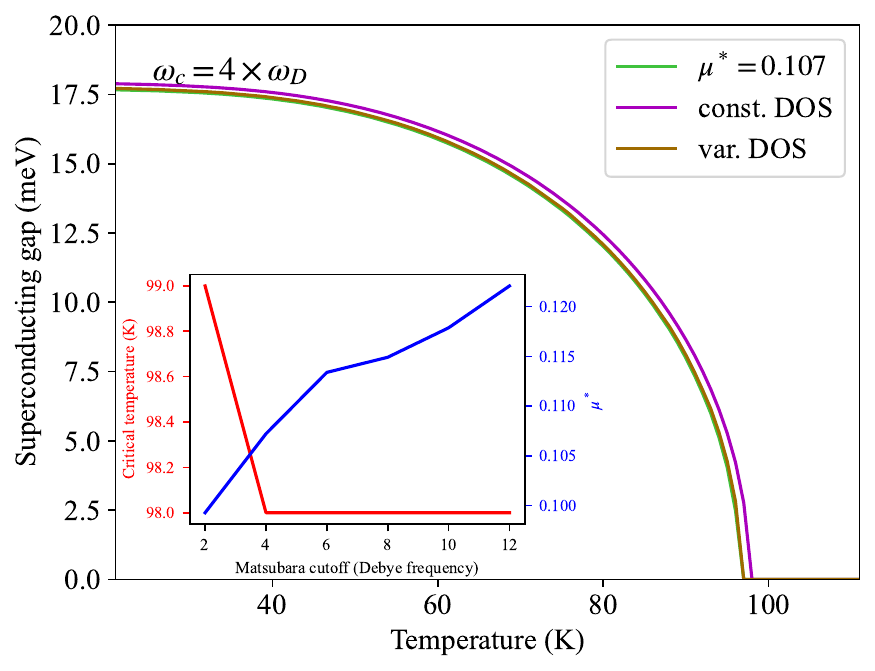}
	\caption{Superconducting gap as a function of temperature calculated with three different approaches: variable density of states DOS Eq.~\ref{eq:ME}, constant density of states Eq.~\ref{eq:constdos} and $\mu^*$ approximation with $\mu^*=0.107$ Eq.~\ref{eq:mustar}. This value is smaller than the one mentioned in the text since here the Matsubara cutoff is 4$\omega _D$. All calculations were performed with Matsubara cutoff of 4$\times$ Debye frequency. The inset shows the dependence of superconducting critical temperature in constant DOS approximation and $\mu^*$ with the Matsubara cutoff. The colors of $y$ axis labels corresponds to the line.}
	\label{fig:delta}
\end{figure}

The superconducting gap obtained within the three approximations discussed above is shown in Fig.~\ref{fig:delta}. No discernible difference is observed between the full solution [Eq.~\ref{eq:ME}] and the constant-DOS approximation [Eq.~\ref{eq:constdos}], reflecting the nearly flat electronic density of states near the Fermi level.

\end{document}